\documentclass[fleqn,usenatbib]{mnras}

\usepackage{newtxtext,newtxmath}
\usepackage[T1]{fontenc}
\usepackage{aecompl}
\usepackage{pdflscape}
\usepackage[a4paper]{geometry}

\usepackage{graphicx}	% Including figure files
\usepackage{amsmath}	% Advanced maths commands

\usepackage{amssymb}	% Extra maths symbols
\usepackage{mathtools}
\usepackage{subcaption}
\usepackage{multirow}
\usepackage{longtable}
\hypersetup{
    colorlinks=true,
    linkcolor=blue,
    filecolor=magenta,      
    urlcolor=cyan,
    pdftitle={Overleaf Example},
   pdfpagemode=FullScreen,
    }
\newcommand{\kms}{km\,s$^{-1}$}

\newcommand{\prin}{$^{\prime\prime}$}

\newcommand{\por}{$\times$}
\newcommand{\hi}{H{\sc i}}

\newcommand{\aprox}{${\sim}$}
\newcommand{\msolar}{$\mathrm{M}_{\odot}$}
\newcommand{\mhi}{$M_{\mathrm {HI}}$}
\newcommand{\flux}{erg\,s$^{-1}$\,cm$^{-2}$\,\AA$^{-1}$}
\newcommand{\angstrom}{\textup{\AA}}

\defcitealias{Lopez-Gutierrez+22}{LG22}
\title[FUV spiral galaxies]
{Tracing Galaxy Evolution in infalling galaxies of Abell 496: From Starburst to Quenching}

\author[M. M. Lopez-Gutierrez et al.]{
M. M. López-Gutiérrez,$^{1,2}$\thanks {E-mail: mmlopez@kias.re.kr}
H. Bravo-Alfaro$^{1}$,
P. T. Rahna$^{3}$,
G. A. Mamon$^{4}$, 
Y. L. Jaff\'e$^{5}$,
\newauthor
L. F. Madrigal-Ayala$^{1}$ \&
E. Acosta-Espinoza$^{1}$
\\
$^{1}$Departamento de Astronom\'ia, Universidad de Guanajuato, 36000, Mexico\\ 
$^{2}$Korea Institute for Advanced Study, 85 Hoegi-ro, Dongdaemun-gu, Seoul 02455, Republic of Korea\\
$^{3}$Centro de Estudios de Física del Cosmos de Aragón (CEFCA), Plaza San Juan 1, E-44001 Teruel, Spain\\
$^{4}$Institut d’Astrophysique de Paris (UMR 7095: CNRS \& Sorbonne Université), 98 bis Bd Arago, 75014 Paris, France\\
$^{5}$Departamento de F\'isica, Universidad T\'ecnica Federico Santa Mar\'ia, Av. Espa\~na 1680, Valpara\'iso, Chile,
}

\date{Accepted XXX. Received YYY; in original form ZZZ}

\pubyear{2025}

\begin{document}
\label{firstpage}
\pagerange{\pageref{firstpage}--\pageref{lastpage}}
\maketitle

\begin{abstract}
During the fall of late-type galaxies into clusters, they can experiment a variety of evolutionary mechanisms according their local environment. Consequently, studying the UV emission and the cold gas of late-type galaxies provide key insights in the evolution of short-lived starburst and galaxy quenching. In this work, we conduted a study of two 28$’$ fields observed with UVIT-AstroSat in the central region of the Abell cluster A496 ($z=0.033$), including \hi\, data from NRAO VLA. We reported 22 cluster members detected in FUV; all of them are detected in \hi, or have upper limits for the \hi-mass. We find our FUV detected galaxies generally have higher specific star formation rates than other star forming galaxies. Most of the FUV galaxies with masses above 10$^9$\msolar\,and showing high sSFR have no close neighbors, pointing at RPS as the dominant mechanism affecting them. In contrast, most of the low-mass FUV objects present at least one companion, suggesting that tidal interactions also play an important role in the triggering of infalling galaxies. Combining the FUV-SFR with the \hi\ properties of the observed galaxies in A496 we identify an evolutionary sequence consisting of five stages: (1) {\it Pre-triggering}, (2) {\it Initial SF-triggering}, (3) {\it Peak of star-formation}, (4) {\it SF-fading}, and (5) {\it SF-quenching}.
During this path, normal gas-rich objects reach a gas-deficiency phase with SFR well below the main sequence. This process, prior to becoming a full passive galaxy, can be accomplished within a few 10$^{8}$\,yr.

\end{abstract}


%%%%%%%%%%%%%%%%%%%%%%%%%%%%%%%%%%%%%%%%%%%%%%%%%%

%%%%%%%%%%%%%%%%% BODY OF PAPER %%%%%%%%%%%%%%%%%%

\section{Introduction}

Our understanding of galaxy evolution in the last billion years has made big progress in the last decades, mainly due to large, multifrequency observational surveys, as well as hydrodynamic cosmological simulations. Explaining the scenario observed at $z=0$ and understanding the gas cycle leading the star formation (SF) history requires considering the initial conditions at the time of galaxy formation and the environmental effects. In the modern paradigm of galaxy evolution, where galaxies can not be considered as isolated systems, {\it nature} and {\it nurture} do not always exclude each other. Both internal and external mechanisms might have a strong influence on the cycle of gas and, as a consequence, on the triggering/quenching of star formation. Although the environment dominates the evolution of low-mass (satellite) galaxies, under extreme conditions, it might shape massive galaxies as well. In the case of spirals, the gas cycle is behind the morphological transformation from blue, star-forming, to red passive objects \citetext{\citealp{Dressler+80, Goto+03, Peng+10}; \citealp*{DeLucia+19}}.
However, a number of key questions remain open. For instance, what are the most important physical mechanisms at play on 
those galaxies being shaped in different environments? 
How does environment affect galaxies in combination with internal mechanisms regulated by their stellar mass?
Is the so-called pre-processing as important as the cluster environment in the systematic transformation of infalling spirals? 
\citep[e.g.][]{Boselli&Gavazzi+06, Boselli+14b, Poggianti+09, Rhee+20, Donnari+21}.   

The well-known morphology-density relation \citep{Oemler+74, Dressler+80, Goto+03} provides clear evidence of the important role played by the local environment in galaxy evolution during the last Gyr. However, as mentioned above, it remains to disentangle what mechanisms are at play, in what density regions they take place, and which galaxies are more vulnerable.  The hydrodynamic mechanisms, such as the ram pressure stripping \citep[RPS,][]{Gunn&Gott+72}, involve the interaction between the intracluster medium (ICM) and the interstellar medium (ISM) of individual galaxies. They are known to disrupt the gas component of the ISM, mainly the \hi, but in some cases the molecular gas as well \citetext{\citealp{Bravo-Alfaro+01}; \citealp*{Kenney+04}; \citealp{Chung+09, Scott+10, Wang+21}}. Other hydrodynamic processes are thermal evaporation \citep{Cowie&Songaila+77}, viscous stripping \citep{Nulsen+82}, and starvation \citep*{Larson+80}. The intensity of the corresponding effects depends on several parameters: the local density of the ICM; the velocity relative to the ICM;  the type of orbit, and the galaxy inclination relative to the movement vector.  On the other hand, tidal interactions are produced by gravitational interactions, either galaxy--cluster \citep{Byrd-Valtonen+90} or a diversity of galaxy--galaxy encounters. Some of these interactions include galaxy harassment \citep{Moore+96} and galaxy--galaxy interactions, going from indirect fly-byes to major mergers \citep{Barnes-Hernquist+92}.  Tidal interactions, unlike hydrodynamic processes, can disrupt both, the stellar disk and the gas components of individual galaxies \citep{Lin+23}; the resulting disruptions are well traced by the old-stellar population and observed in NIR \citep{Venkatapathy+17}. These mechanisms are known to cut the supply of external gas and produce the so-called galaxy starvation \citep[][and references therein]{Cortese+21}.

One major question in galaxy evolution during the last Gyr concerns the transformation of normal, star-forming late-types into red, passive objects. It is known that for satellite galaxies such process might include a star-burst stage that can be produced by RPS as well as by mergers \citep{Cortese+21}. During this starburst phase, the gas is exhausted due to its transformation into stars and, for fast infalling cluster galaxies, by ICM-sweeping as well. Therefore, it remains to distinguish whether hydrodynamical or gravitational mechanisms dominate the cold gas perturbations. It is worth noting that in certain cases both types of process might work in parallel. Disruptions of the ISM in the form of asymmetries, as well as conversion of atomic into molecular gas, are observed in mergers as well as in ram-pressure stripped galaxies  
\citetext{\citealp*{Kenney+15}; \citealp{George+18, Moretti+20, Roberts&Parker+20, Boselli+21}}. The more advanced stages of gas exhaustion are traced by large tails and high values of \hi-deficiency \citetext{\citealp{Bravo-Alfaro+01}; \citealp*{Kenney+04}; \citealp{Chung+09, Scott+10, Scott+12, Wang+21, Lopez-Gutierrez+22}}. The quenching mechanisms and the corresponding timescales exhibit significant variations due to a multitude of factors, including the stellar mass, the local ICM-density, the presence of companions, etc. Nevertheless, observations and simulations suggest that quenching should occur at time-scales not much longer than the full gas depletion \citep{Cortese+21,Salinas+24}.

Our goal is to unveil the details of the evolutionary sequence transforming normal, gas-rich late-types into red passive objects, looking for constraints on the star-formation quenching time-scale.  With this aim we combine data of one of the best tracers for interactions, the \hi-imaging, with a direct indicator of star formation, the FUV emission at 1000$-$2000\,\angstrom.  FUV emission is produced by intermediate-mass stars, having 2 to 5\,\msolar, and whose lifetime is below 1\,Gyr \citep{Kennicutt+98}. With this strategy, we expect to draw a coherent evolutionary track involving the triggering of the star-formation burst and its eventual extinction. Similar methods have proven to be useful in the study of cluster infalling galaxies \citetext{\citealp*{Mahajan+11}; \citealp{Jaffe+15, Yoon+17, Bellhouse+21, Mun+21, Wang+20, Wang+21}}.

In this paper, we study a sample of late-type, bright UV galaxies in the nearby cluster A496, observed with the Astrosat-UVIT. We complement this with VLA-\hi-data being part of a survey covering the full cluster volume. We reported 22 FUV galaxies being cluster members; all of them are either detected in \hi, or dispose of upper limits for the \hi-mass. We discuss the location of the FUV galaxies on the projected phase-space diagram (PPS), as a first approach to tracing their orbital histories. We analyze the FUV (morphology and SFR$_\mathrm{FUV}$) and \hi\ properties of this sample, as a function of their stellar mass, relative velocity, and local environment. The latter is quantified by the ICM density as well as the presence of companions. We propose an evolutionary sequence based on the FUV and \hi, tracing the stages from a normal main-sequence galaxy until an advanced quenched phase.

This paper is organized as follows. In Sect. \ref{secc:sample_observations}, we present the data collected for this work, the UVIT-AstroSat, and the \hi-data. In Sect. \ref{secc:results}, we estimate the SFR based on the FUV and we provide the observational properties, including those coming from the \hi. We estimate the cluster escape velocity and the RPS threshold curve to be plotted on the PPS diagram.  In Sect. \ref{secc:discussion} we approach the orbital histories of the galaxies based on their position on the PPS diagram. We seek possible correlations of the combined FUV-\hi\ properties, with parameters such as the relative cluster-velocity, stellar mass, 
and the presence of neighbor galaxies. We propose a five-step evolutionary sequence, from normal main-sequence galaxies to red, gas-poor, and passive objects. Our conclusions are summarized in Sect. \ref{secc:conclusions}.

Throughout this work we assume a \cite{Chabrier+03} initial mass function (IMF), and a $\Lambda$CDM cosmological model with $\Omega_\mathrm{m}$\,=\,0.3, $\Omega_\Lambda$\,=\,0.7, and $H_0=70$~\kms~Mpc$^{-1}$ to obtain distances and luminosities.

\section{The datasets and observations}
\label{secc:sample_observations}

\subsection{The cluster Abell\,496} \label{secc:A496}

A496 is a nearby cluster ($v_{\mathrm{cl}}=$9\,892\,\kms) with relatively low-velocity dispersion of $\sigma_{\mathrm{cl}}=688$\,\kms \citep{Caretta+23}.  Being dominated by a cD galaxy, this cluster is classified as type I in the Bautz-Morgan system \citep{Molinari+97}. The mass $M_\mathrm{200}$, which approximates the virial mass, corresponds to the mass within the sphere of radius $R_\mathrm{200}$, where the mean mass density is 200 times the critical density of the Universe at that redshift. \cite{Markevitch+99} estimate $M_\mathrm{500}$ for 
A496, obtaining a value of 3.9\,\por\,$10^{14}$\,\msolar, while $M_\mathrm{200}$ is approximately 3.5\,\por\,$10^{14}$\,\msolar\ and the X-ray luminosity is 3.8\,\por\,$10^{44}$ erg\, s$^{-1}$ \citep[][hereafter LG22]{Lopez-Gutierrez+22}.

\begin{figure}
   \centering
  \includegraphics[width=\columnwidth]{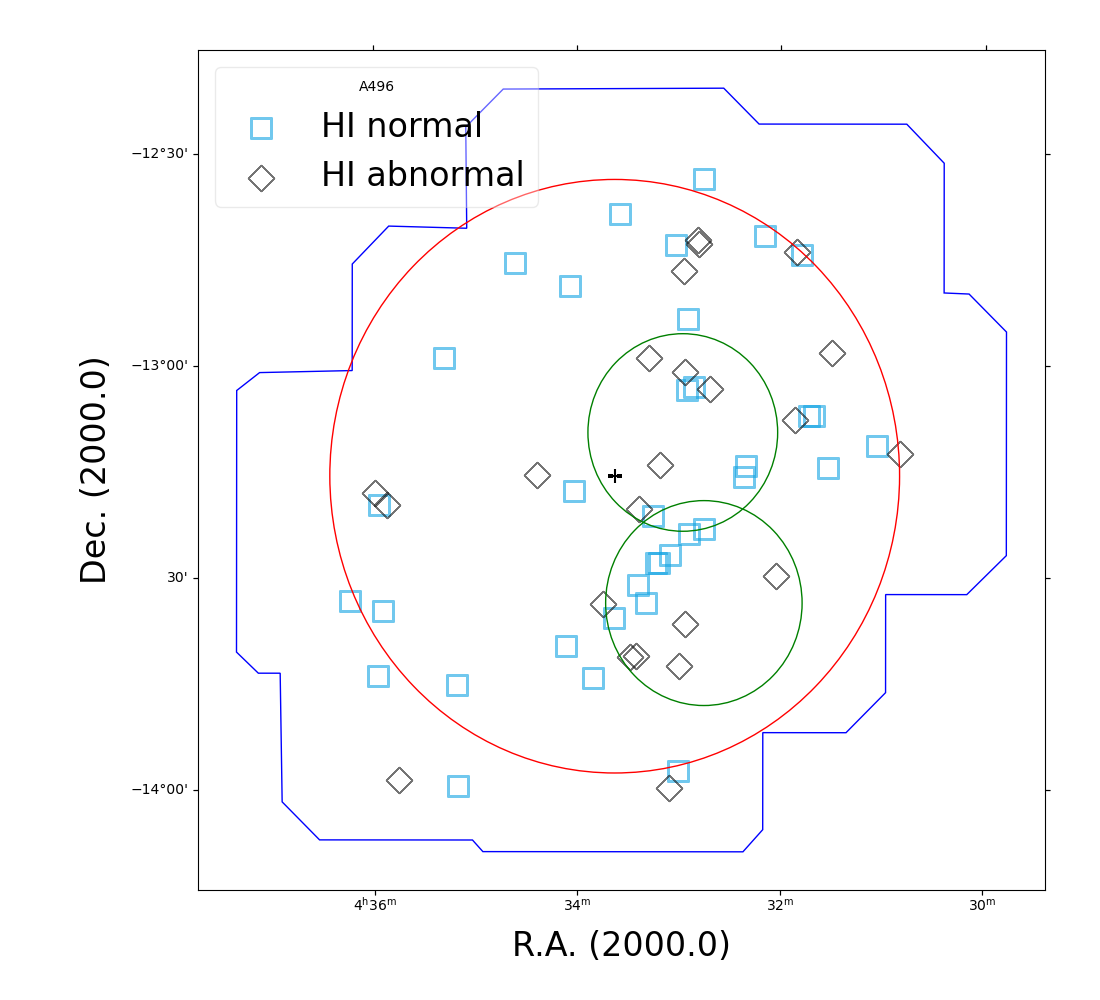}  
  \caption{Two degrees region around A496 showing the 1\,$R_\mathrm{200}$ (red circle). The two UVIT fields are shown with green circles and the VLA-\hi\ survey is indicated with a blue polygon. The \hi-detected galaxies are shown with blue squares (\hi-normal) and black diamonds (\hi-abnormal), see Sect.\,\ref{secc:hi_and_opt_data}. The distribution of cluster members (not shown here) is rather regular, as seen in figure 3 of \citetalias{Lopez-Gutierrez+22}. The cluster center is indicated with a black cross.}
\label{Fig_A496_distribution}
\end{figure}

We selected the cluster A496 because it shows some intriguing features. On one hand, its regular morphology seen in optical and X-rays suggests an evolved system dominated by early-type galaxies \citep{Durret+99, Boue+08a}. The brightest cluster galaxy (BGC) position is coincident with the X-ray maximum \citep{Piffaretti+11}, supporting the idea that this is a relaxed system. However, \citetalias{Lopez-Gutierrez+22} reported A496 as \hi-rich cluster that includes a sample of very low-mass galaxies with high gas content (see Figure \ref{Fig_A496_distribution}).
Furthermore, a significant fraction of blue, star-forming galaxies with high amounts of \hi\ gas, are projected near the cluster core. The typical trend of \hi-deficiency increasing towards the cluster center \citep{Cayatte+90} is not observed in this system. Even a slight anti-correlation was reported by \citetalias{Lopez-Gutierrez+22}.

Concerning the large-scale structure, A496 was thought to be an isolated cluster, but this is at odds with the apparent refueling of blue, \hi-rich galaxies observed in the cluster. \citetalias{Lopez-Gutierrez+22} inspected the 6dF survey around A496, unveiling a large-scale filament running along a NW-SE axis. The same authors reported one single substructure, coincident with the orientation of that filament, 68$’$ ($\sim 2.7$\,Mpc) NW of the cluster center. This might contribute to the replenishment of late-type galaxies but is hardly associated with the blue, \hi-rich objects observed in the central region of A496.

\subsection{UVIT FUV data}\label{subsec:FUV}

This work utilizes FUV imaging carried out with the %0.4m
0.38m twin telescopes, one for the FUV channel and another one is for the NUV channel, Ultra Violet Imaging Telescope (UVIT) on board AstroSat \citep{Kumar+12}.  The GALEX survey presents a large blind zone around A496, therefore, the present data should constitute the first UV imaging on this cluster. Originally, UVIT observed two UV windows simultaneously: FUV (130-180nm) and NUV (200-300nm). However, NUV stopped working after March 2018. Our survey was performed during two observing cycles in November 2021 and September 2022 (see Table \ref{tab_FUV_fields}). \\

We selected two 28$’$ diameter fields, both in the inner cluster regions, where a high density of blue, \hi-rich galaxies was reported by \citetalias{Lopez-Gutierrez+22} (see Figure \ref{Fig_A496_distribution}). The UVIT fields located at center NW and center SW, hereafter named for short, C-NW and C-SW, have a total integration time of 14\,630\,s and 17\,085\,s, respectively, and a spatial resolution of $\sim$ 1.8$''$. Table \ref{tab_FUV_fields} compiles relevant information on the two observed FUV fields. The 28$’$ FoV corresponds to $\sim 1.1$\,Mpc at the cluster distance.
The C-SW field suffers from a noisy edge, mainly affecting the eastern side of the field forcing us to discard a region of 2$’$ in the most external zone. The two observed UVIT fields lie within the 1\,$R_\mathrm{200}$ radius with a surveyed area going as far as $\sim 0.8$\,Mpc to the NW, and up to 1.5\,Mpc to the SW (see Figure \ref{Fig_A496_distribution}).

The data pipeline for converting satellite data into
Level 1 and Level 2 data were prepared by the Indian Space Research Organization (ISRO). After retrieving Level 2 data from the official pipeline, we performed co-addition and astrometry separately. The astrometry on the UVIT level 2 processed data was performed using the \textsc{astrometry.net}\footnote{https://astrometry.net/} software and JUDE\footnote{https://github.com/jaymurthy/JUDE} scripts \citep{Murthy+17}. The final co-addition of the  UVIT images were done separately using Curvit\footnote{https://github.com/prajwel/curvit}, an open-source Python package to generate light curves from UVIT data \citep{Joseph+21}. We then applied the SExtractor\footnote{https://www.astromatic.net/software/sextractor/} \citep{Bertin&Arnouts+96} software using the default parameters, modifying the detection threshold from \textsc{DETECT\_THRESH}$\,=1.2$ to \textsc{DETECT\_THRESH}$\,=3.0$, which corresponds to a 3$\sigma$ detection level in both fields, to obtain candidates for FUV-bright objects. We obtained 237 and 160 detections in the C-NW and C-SW images, respectively. Fluxes for each object were estimated by applying two different methods to define the corresponding apertures; one based on tools offered in ds9 software and a second through our Python routine based on \textsc{photutils.aperture} package. Fluxes have a good agreement with each other.  \\

\begin{table}
    \centering
    \caption{Log of FUV-UVIT observations}
    \label{tab_FUV_fields}
    \begin{tabular}{lccccc}
  \hline
    Field & R.A. &Dec & Obs  & Exposure  \\
     ID &  (J2000) & (J2000) & date & (s)  \\
    (1) & (2)&(3) &(4) & (5) \\ 
    \hline
   C-NW &04 33 00.7 & $-$13 10 21.5 & Nov. 2021 
    &14\,630\\
    C-SW  & 04 32 53.2 & $-$13 31 22.3 & Sep. 2022 &17\,085\\
   \hline
\end{tabular}
\end{table}

\begin{table*}
    \centering
    \caption{Observational parameters of the FUV galaxies in A496}
   \label{tab_A496_FUV_galaxies}
    \resizebox{\textwidth}{!}{%
    \tabcolsep=3pt
    \begin{tabular}{rclcccclcccc}
\hline 
    ID& Name  & R.A.  & Dec & Field & \textit{g} & Redshift & $v_{\mathrm{LOS}}$ & \multicolumn{2}{c}{UVIT FUV} & \multicolumn{2}{c}{FUV Flux} \\
    &   & (J2000) &(J2000) & & mag  & & \kms & $m$(AB)$_\mathrm{corr}$ & $M$(AB)$_\mathrm{corr}$& \flux & Jy\\
   (1) &(2)&(3)&(4)&(5)&(6)&(7)&(8)&(9)&(10)&(11)&(12)\\
   \hline
1 & 1054 & 04 32 02.8 & $-$13 29 48.2 & C-SW & 16.7 & 0.0343 & 10\,289 & 17.81 & $-$18.01 & 1.39E$-$15 & 2.72E$-$04 \\
2 & 1169 & 04 32 20.4 & $-$13 14 11.0 & C-NW & 16.4 & 0.0300 &  \hspace{4pt}8\,990 & 17.73 & $-$18.10 & 2.31E$-$15 & 2.95E$-$04 \\
3 & 1179 & 04 32 21.8 & $-$13 15 48.4 & C-NW & 17.6 & 0.0335 & 10\,057 & 18.61 &$-$17.22 & 1.01E$-$15 & 1.31E$-$04 \\
4 & 1275 & 04 32 41.8 & $-$13 03 21.1 & C-NW & 19.1 & 0.0354 & 10\,608 & 20.30 & $-$15.52& 2.33E$-$16 & 2.75E$-$05 \\
5 & 04.56-13.38 & 04 32 45.3 & $-$13 23 08.0 & C-SW &  21.1 & $-$ &  10\,694$^{*}$ & 20.02 & $-$15.81 & 2.17E$-$16 & 7.58E$-$05 \\
6 & 1338 & 04 32 51.1 & $-$13 03 07.0 & C-NW & 18.9 & 0.0349 & 10\,474 & 19.80 & $-$16.02 & 3.55E$-$16 & 4.37E$-$05 \\
7 & 1354 & 04 32 54.2 & $-$13 23 50.5 & C-SW & 17.2 & 0.0331 &  \hspace{4pt}9\,932& 18.44 & $-$17.38 & 8.39E$-$16 & 1.53E$-$04\\
8 & \hspace{7pt}198$^a$ & 04 32 56.1 & $-$13 36 38.5 & C-SW & 15.7 & 0.0379 & 11\,349 & 17.46 & $-$18.37 & 1.39E$-$15 & 3.77E$-$04 \\
9 & 1366 & 04 32 56.2 & $-$13 00 56.9& C-NW &  17.5 & 0.0285 & \hspace{4pt}8\,553 & 17.35 & $-$18.48 & 3.25E$-$15 & 4.18E$-$04   \\
10 & 1386 & 04 32 59.8 & $-$13 42 32.6 & C-SW & 17.0 & 0.0352 & 10\,557 & 17.70 & $-$18.12 & 7.51E$-$16 & 3.02E$-$04 \\
11 & 1418 & 04 33 05.3 & $-$13 26 54.9 & C-SW & 18.4 & 0.0349 &  10\,473$^{*}$ & 18.65 & $-$17.17 & 6.48E$-$16 & 1.25E$-$04 \\
12 & 1457 & 04 33 10.9 & $-$13 14 10.4 & C-NW & 17.8 & 0.0300 &  \hspace{4pt}8\,991 & 19.45 & $-$16.38 & 4.05E$-$16 & 6.04E$-$05 \\
13 & 1461 & 04 33 11.4 & $-$13 28 01.7 & C-SW & 18.1 & 0.0356 & 10\,670 & 19.10 & $-$16.73 & 3.96E$-$16 & 8.33E$-$05 \\
14 & 1474 & 04 33 13.7 & $-$13 27 56.2 & C-SW & 18.9 & 0.0339 & 10\,159 & 19.25 & $-$16.58 & 3.51E$-$16 & 7.27E$-$05 \\
15 & 1482 & 04 33 15.1 & $-$13 21 19.0 & C-NW & 18.0 & 0.0340 & 10\,207 & 18.75 & $-$17.07 & 6.35E$-$16 & 1.14E$-$04 \\
16 &\hspace{7pt}225$^a$ & 04 33 17.4 & $-$12 59 01.0 & C-NW & 15.6 & 0.0361 & 10\,826 & 19.06 & $-$16.77 & 6.03E$-$16 & 8.65E$-$05 \\
17 & 1512 & 04 33 19.6 & $-$13 33 38.8 & C-SW & 19.2 & 0.0343 & 10\,268 & 19.17 & $-$16.65 & 2.22E$-$16 & 7.81E$-$05 \\
18 & 1547 & 04 33 23.3 & $-$13 20 21.1 & C-NW & 16.9 & 0.0340 & 10\,188 & 18.06 & $-$17.76 & 1.18E$-$15 & 2.17E$-$04 \\
19 & 1549 & 04 33 23.9 & $-$13 31 02.9 & C-SW & 16.7 & 0.0341 & 10\,228 & 17.04 & $-$18.78 & 2.42E$-$15 & 5.54E$-$04 \\
20 & 1565 & 04 33 25.4 & $-$13 41 09.1 & C-SW & 16.7 & 0.0362 & 10\,866 & 17.52 & $-$18.30 & 9.02E$-$16 & 3.56E$-$04 \\
21 & 1584 & 04 33 29.0 & $-$13 41 14.7 & C-SW & 18.2 & 0.0315 &  \hspace{4pt}9\,443 & 18.53 & $-$17.30 & 3.33E$-$16 & 1.41E$-$04 \\
22 & 1656 & 04 33 38.5 & $-$13 35 46.3 & C-SW & 17.0 & 0.0373 & 11\,194 & 17.06 & $-$18.76 & 1.62E$-$15 & 5.43E$-$04 \\
\hline
\multicolumn{12}{l}{Column (1): Sequence ID number.}\\
\multicolumn{12}{l}{Column (2): [SDG99]-SRC galaxy name \citep{Slezak+99}; ($^a$): [DFL99] galaxy name \citep{Durret+99};}\\
\multicolumn{12}{l}{04.56-13.38 is an object not previously cataloged in NED.}\\
\multicolumn{12}{l}{Columns (3) and (4): R.A. Dec. coordinates from NED. }\\
\multicolumn{12}{l}{Column (5): UVIT FUV Fields, C-NW and C-SW.}\\
\multicolumn{12}{l}{Column (6): The $g$ magnitude taken from \citetalias{Lopez-Gutierrez+22}.}\\
\multicolumn{12}{l}{Column (7): Redshift from our redshift membership catalog (see Sect.\,\ref{subsec:FUV}).}\\
\multicolumn{12}{l}{Column (8): Optical velocity from our redshift membership catalog.  ($^{*}$): heliocentric \hi-velocity.}\\
\multicolumn{12}{l}{Columns (9) and (10): FUV apparent and absolute AB magnitude.}\\
\multicolumn{12}{l}{Columns (11) and (12): FUV flux (see Sect. \ref{subsec:FUV}).} 
\end{tabular}
}%end resizebox
\end{table*}

We carried out a cross-match between the FUV detections
and the catalog of \hi\ galaxies that are members of A496, reported by \citetalias{Lopez-Gutierrez+22}.  This helps eliminate foreground and background objects and, above all, restricts our sample to late-types  disposing of information on their gas component, including those only marginally detected (see next section).
This procedure delivered a final sample of 22 FUV galaxies within the two observed fields in A496. The spatial distribution of this sample is given in Figure \ref{Fig_A496_distribution}. Table \ref{tab_A496_FUV_galaxies} provides some observational parameters for these objects and their FUV flux data.

The photometric calibration was done by measuring the FUV emission in the UVIT frames,
from measuring the counts per second (CPS) over an ellipse aperture covering all the FUV emissions. Following \cite{Rhee+17, Das+21}, the flux is calculated as:

\begin{equation}\label{eq:flux_UV}
   \mathrm{Flux}\,[\mathrm{erg\,s^{-1}\,cm^{-2}}\,\angstrom^{-1}]\,=\,\mathrm{CPS}\,\times\, (\mathrm{UC})
\end{equation}

\noindent
where UC is the unit conversion factor, being 3.09\,\por\,10$^{-15}$[\flux] for the FUV filter F148W (CaF2-1)  \citep{Tandon+17}. Hence, we estimate the apparent magnitude in the system AB using the formula m(AB)$\,=\,-2.5$\,log($\mathrm{CPS}$)+$\mathrm{ZP}$ where zero point (ZP) is 18.016 $\pm$ 0.01 mag \citep{Rahna+17}, see magnitudes in column (9) of Table \ref{tab_A496_FUV_galaxies}.  In order to estimate the corresponding magnitude limits we took a conservative value of 3$\sigma$ obtaining 25.66 and 25.61 mag~arcsec$^{-2}$ for C-NW and C-SW, respectively.

The next step is to correct the absolute magnitude for galactic extinction and $K$-correction. Following \cite{Mahajan+22}, we have taken the $E(B-V)$ values from NASA/IPAC Galactic Dust Reddening and Extinction\footnote{https://irsa.ipac.caltech.edu/applications/DUST/}. This provides, for specific coordinates, the $E(B-V)$ reddening and two types of extinction: $A$v$_\mathrm{SFD}$ \citep*{Schlegel+98}, and $A$v$_\mathrm{S\&F}$ \citep{Schlafly+11}. We estimate the $A_\mathrm{FUV}=8.06\,E(B-V)$ \citep{Bianchi+11}, using the $A$v$_\mathrm{SFD}$ extinction and assuming a visual extinction to reddening ratio $A\mathrm{v}/E(B-V)=3.1$. We use SExtractor to carry out the photometric analysis of the $u$ band within the CFHT image dataset. Despite the low redshift we decided to implement a $K$-correction taking advantage of the calculator provided by GAISh\footnote{http://kcor.sai.msu.ru}, which uses the FUV$-u$ color for the magnitude correction of the FUV galaxies. As expected, the $K$-corrections are low, between $-$0.05 to 0.0 mag.

We calculate the corrected apparent magnitude $m$(AB)$_\mathrm{corr}$ and absolute magnitude $M$(AB)$_\mathrm{corr}$ assuming a luminosity distance $D_\mathrm{L}$ of 146 Mpc for the cluster.  Finally, the flux in Jy (given in column (12) of Table \ref{tab_A496_FUV_galaxies}) is obtained as follows:
\begin{equation}
\mathrm{Flux}\,[\mathrm{Jy}]\,=\,10^{(m\mathrm{(AB)}_\mathrm{corr} - 8.9)/2.5)}
\end{equation}

\subsection{\hi\ and optical data} \label{secc:hi_and_opt_data}

We revisited the \hi\ survey of A496 carried out with the NRAO VLA in C-configuration and described by \citetalias{Lopez-Gutierrez+22}. The total FoV after mosaicing (blue polygon in Figure \ref{Fig_A496_distribution}) is roughly 100\,arcmin size, i.e. beyond 1\,$R_\mathrm{200}$. The observed velocity window covers $\sim 2\,000$ \kms, which is more than three times the velocity dispersion of A496: $\sigma_{\mathrm{cl}}$ =\,688 km~s$^{-1}$. We use \hi\ maps and velocity fields with a spatial resolution of 24$''$\por 17$''$, corresponding to a physical resolution of 16 kpc \por 11 kpc.

Recently, \citetalias{Lopez-Gutierrez+22} proposed to distinguish normal from abnormal \hi\ galaxies (blue squares and black diamonds in Figure \ref{Fig_A496_distribution}, respectively) following a number of quantitative criteria.  The \hi-abnormal objects were classified into four types:
(a) \hi-deficient (D): they have lost more than 60\% of their gas, implying a deficiency Def$_\mathrm{HI}$\,$\gtrsim$\,0.3; (b) \hi-asymmetries (A):  objects showing a spatially resolved gas asymmetry; (c)  spatial offset (P): this implies that the \hi\ and the optical centroids are a separated by $\gtrsim$  20 kpc (30$''$) from each other; this is roughly 1.5 times the VLA beam applied in our survey; (d) velocity offset (V): galaxies having a difference $\gtrsim$ 135 \kms\ between the optical and \hi\ velocities (this value is equivalent to 3 times the channel width of our \hi\ data cubes).

Finally, we used public $u$ $g$ $r$ $z$-band images taken with the 3.6m Canada France Hawaii Telescope (CFHT). The observed fields are  $1\times 1$ deg$^2$ in size and cover the regions under study in this work. Accurate photometry was obtained using SExtractor \citep{Bertin&Arnouts+96}, with a zero point of 30.0 mag and a pixel scale of 0.187 $''$/pixel. Stellar masses were estimated following \cite{Mahajan+18}, using {\it r}-band absolute magnitude from CFHT image. More details on these frames are given by \cite{Boue+08a}.  We combine these data with a spectroscopic catalog of member galaxies described by \citetalias{Lopez-Gutierrez+22}.

\begin{figure}
\centering
 \includegraphics[width=1.0\columnwidth]{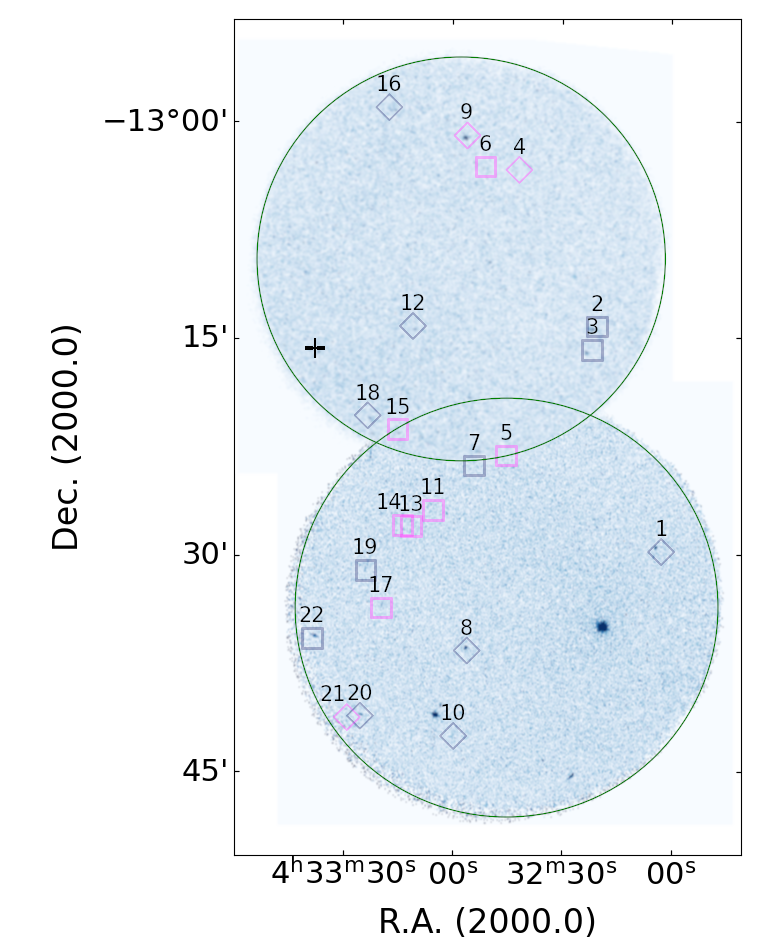} 
    \caption{UVIT FUV image of A496 with the 22 detected galaxies. FoV images of UV fields are shown in bluescale. Squares and diamonds represent \hi\ normal and abnormal galaxies, respectively, with color indicating the stellar mass: dark blue for masses above 10$^9$\,\msolar, and magenta for masses below this threshold. The numbers correspond to the IDs listed in Table\,\ref{tab_A496_FUV_galaxies}. The bright source located SE from galaxy number 1 is a star. The cluster center is indicated by a black cross.}
\label{Fig_A496_distribution_FUV_galaxies}
\end{figure}

\begin{figure}
\centering
    \begin{subfigure}[tbp]{0.44\textwidth}
\includegraphics[width=\columnwidth]{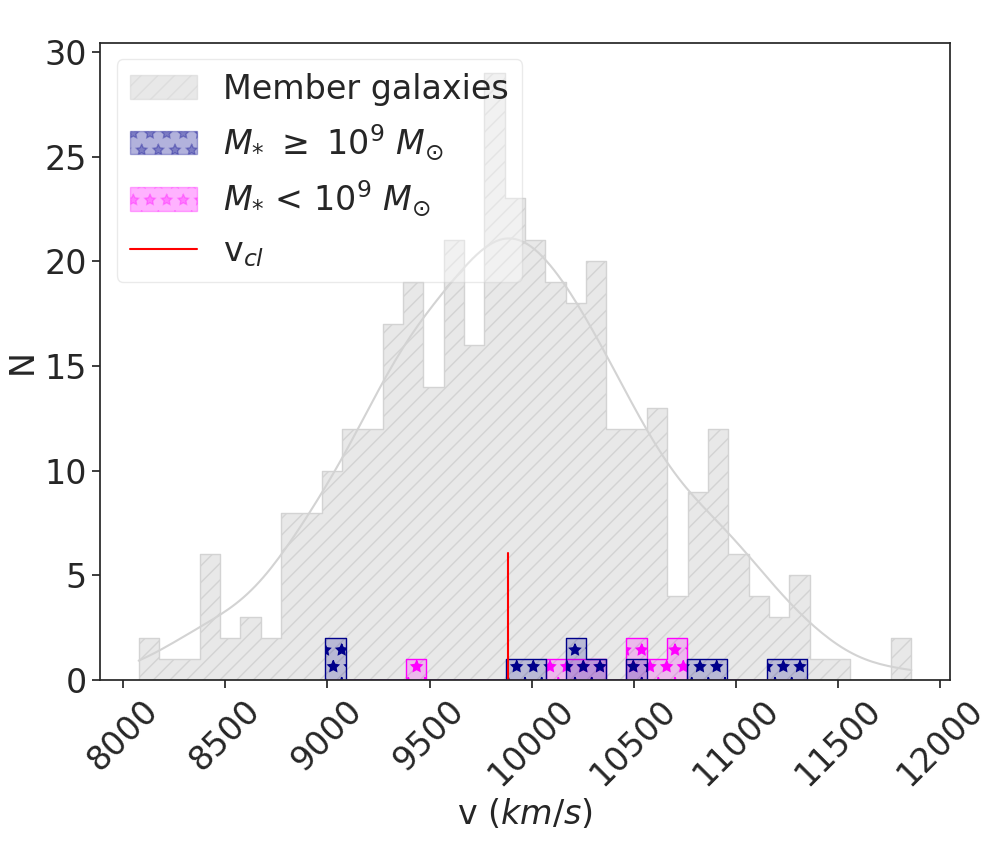}
\end{subfigure}\\
\begin{subfigure}[tbp]{0.44\textwidth}
\includegraphics[width=\columnwidth]{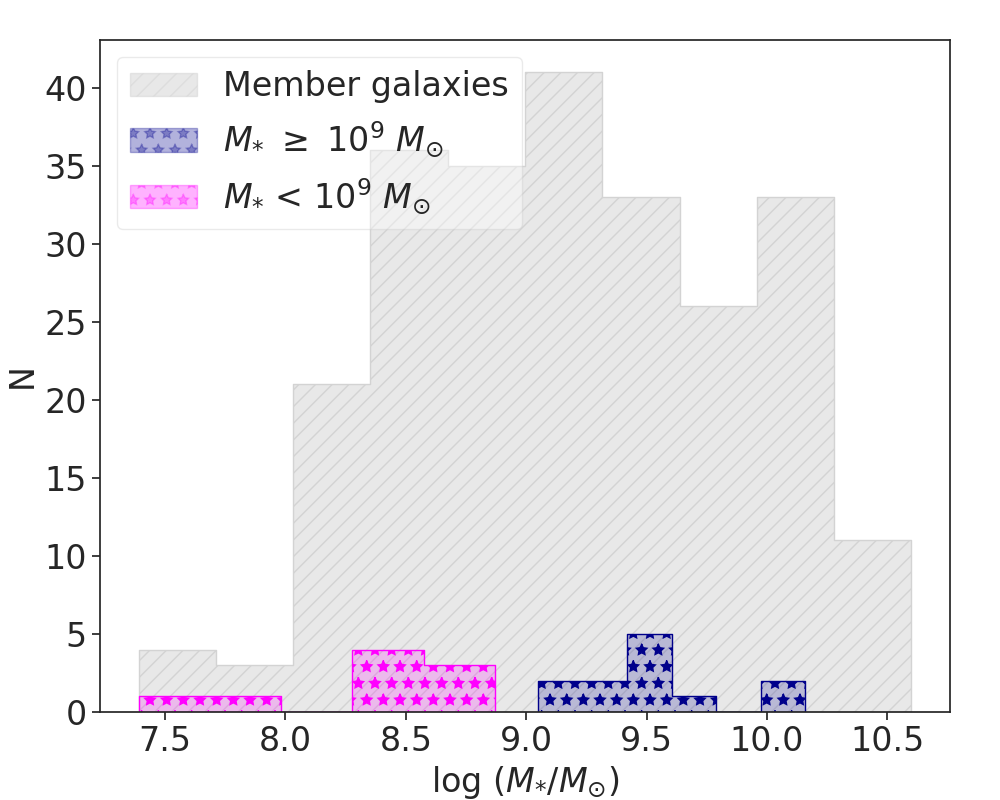}
    \end{subfigure}
\caption{The line-of-sight velocity (upper panel) and mass distributions (bottom panel) of the FUV galaxies in A496, shown in dark and magenta depending on their stellar mass. The full distribution of member galaxies is indicated in light gray.}
\label{Fig_Histograma_A496_Mass_and_Velocity}
\end{figure}

%%%%%%%%%%%%%%%%%%%%%%%%%%%%%%%%%%%%%%%%%%%%%%

\section{Results}\label{secc:results}

\subsection{The FUV galaxies in A496 }\label{subsec:results_FUV}
 
We report 22 cluster member galaxies detected in FUV within the two observed UVIT fields. The main observational parameters (\hi, optical, and FUV) are given in Table \ref{tab_A496_galaxies_FUV_fields}. All but three of these objects are brighter than $m_\mathrm{FUV}=19.5$, with the full range of apparent magnitudes going from  17.0 to 20.3.  The stellar masses for this sample, given in log($M_{*}$/M$_{\odot}$), are in the range 7.4 to 10.2. We follow \citetalias{Lopez-Gutierrez+22} who applied the value of log($M_{*}$/M$_{\odot})=9.0$ to separate very low-mass galaxies from regular spirals. We will study some environmental effects of each sub-sample separately. Figure \ref{Fig_A496_distribution_FUV_galaxies} shows a zoomed view of the distribution of FUV galaxies throughout the observed fields in A496.

The distribution of radial velocities and stellar masses of the FUV galaxies are shown in Figure \ref{Fig_Histograma_A496_Mass_and_Velocity}, contrasted with the full sample of cluster members (grey background). A histogram displaying only the member galaxies inside the UVIT fields has a very similar distribution and it is not shown. 
Interestingly, with very few exceptions, the FUV galaxies lie on the high-velocity side of the Gaussian shown in the top panel of Figure\,\ref{Fig_Histograma_A496_Mass_and_Velocity}. Two subsamples can be seen, centered at roughly 10\,150\,\kms and 10\,650\,\kms. Concerning the stellar mass of the FUV galaxies, ten of the 22 objects lie below the threshold log($M_{*}$/M$_{\odot})=9.0$. Two extreme cases are below, log($M_{*}$/M$_{\odot})<8.0$.

\begin{table*}
    \centering
    \caption{Optical, FUV and \hi\ data for the studied galaxies in A496}
   \label{tab_A496_galaxies_FUV_fields}
     \resizebox{\textwidth}{!}{%
     \tabcolsep=3pt
    \begin{tabular}{rcccrrrrrrcccccc}
\hline 
 ID & Name & Morph &	color  & Diam.  & $\Delta \,v_{\mathrm{LOS}}$ & \multirow{2}{*}{$\frac{\Delta \,v_{\mathrm{LOS}}}{\sigma_{\mathrm{cl}}}$}&	$M_{*}$ &	\mhi &  SFR$_\mathrm{FUV}$& log(sSFR$_\mathrm{FUV}$)&  \hi & FUV & PPS& Comp. & Evol.\\
  &   &  type & ($g-r$)& kpc & \kms &    &  $10^9\,$\msolar   & $10^9\,$\msolar   &  \msolar\,yr$^{-1}$ & yr$^{-1}$ & morph & morph & zone & & stage\\
   (1) &(2)&(3)&(4)&(5)&(6)&(7)&(8)&(9)&(10)&(11) &(12)&(13)&(14)&(15)&(16)\\
   \hline
1  & 1054	& S*& 0.464 & 14.29 & 397 &0.58	&3.39   &1.48  &0.75 & $-$9.7 & P, V & pec & inf & I & 3\\
2  & 1169	& Sb& 0.401 & 32.71 & $-$902 &$-$1.31	&4.47   &8.94  &0.81 & $-$9.7  & N &  pec & rps & I &  2\\
3  & 1179	& S*& 0.371 & 13.46 & 165 & 0.24 &1.26	&1.42   &0.36 & $-$9.5 & N & pec & vir & I &  2\\
4  & 1275	& S*& 0.349 &  7.32 & 716 &1.04 &0.23	&0.95	&0.08 & $-$9.5  & V & pec & inf & C &  1\\
5  & 04.56-13.38 & I*& 0.296 & 7.47 & 802 &1.17 &0.03   &0.57  &0.10 &$-$8.4  & N & pec & rps & I & 2\\
6  & 1338	& S*& 0.320 &  7.89 &582 &0.85 &0.27	&1.12	&0.12 & $-$9.4 & N &  pec & inf & C &  1\\
7  & 1354	& S*& 0.498 & 17.64 & 40 &0.06	&2.24	&3.60	&0.42 & $-$9.7  & N &  N & vir & I &  1\\
8  & \hspace{5pt} 198$^{a}$ & Sc& 0.709	& 13.35 & 1\,457 &2.12	&14.50  &2.50  &1.04 & $-$10.1 & V & pec & inf & I &  2\\
9 & 1366 & S* & 0.063& 18.61 & $-$1\,339 & $-$1.95 & 0.91 & <0.30 & 1.15 & $-$8.9 & D & pec & inf & I & 3 \\
10  & 1386	  & S*& 0.739	& 23.25 & 665 &0.97 &3.72	&3.06  &0.83 & $-$9.7 & A, P &  pec & inf & I &  4\\
11 & 1418	& S*& 0.472	& 14.74 & 581 &0.84 &0.56   &1.21	&0.35 & $-$9.2 & N &  N  & vir & C & 2\\
12 & 1457	& S*& 0.495 & 15.48 & $-$901 &$-$1.31	&1.12	&1.80	&0.17 & $-$9.8 & P &  pec & rps & I & 1\\
13 & 1461	& S*& 0.461 & 11.71 & 778 &1.13	&0.74	&2.84   &0.23 & $-$9.5 & N & pec &rps & C & 1\\
14 & 1474 & S*& 0.415	&  9.46 & 267 &0.39 &0.30   &1.06   &0.20 & $-$9.2 & N & pec & inf & C & 1\\
15 & 1482 & S*& 0.317 & 14.20 & 315 &0.46	&0.68	&0.81	&0.32 & $-$9.3 & N & pec & vir & C &  2\\
16 & \hspace{5pt} 225$^a$& Sb& 0.674 & 30.32 & 934 &1.36 &14.50	&0.62  &0.24 & $-$10.8  & D & N & rps & I &  5\\
17 & 1512	& I*& 0.358 &  7.03 & 376 &0.55	&0.21	&0.71  &0.22 & $-$9.0 & N & pec & vir & C &  2\\
18 & 1547	& Sc& 0.504 & 21.67 & 296 &0.43 &3.02   &3.76  &0.60 & $-$9.7 & P & pec & vir & C &  1\\
19 & 1549 & S & 0.590 & 35.23 & 336 &0.49 &3.89	&10.43 &1.53 & $-$9.4 & N & pec & vir & C & 2\\
20 & 1565	& Sc& 0.524	& 18.74 & 974 &1.42	&3.55	&3.14  &0.98 & $-$9.6  & P & pec & inf & I &  3\\
21 & 1584	& I*& 0.273	&  9.82 & $-$449 &$-$0.65	&0.05	&0.42	&0.39 & $-$8.1 & D & pec & inf & I &  4\\
22 & 1656   & S*& 0.411	& 28.07 & 1\,302 &1.89  &2.40	&0.89 	&1.50 & $-$9.2 & -- & pec & inf & I & 3\\
%1080	&04 32 06.5	&-13 04 59	& Sc& 0.737 &1.40   &9.49   &< 0.30& - &-	&-	&-&0.59 &Def\\
%1443	&04 33 08.8	&-13 02 36	&Sbc& 0.651 &0.46   &9.42   &< 0.30& - &-    &- &- & 0.36 &Def\\
%1520	&04 33 20.2 &-13 26 22	&SBb& 0.707 &0.08   &9.72	&< 0.30&  - &-&-&-&0.27 &Def\\
\hline
\multicolumn{16}{l}{Column (1): Sequence ID number.}\\
\multicolumn{16}{l}{Column (2): [SDG99]-SRC galaxy name \citep{Slezak+99}; ($^a$): [DFL99] galaxy name \citep{Durret+99}.}\\
\multicolumn{16}{l}{Column (3): The morphological type from NED. ($^*$) indicates our own morphological classification.}\\
\multicolumn{16}{l}{Column (4): The ($g-r$) 
color index taken from \citetalias{Lopez-Gutierrez+22}.}\\
\multicolumn{16}{l}{Column (5): The major axis measured in $r$-band as a proxy for the galaxy diameter. They are estimated by using the angular cluster }\\
\multicolumn{16}{l}{distance of 137 Mpc.}\\
\multicolumn{16}{l}{Column (6): Relative velocities compared to $v_\mathrm{cl}$.}\\
\multicolumn{16}{l}{Column (7): The relative velocities normalized by the cluster velocity dispersion.}\\
\multicolumn{16}{l}{Column (8): The stellar mass is estimated according to \cite{Mahajan+18}.}\\
\multicolumn{16}{l}{Column (9): The \hi-mass following \cite{Haynes-Giovanelli+84}; uncertainties are below 10\%. [SDG99]-SRC\,1366 is below the detection} \\
\multicolumn{16}{l}{limit (see text.)}\\
\multicolumn{16}{l}{Column (10): The SFR obtained from the FUV emission (see Sect. \ref{subsec:FUV}).}\\
\multicolumn{16}{l}{Column (11): The logarithm of specific star formation rate (sSFR), defined as SFR/$M_{*}$ (see Sect. \ref{subsec:results_SFR-FUV}).}\\
\multicolumn{16}{l}{Column (12): \hi\ disruption code, either normal (N) or different types of abnormal: \hi-deficient (D), \hi-asymmetries (A), spatial offset (P)}\\ 
\multicolumn{16}{l}{ and velocity offset (V) (see the text for more details). The velocity coverage is incomplete for [SDG99]-SRC\,1656 and no code is allocated. }\\
\multicolumn{16}{l}{Column (13): Code for FUV morphology: normal (N) and peculiar (pec).}\\
\multicolumn{16}{l}{Column (14): Code indicating the region where a galaxy falls within the PPS: In-falling (inf), virialized (vir), ram-pressure stripping (rps).}\\
\multicolumn{16}{l}{olumn (15): Code indicating whether a galaxy has a nearby companion (C) or is isolated (I).}\\
\multicolumn{16}{l}{Column (16): Code representing the evolutionary stage, ranging from 1 to 5 (see Sect.\,\ref{secc:discussion}).}
\end{tabular}
}%end resizebox
\end{table*}

\begin{figure}
   \centering
       \includegraphics[width=\columnwidth]{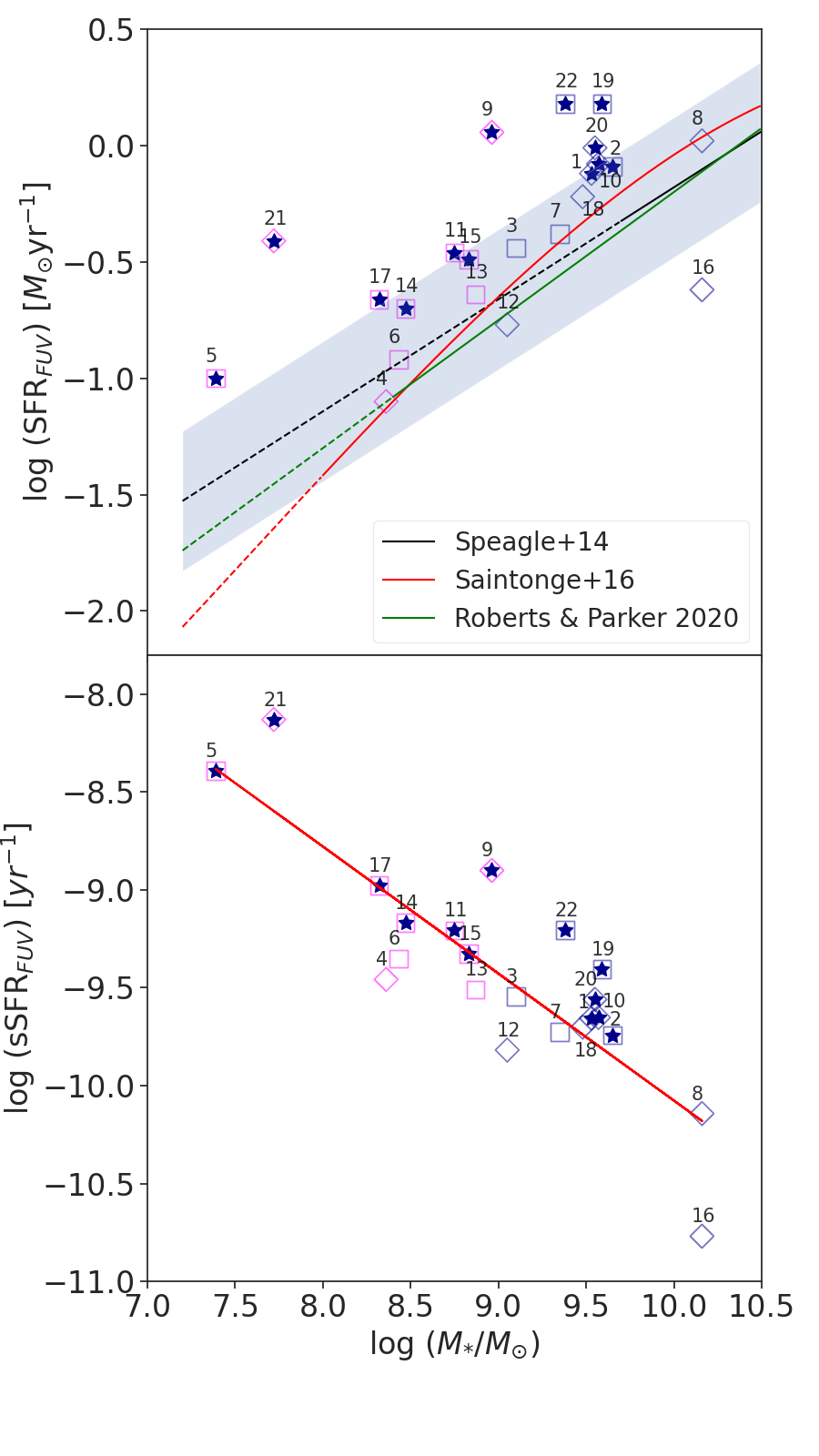} 
    \caption{The SFR upper panel and sSFR (bottom panel) vs. stellar mass diagram of the FUV galaxies. IDs come from Table\,\ref{tab_A496_FUV_galaxies} and symbols are the same as in Figure\,\ref{Fig_A496_distribution_FUV_galaxies}. Blue stars indicate the objects having {\it high} star formation activity (see text).
The upper panel shows three star-forming main sequences (MS) at $z=0$. The black solid line corresponds to the MS defined by \protect\cite{Speagle+14} with the typical scatter (\aprox 0.3 dex) shown in light grey. For comparison, the red and green lines represent the MS relations from \protect\cite{Saintonge+16} and \protect\cite{Roberts&Parker+20}, respectively. Dashed lines indicate extrapolations of these relations toward lower stellar masses. Bottom panel. The specific star formation rate (sSFR) versus stellar mass. The linear regression fit is shown in red:
    $y = -0.6491\,x -3.5845$.}
    \label{Fig_SFR_and_sSFR_M_stellar}
\end{figure}

\subsection{Estimating the star-formation activity }\label{subsec:results_SFR-FUV}

We computed the SFR of the 22 members galaxies of A496, detected with UVIT, based on their FUV luminosity. Our goal is to distinguish from different levels of star formation activity and to unveil possible effects of the local environment in the increasing and quenching of SF.  We estimated the SFR$_{\mathrm{FUV}}$ by following \cite{Salim+07}:

\begin{equation}\label{eq:sfr}
\mathrm{SFR}_{\mathrm{FUV}}\,[\mathrm{M}_{\odot}\,\mathrm{yr}^{-1}]\,=\,1.08\,\times\,10^{-28} L_\mathrm{FUV}\,[\mathrm{erg\,s^{-1}\,Hz^{-1}}]
\end{equation}

The upper panel of Figure\,\ref{Fig_SFR_and_sSFR_M_stellar} presents the relationship between the star formation rate (SFR) and stellar mass for the 22 FUV galaxies. We present various fits to the main sequence (MS) of star-forming galaxies, including those reported by \cite{Speagle+14} (black solid line), \cite{Saintonge+16} (orange line), and \cite{Roberts&Parker+20} (yellow line). These fits exhibit a reasonable good agreement. We trace the corresponding extrapolations to low-mass values in order to cover our full sample. 
For subsequent analysis of the star formation activity we will adopt the fit by \cite{Speagle+14}. If we consider the 0.3\, dex scatter proposed by \cite{Speagle+14} we see that, with only one exception, all FUV galaxies are star-forming objects. According to this, we classify the objects near the upper border of the MS zone, and those clearly above, as galaxies with an increased level of SF activity. Being hard to settle on a universally accepted value defining a {\it starburst}, we do not apply this term to the galaxies in our sample (values above 0.1 Gyr$^{-1}$ are often considered indicative of a starburst galaxy). Nonetheless, we highlight that we obtained values log(sSFR$_\mathrm{FUV}$) $\geq -9.7$\,yr$^{-1}$, equivalent to sSFR $\geq$ 0.2\,Gyr$^{-1}$, for 19 of the 22 objects in our sample (see Table\,\ref{tab_A496_galaxies_FUV_fields}). We applied a binomial test to estimate the significance of this result, considering the null hypothesis of a random distribution against the probability of 19 successes in 22 events, with both results having 50\% chance to occur (i.e. falling {\it above} or {\it below} the median trend). The extremely low probability obtained, P\aprox0.0004, implies a highly unlikely outcome under the null hypothesis. In the low-SF regime, we use the threshold proposed by \cite{Salim+16}, log(sSFR)$<-11$ yr$^{-1}$, to distinguish the passive galaxies. Table \ref{tab_A496_galaxies_FUV_fields} and Figure\,\ref{Fig_SFR_and_sSFR_M_stellar} show that only the very red spiral [DFL99]\,225 (ID 16) is close to this value.

We investigate the relation between stellar mass and SF activity, which reveals a certain trend where galaxies with low stellar masses appear, systematically, more active in star formation (see upper panel in Figure\,\ref{Fig_SFR_and_sSFR_M_stellar}). In principle, this is in agreement with the evidence for galaxy downsizing; while more massive galaxies have formed their stars earlier and over shorter periods of time, their low-mass counterparts continue forming stars at the present epoch and have done so over a more extended period. A more direct representation of the relationship between specific star formation rate (sSFR) and stellar mass is shown in %Figure \ref{fig:sSFRvsMstellar}
the bottom panel of Figure \ref{Fig_SFR_and_sSFR_M_stellar}, highlighting this trend.
%Again, the two objects with the highest 
Table \ref{tab_A496_galaxies_FUV_fields} shows that the  highest sSFR values (log(sSFR)$> -8.4$\, yr$^{-1}$), those with ID 5 and 20, are by far the less massive in our sample (log($M_{*}$/M$_{\odot}) < 8.0$). It is worth noting that further studies based on larger samples of low-mass galaxies are needed to confirm this trend.

\subsection{FUV and \hi\ morphology} \label{subsecc:FUV_HI_morphology}

As mentioned earlier, the presence of blue, \hi-rich galaxies projected onto the central regions of A496 constitutes a very puzzling issue. Despite a relaxed morphology, \citetalias{Lopez-Gutierrez+22} reported this cluster as an \hi\ rich system with 58 detections, half of them with stellar masses below 10$^9$\msolar.  These authors reported A496 as the first cluster showing an anti-correlation between the cluster-centric radius and the \hi-content; i.e. the deficient and abnormal galaxies lie, on average, at larger projected radii compared with \hi-normal objects.  This is in contradiction with all the \hi\ observations made in other clusters, as well as with RPS models, all showing that galaxies at high ICM densities undergo strong gas-stripping \citep[e.g.][and references therein] {Bravo-Alfaro+00, Bravo-Alfaro+09, Tonnesen+19}.  In the next sections we will explore the hypothesis of the FUV galaxies infalling through orbits having large velocity components along the LOS.

\begin{figure}
    \centering
\includegraphics[width=\columnwidth]{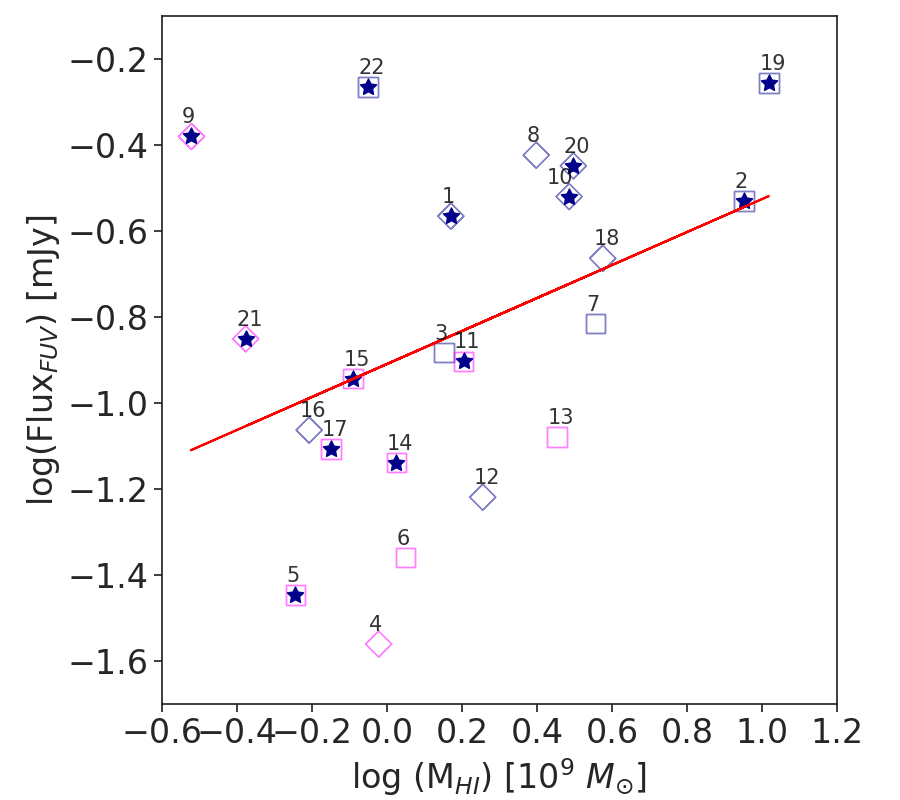}
    \caption{The relation between FUV flux and \hi-mass. The red line indicates the linear regression fit: 
    log(flux$_\mathrm{FUV})\,=$ 0.3835\,log(\mhi)$-$0.9093.  The corresponding correlation factor is close to 0.5. The galaxy IDs are taken from Table \ref{tab_A496_FUV_galaxies}.}
    \label{Fig_HI_mass_vs_Flux}
\end{figure}

With one exception all the FUV objects reported in the present work were previously detected in \hi\ and reported by \citetalias{Lopez-Gutierrez+22}. The galaxy [SDG99]-SRC\,1366 (ID 9 in Table\,\ref{tab_A496_FUV_galaxies}) is only marginally detected due to its central velocity matching with the edge of the \hi-data cube coverage. Half of the gas disk was expected to emerge in our data, yet still, only reduced \hi\ emission was found at the galaxy position, not reaching the 6$\sigma$ detection limit defined by \citetalias{Lopez-Gutierrez+22}. This information allows, however, to classify this galaxy as \hi\ deficient. The full atlas of FUV images is shown in Figure\,\ref{Fig_FUV_HI_images} (left panels); the corresponding $ugr$-images are overlayed with the \hi\ contours (right panels). We measured the trend between the \hi-mass and the FUV flux, finding a relatively high correlation factor, close to 0.5, illustrated in Figure\,\ref{Fig_HI_mass_vs_Flux}. This agrees with other authors who have found that \hi\ is a reliable tracer of the star formation associated with the UV emission \citep{Catinella+18}. In this respect, the Kennicut--Schmidt (KS) relation \citep{Schmidt+59, Kennicutt+98}  describes the dependence between the surface densities of SFR and \hi-gas content. This is known to be weaker compared with the dependence on molecular gas. Nevertheless, in \hi-dominated regions, like spirals outskirts, the correlation $\Sigma_\mathrm{SFR}$ = $\Sigma_\mathrm{HI}^n$ may have a power-law index $n \sim 1$. Combining UV with \hi\ allows for probing the KS relation at least in low-density environments. 
%such as the outer regions of spirals, where molecular gas is scarce. 
We therefore confirm that the \hi\ content is a fair tracer of star formation activity, particularly useful in the absence of more direct tracers like ionized/molecular gas.

\subsection{Projected phase-space view}\label{secc:pps}

\begin{figure}
    \centering
    \includegraphics[width=\columnwidth]{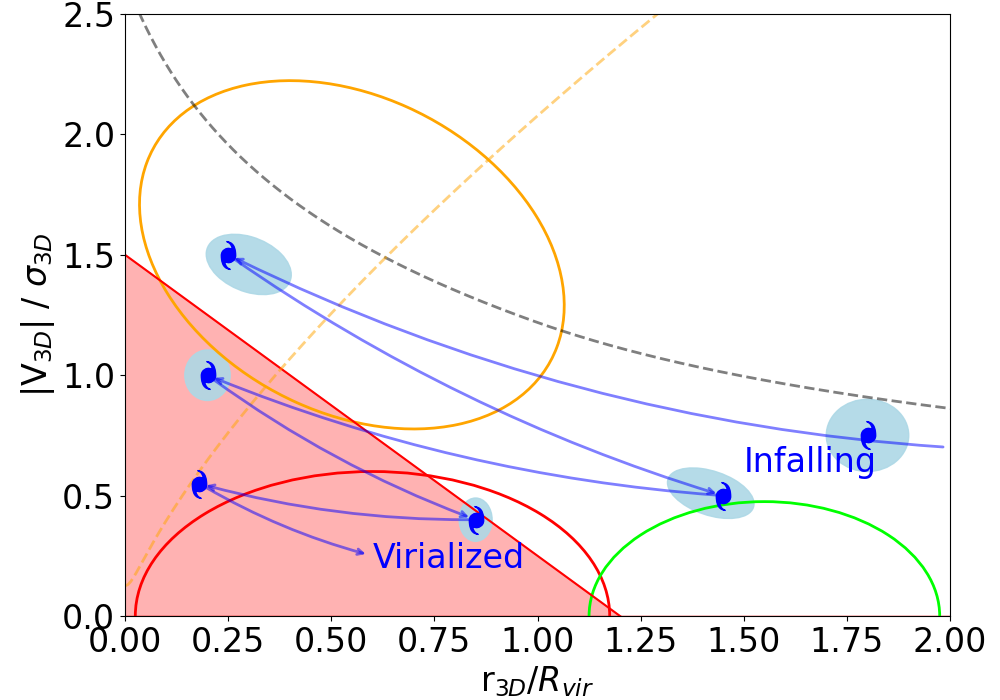}
    \caption{Schematic orbit of a galaxy from infalling to the virialized zone in phase space. Individual blue-shaded ellipses represent the gas component that is diminished as the galaxy moves to denser regions. The black dashed line represents the cluster escape velocity, and the red straight line indicates the virialized region. The orange dotted line is the boundary where RPS balances the anchoring force (see text). The orange, red, and green ellipses illustrate the RPS, virialized, and backsplash regions, respectively (based on \protect\cite{Rhee+17} and \protect\cite{Yoon+17}).}
    \label{Fig_schematic}
\end{figure}

We take advantage of the fact that, during the first infall, galaxies have fairly radial orbits. Only after the galaxies spend enough time in the cluster and as they get closer to denser regions their orbits become more isotropic \citep{Vollmer+01, Hwang+08, Mamon+19}.  The projected phase-space \citep[PPS,][]{Jaffe+15, Yoon+17} diagram helps to infer the assembly histories of clusters as well as the orbital histories of their member galaxies. 
They also permit the identification of different regions where galaxies can be located: infalling, virialized, ram-pressure stripped, and backsplash. Both, simulations and observations \citetext{\citealp*[e.g.][]{Mahajan+11}; \citealp{Yoon+17}} suggest that galaxies on first passage come from large distances with low initial velocities. As they approach the cluster center during their first passage they increase their velocities up to $\sim$ 1000 \kms\ \citep{Rhee+17}. Figure \ref{Fig_schematic} illustrates a typical galaxy trajectory, from the first infall toward the virialized zone. During different crossing times, the galaxies lose energy, which is represented as a diminution of the relative velocity, to finally settle into the virialized zone.

In order to trace the PPS diagram for A496, and plot the position of the FUV galaxies, we need to calculate the corresponding escape velocity. First, we obtain the corresponding mass of the cluster, $M_\mathrm{200}$, following \cite{Barsanti+18}:

\begin{equation}
 M_{200} = \left( \frac{\sigma_{\mathrm{cl}}}{1090\, \mathrm{km\,s^{-1}}}\right)^3\,\frac{1}{\sqrt{\Omega_\Lambda\,+\,\Omega_\mathrm{m}\,(1+z)^3}}\,h^{-1}_{100}\,10^{15} \, \mathrm{M_{\odot}}
\end{equation}

We obtained $M_\mathrm{200}$ is 3.5\,\por\,10$^{14}$\,\msolar. Afterward, we calculated $R_\mathrm{200}$ following \cite{Poggianti+06} and \cite{Mamon+13}, in order to maintain consistency with the Barsanti equation for $M_\mathrm{200}$. These authors employ NFW models obtaining $\sigma_\mathrm{v}$ / $V_\mathrm{200}$ $\simeq$ 0.65. We then expressed $R_\mathrm{200}$ as:

\begin{equation}
 R_{200} = 1.49\,\frac{\sigma_{\mathrm{cl}}}{1000\, \mathrm{km\,s^{-1}}}\,\frac{1}{\sqrt{\Omega_\Lambda\,+\,\Omega_\mathrm{m}\,(1+z)^3}}\,h^{-1}_{100} \, \mathrm{Mpc}
\end{equation}

Using this, we obtained a value of $R_\mathrm{200}=1.44$\,Mpc.
Finally, we followed the strategy described by \cite{Jaffe+15} to estimate the escape velocity, $v_\mathrm{esc}$. The $r_\mathrm{3D}$ scales with the projected $R$, and the 3D escape velocity is related to the line-of-sight (LOS) value by a conversion factor of $\sqrt{3}$. 

\begin{figure}
\includegraphics[width=\columnwidth]{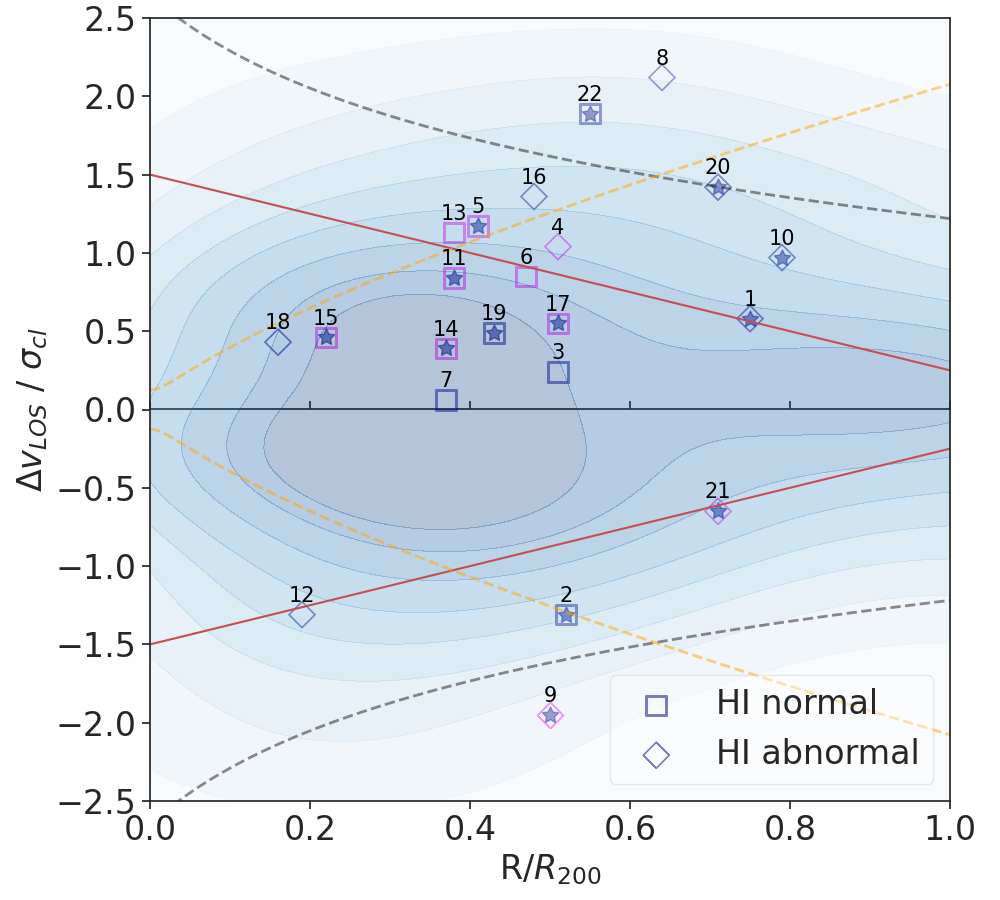}
% \end{subfigure}
    \caption{Projected phase space distribution for A496; the blue density contours show the distribution of member galaxies. The position of FUV galaxies is displayed with symbols and IDs as in Figure\,\ref{Fig_HI_mass_vs_Flux}. Solid red lines draw the virialized zone and the dotted grey curves indicate the escape velocity. The yellow lines show the region where RPS and the anchoring force balance each other.
    }
    \label{Fig_PPS_perturbance}
\end{figure}

With this in mind we draw the ram-pressure stripping region by using the \cite{Gunn&Gott+72} relation:

\begin{equation}\label{eq:rps}
    P_\mathrm{ram} = \rho_\mathrm{ICM}\, v_\mathrm{rel} ^2
\end{equation}

\noindent
where $v_\mathrm{rel}$ is the radial component of the galaxy velocity relative to the cluster, and $\rho_\mathrm{ICM}$ is the local density of the ICM.  We estimate the latter using the hydrostatic-isothermal $\beta$-model proposed by \citet{Cavaliere+76}, expressed as \begin{math} \rho(r) = \rho_{0} [1 + (r/r_{\mathrm{c}})^{2}]^{-3\beta/2} \end{math}, where the core radius is $r_{\mathrm{c}} = 30$ kpc, the central density is $\rho_{0} = 0.0407$ cm$^{-3}$, and $\beta = 0.484$ \citep{Chen+07}. The RPS line in Figures \ref{Fig_schematic} and \ref{Fig_PPS_perturbance} draws the zone where RPS and the gravitational anchoring force (per surface unit) are balanced. This anchoring {\it pressure} is given by: \begin{math}\Pi_\mathrm{gal} = 2\pi \,G \,\Sigma_{\mathrm {s}} \,\Sigma_{\mathrm {g}} \end{math}, where $\Sigma_{\mathrm{s}}$ and $\Sigma_{\mathrm{g}}$ are the surface densities of the stellar and gaseous disks, respectively. We use the value of 
$\Pi_{\mathrm{gal}}=5.0$\por10$^{-13}$\, N\,m$^{-2}$, obtained by \citetalias{Lopez-Gutierrez+22}, as representative for spirals with masses $\geq 10^9$\,\msolar. This value constitutes a logic upper limit for galaxies having stellar masses below that threshold.

Figure \ref{Fig_PPS_perturbance} presents a PPS diagram for the central region of A496, with grey-blue contours indicating the distribution of the cluster members. This plot confirms the striking asymmetry in the velocity distribution of the FUV galaxies which are biased to velocities larger than the systemic value of A496 (Figure\,\ref{Fig_Histograma_A496_Mass_and_Velocity}). We prove statistically this result in the next section. Squares and diamonds in this figure distinguish normal from disrupted objects in \hi. The blue contours confirm the high concentration of member galaxies in the cluster core, in agreement with a relaxed system (\citetalias{Lopez-Gutierrez+22} and references therein).  However, as mentioned above, the fraction of FUV galaxies with normal \hi\ content, projected within (or close to) the virialized zone, remains to be explained.

\citetalias{Lopez-Gutierrez+22} quantified the RPS as a function of the cluster radius and they have found that, with few exceptions, all the FUV galaxies would be stripped if the projected radius is used as a proxy for the physical distance $r$.  In this respect it is possible to approach, at least statistically, the real cluster-centric distance from the projected radius $R$. For instance, \cite*{Mahajan+11} reported the relation between these two parameters on the basis of dark-matter cosmological simulations. We use their figure 15 which provides, for different bins of absolute LOS velocity, the distribution of the 3D physical radii as a function of the projected distance. Our sample of FUV galaxies are contained in the range $R / R_v \leq$\,0.5, and are within the velocity bins 0--1$\sigma_\mathrm{v}$, 1--2$\sigma_\mathrm{v}$. \cite*{Mahajan+11} obtained a relation between the real and projected radii, $r \sim$1.3\,$R$. We take this into account in the next section where we explore different hypothesis accounting for the blue, \hi-rich population projected onto the virialized zone of A496.

\section{Discussion}\label{secc:discussion}

Here we explore the projected radial positions of the FUV galaxies and their velocities relative to the cluster, to ascertain whether their orbital trajectories align more closely with radial or tangential orbits.  
%First, we trace infalling time curves along a PPS diagram as a test for the infalling stage where the FUV galaxies could be found.  Alternatively, 
We tackle the peculiar distribution of the star-forming, \hi-rich galaxies, projected onto the center of A496. This cluster is, to our knowledge, the only one that shows an anti-correlation between the cluster-centric distance and the \hi-deficiency (see \citetalias{Lopez-Gutierrez+22}).  We use the location of the FUV galaxies in PPS diagrams, as well as their combined properties of star formation and gas content, to elucidate the path followed by normal late-types while evolving to the quenching stage.

\subsection{Infering the orbital histories of FUV galaxies}
\label{secc:orbital}

In clusters, the star-forming galaxies are preferentially located in the outer regions  \citep{Balogh+97, Balogh+04}. As galaxies travel through a massive cluster, they experience different environments of low and high densities.  During their first cluster crossing, spirals can experience a significant reduction in star formation activity. Some massive galaxies in the virialized zone may still retain a considerable amount of gas \citep{Oman&Hudson+16, Rhee+20, Oman+21} and a number of first infallers can be found in that region, as shown by \cite{Rhee+17}. The subsequent quenching is expected to occur during (or shortly after) the first pericenter passage. The timescale for this quenching process remains an unresolved question due to the complexity of the physical process involved.  It strongly depends on the galaxy mass as well as on the location where the gas sweeping occurs: either within the cluster ICM or previous to the infall, through the so-called pre-processing within smaller galaxy systems \citep{Poggianti+09, Rhee+20, Sengupta+22}.

We used two strategies based on PPS diagrams to infer the time since infall \citep{Haines+15, Oman&Hudson+16,Rhee+17}. First we follow \cite{Pasquali+19} to trace the infall time $T_\mathrm{inf}$ since a galaxy crossed for the first time the virial radius of its parent host. Second, we estimate the fraction of virialized galaxies following \cite*{Mahajan+11}, who traced the distribution of virial, infall and backsplash populations, based on simulations.

\begin{figure}
    \centering
\includegraphics[width=\columnwidth]{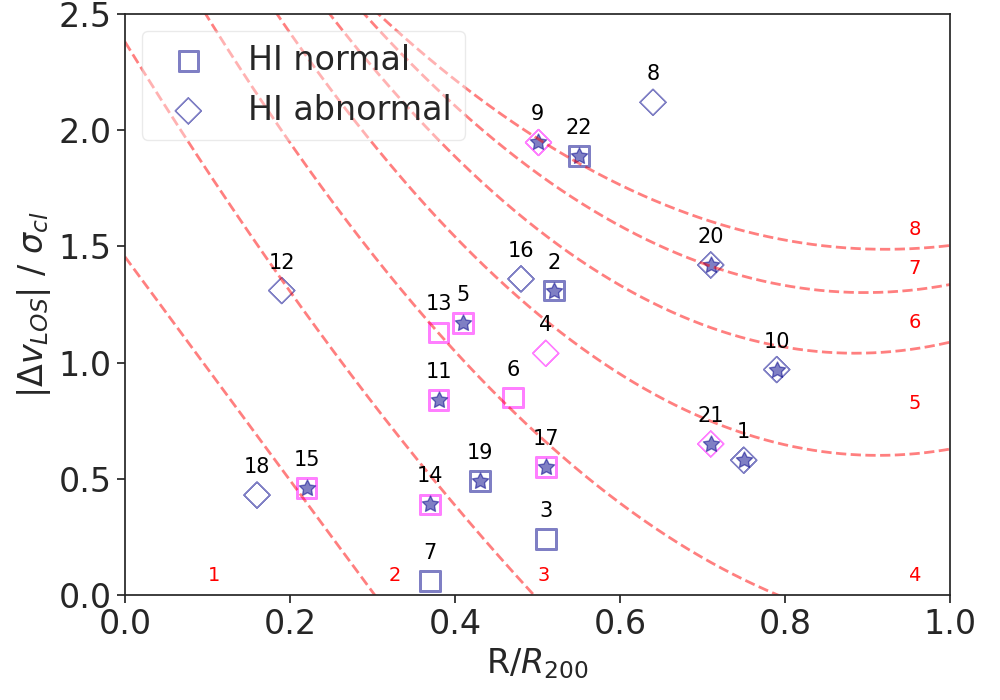}
    \caption{Projected phase space diagram showing  eight zones of $T_\mathrm{inf}$ within 1\,$R/R_\mathrm{200}$, defined by \protect\cite{Pasquali+19}. Each region corresponds to a different time since infall, indicated with dashed lines and red numbers running from 1 to 8 with decreasing infall lookback times given in Table\, \ref{Table_infalling_time}. The FUV galaxies are shown with symbols as in Figure\,\ref{Fig_A496_distribution_FUV_galaxies}.    }
    \label{Fig_infall_time}
\end{figure}

Phase-space diagrams, combined with observed galaxy properties, have been widely used as tools to investigate galaxy evolution \citep[e.g.,][]{Yoon+17}. \cite{Pasquali+19} used the location of satellite galaxies within PPS diagrams to study average observational properties, included the sSFR, as a function of environmental effects. With the help of cosmological simulations these authors define the mean infall time for different zones of the phase space (due to the mixing of infall times within each region of the PPS, it is not possible to assign a unique infall time to individual galaxies). The redshift range in that work lies between $z=\,$0.01 and 0.2; the satellites span stellar masses from 10$^9$ to 10$^{11.5}$\,\msolar, and the haloes are between 10$^{12}$ and 10$^{15}$\,\msolar. Abell496 and the sample of {\it massive} FUV galaxies in our sample, fit within these parameters.
Figure\,\ref{Fig_infall_time} shows eight zones corresponding to different times since infall (taken from \cite{Pasquali+19}, see Table\,\ref{Table_infalling_time}); the positions of the FUV galaxies in A496 are displayed in that plot.

The virialized region mostly encompasses galaxies that have resided within the cluster for some time, at least 3.89 Gyr (Zone 4 in Table\,\ref{tab:my_label}). Intriguinly, we have found some low-mass ($<$10$^9$\msolar) objects projected onto (or close to) the virialized zone, i.e. between Zone\,1 and Zone 3 in Figure \ref{Fig_infall_time}. We remind that models predict that low-mass galaxies would get easily gas stripped and quenched, previous to their first visit to the pericenter \citep{Boselli+14b, Donnari+21}. The low stellar masses and the normal \hi\ content of several FUV galaxies projected onto regions 1-to-4 (Figure\,\ref{Fig_infall_time}), strongly suggest that most of these objects undergo early stages of their first infall. The presence of these galaxies close to the cluster core (see Figs.\,\ref{Fig_A496_distribution_FUV_galaxies} and \ref{Fig_PPS_perturbance}) would be explained by projection effects. These objects would have orbits dominated by a motion along the observer LOS, falling into the cluster from the foreground.

\begin{table}
    \centering
     \caption{The mean infall time and the standard deviations taken from \protect\cite{Pasquali+19}.}
     \label{Table_infalling_time}
    \begin{tabular}{ccc}
    \hline
    Zone &    $T_\mathrm{inf}$ & $\sigma$\,($T_\mathrm{inf}$)\\
     & Gyr  & Gyr \\
       \hline
     1 & 5.42 & 2.51 \\
     2 & 5.18 & 2.60 \\
     3 & 4.50 & 2.57 \\
     4 & 3.89 & 2.34 \\
     5 & 3.36 & 2.36 \\
     6 & 2.77 & 2.29 \\
     7 & 2.24 & 1.97 \\
     8 & 1.42 & 1.49 \\
     \hline
    \end{tabular}
    \label{tab:my_label}
\end{table}

However, this hypothesis is not totally supported by previous simulations. For instance, \cite*{Mahajan+11} obtained the projected distribution of virialized and infall galaxies within their mock cluster built from the stack of 93 regular clusters (see figure\,10 of that work). Their predictions do not exclude that some objects are located within the inner cluster zones, just as seen in Figure\,\ref{Fig_infall_time}.  
Unfortunately, the small sample of FUV galaxies available in this work does not permit us to resolve the question. Distinguishing between the two hypothesis requires a larger sample of FUV galaxies distributed across larger cluster volumes, as well as more mock cluster simulations that are directly comparable to the system under study. We will tackle this issue in future work.

\subsection{Evolutionary sequence from star forming to quenching} \label{subsecc:evol_seq}

Next we measure the environmental effects exerted on the ISM and on the star formation activity of the studied sample. With this aim we applied the criteria described in Sect.\,\ref{secc:hi_and_opt_data} to separate \hi-normal from abnormal objects.  Concerning the FUV, we carried out a visual inspection of the morphology in order to distinguish peculiar objects. Our criteria include the presence of asymmetries, strong knots and a SFR above the main sequence (see upper panel in Figure\,\ref{Fig_SFR_and_sSFR_M_stellar}). Combining \hi\ and FUV helps to avoid mistaking a normal star-forming object for an environment-disrupted one. Very interestingly, Table\,\ref{tab_A496_galaxies_FUV_fields} shows that the majority of galaxies in our sample displays abnormalities in the FUV, while
less than half exhibit strong \hi\ disruptions. This suggests a shorter time-scale for triggering the star formation, compared with the time to remove a large fraction of cold gas.  When the burst of star formation occurs by external mechanisms, it is the outer \hi\ disk that is affected, accounting for this fact.

We use FUV and \hi\ properties as a first approach estimating the time elapsed since the disrupting event at the origin of the observed rising of star formation activity. The FUV emission is produced by the hottest and youngest massive stars, with lifetimes of a few million years.  Several generations of such stars can be produced, depending on the original gas reservoir, the SF-efficiency, and the gas replenishment. Typically, these {\it starburst} phases have short time-scales of \aprox 10$^{8}$\,yr \citep{Larson&Tinsley+78}. On the other side, the \hi\ stripping time-scales for cluster galaxies are longer, of several  10$^8$\,yr \citep{Boselli&Gavazzi+06, Cortese+21, Smith+22, Salinas+24}. Observations and simulations suggest that the subsequent SF-quenching will occur rather quickly once the gas has been stripped \citep{Vollmer+08}.  Therefore, if we consider that most of the FUV galaxies show a rather normal gas content, we consider that a few times 10$^8$\,yr constitutes a reasonable upper limit to the infall time of the FUV galaxies in A496. This is one order of magnitude shorter than the values given in Table\,\ref{Table_infalling_time}, suggesting that the \hi\ depletion is in its early stages, with the exception of [DFL99]\,225, the most processed object in the FUV sample.

\begin{figure*}
\centering
\begin{subfigure}[tbp]{0.23
\textwidth}
 \includegraphics[width=\columnwidth]{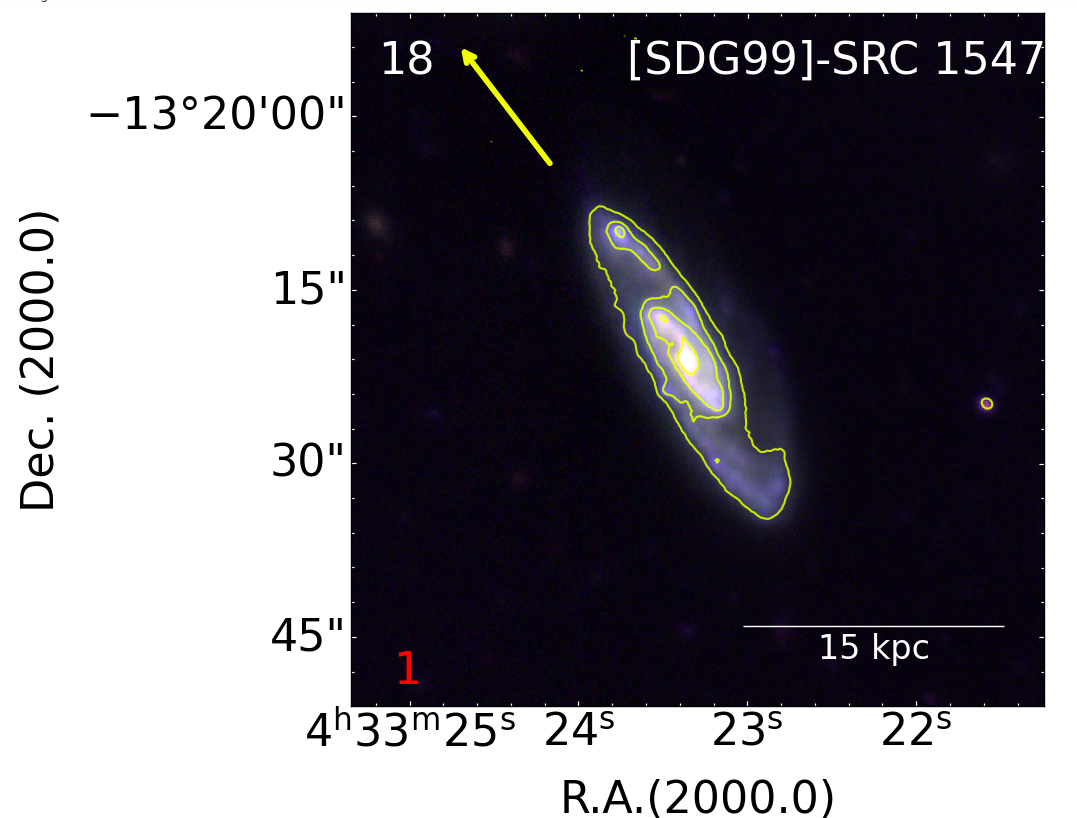}
  \end{subfigure}
  \begin{subfigure}[tbp]{0.23\textwidth}
 \includegraphics[width=\columnwidth]{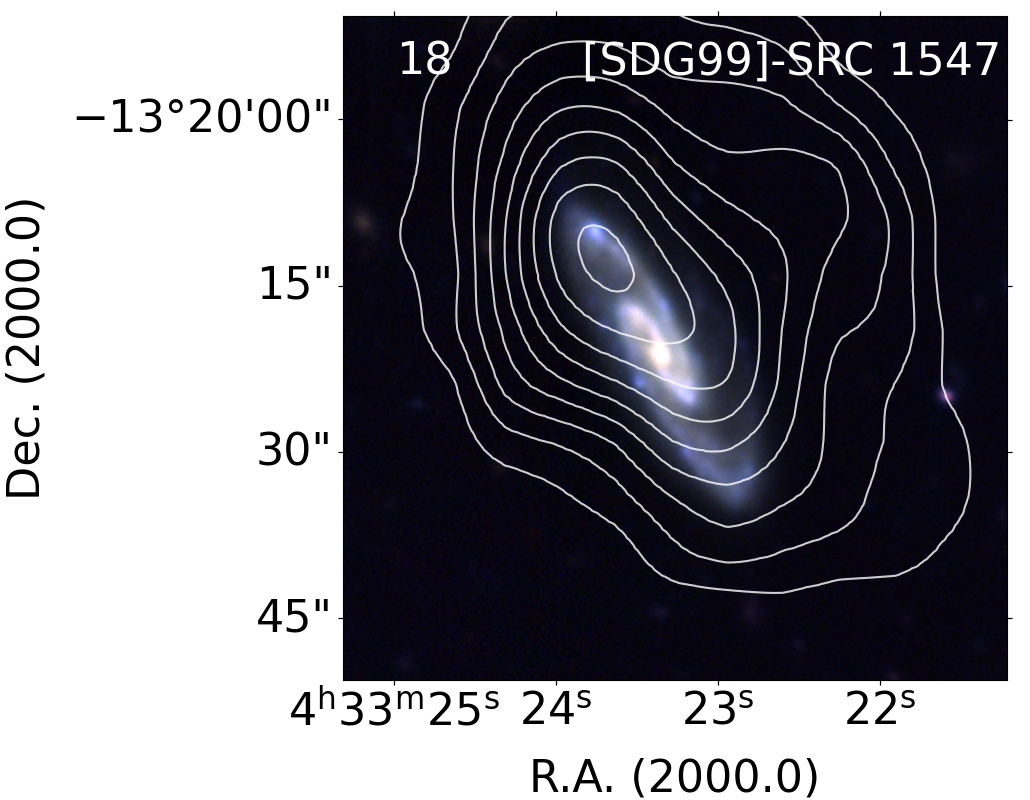}
  \end{subfigure}
   \begin{subfigure}[tbp]{0.23\textwidth}
 \includegraphics[width=\columnwidth]{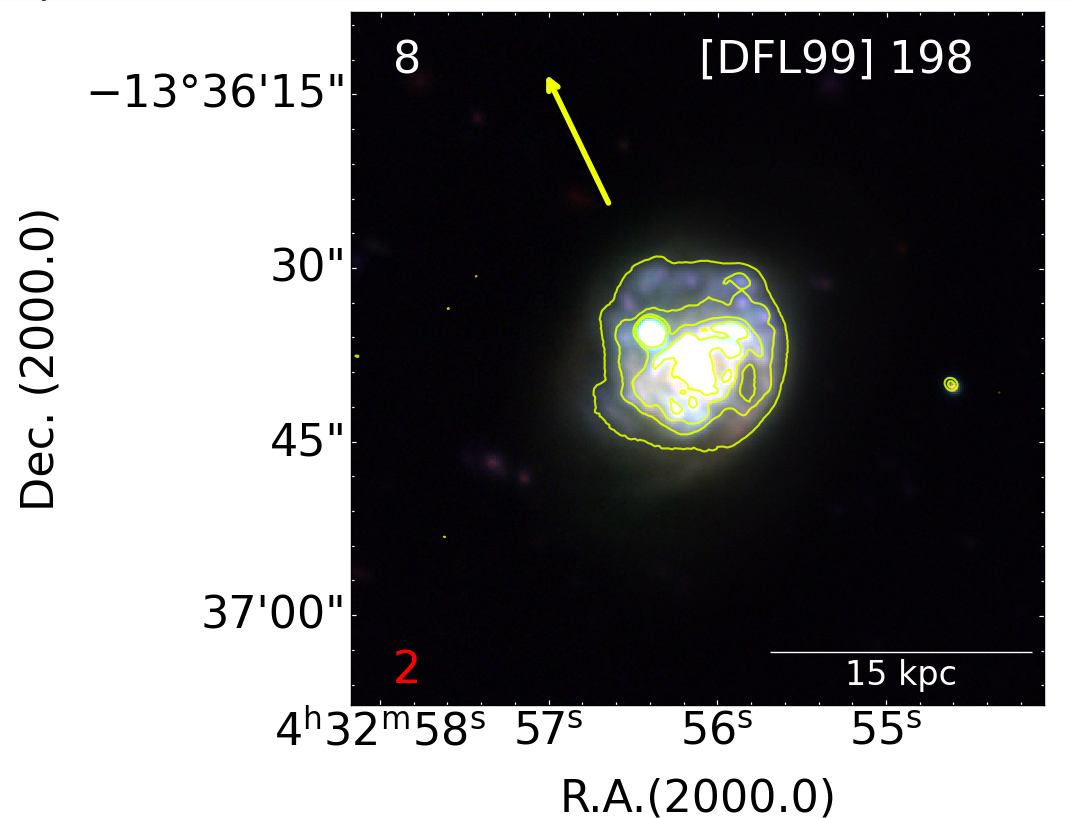}
  \end{subfigure}
  \begin{subfigure}[tbp]{0.23\textwidth}
 \includegraphics[width=\columnwidth]{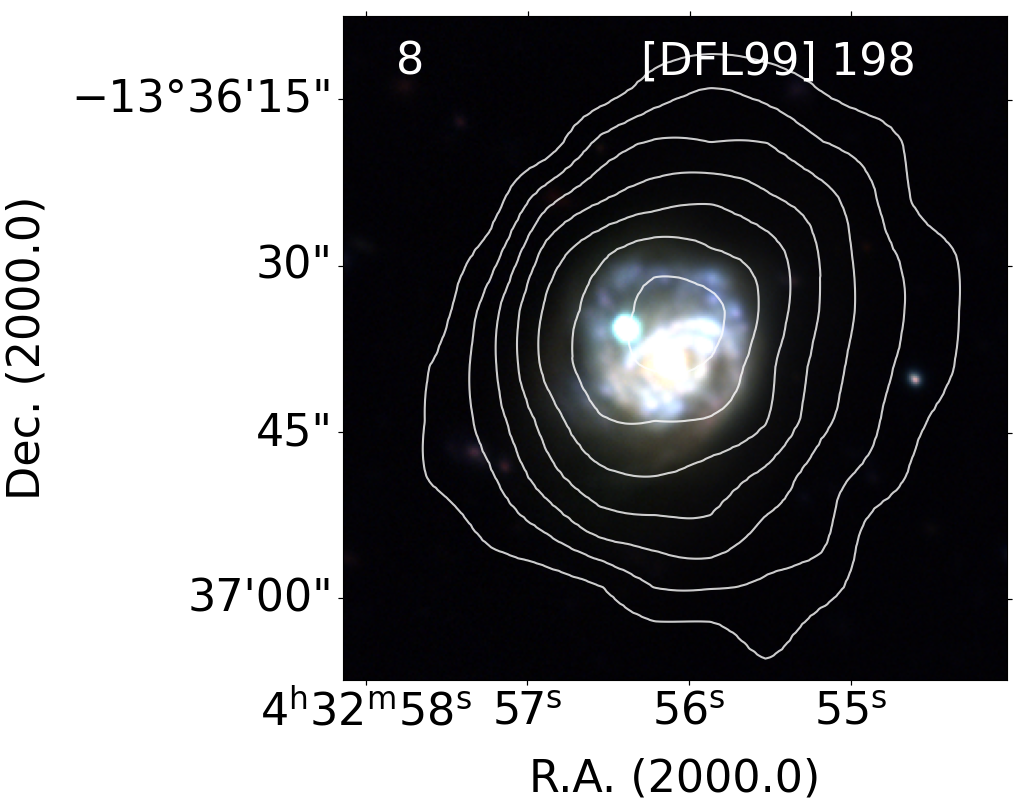}
  \end{subfigure}
  \\
   \begin{subfigure}[tbp]{0.23\textwidth}
 \includegraphics[width=\columnwidth]{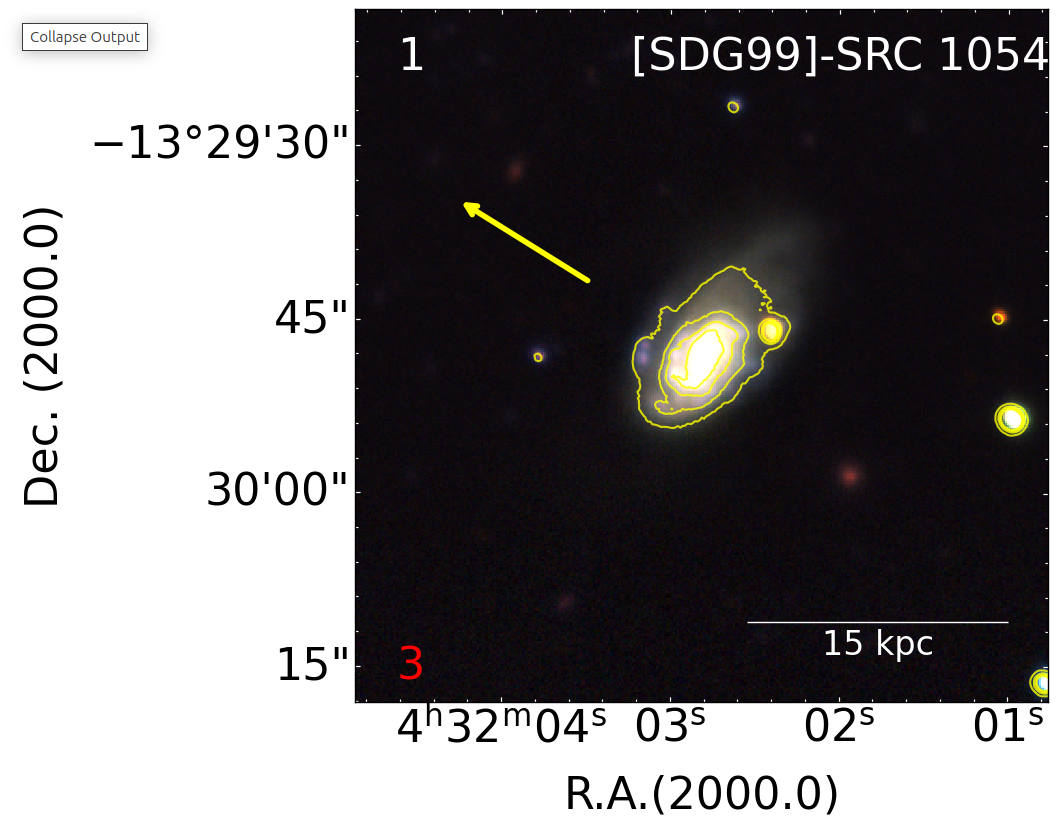}
  \end{subfigure}
  \begin{subfigure}[tbp]{0.23\textwidth}
 \includegraphics[width=\columnwidth]{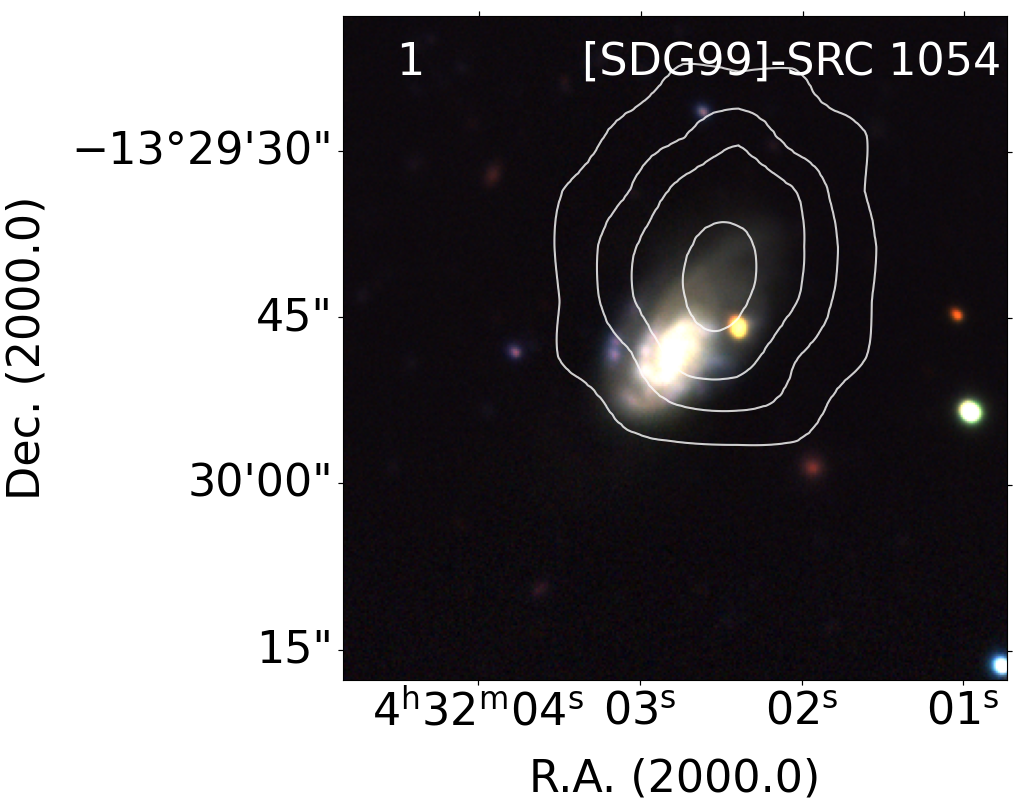}
  \end{subfigure}
   \begin{subfigure}[tbp]{0.23\textwidth}
 \includegraphics[width=\columnwidth]{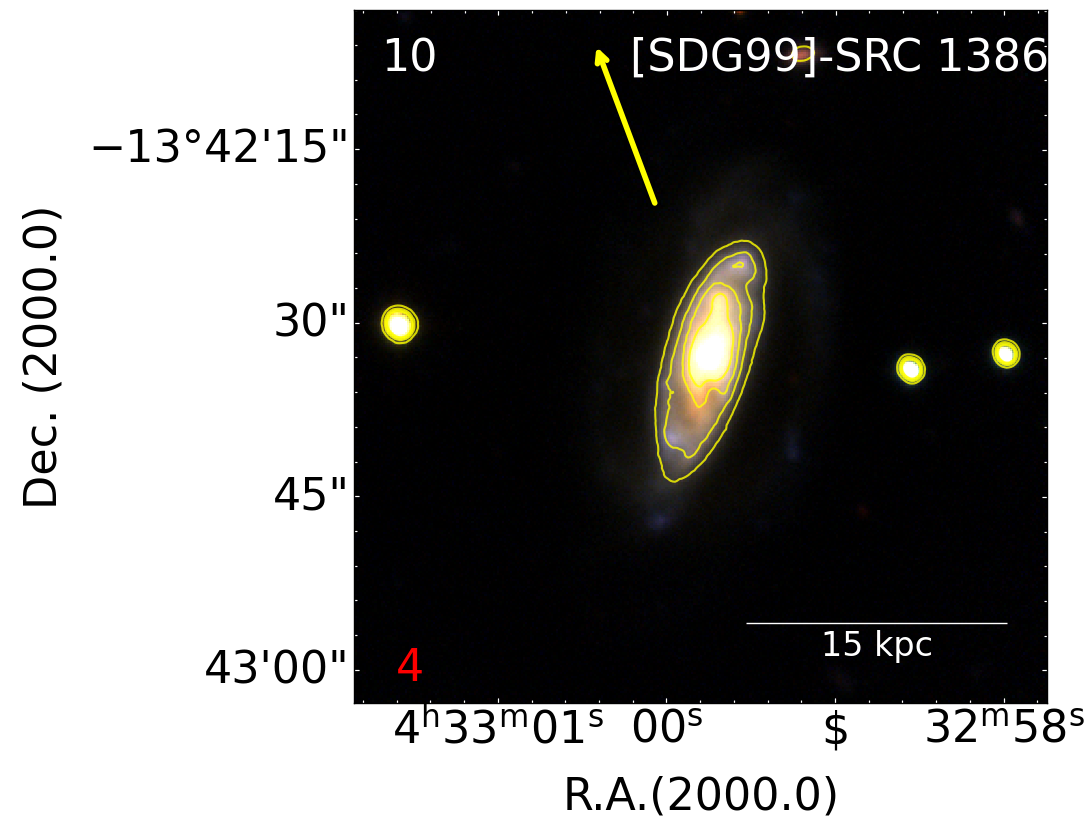}
  \end{subfigure}
  \begin{subfigure}[tbp]{0.23\textwidth}
 \includegraphics[width=\columnwidth]{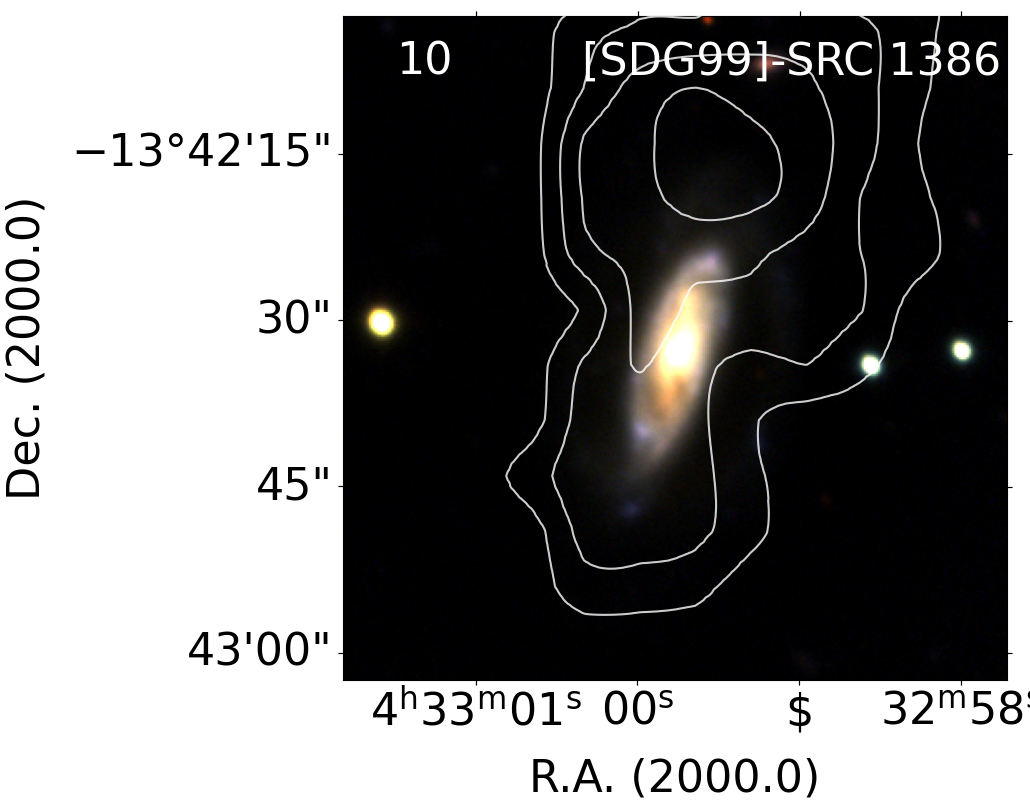}
  \end{subfigure}
   \caption{Examples of galaxies in the first four evolutionary stages of SF triggering/quenching. The UVIT FUV surface density (left panels) and the \hi\ maps (right panels) are overlaid on optical. The galaxy names are displayed in the upper-right corner, the galaxy IDs appear in the upper-left corner, and the evolutionary stage number is indicated in red in the bottom-left corner as shown in Tables\,\ref{tab_A496_FUV_galaxies} and \ref{tab_A496_galaxies_FUV_fields}. The yellow contours indicate the FUV emission and the white contours show the \hi\ distribution, both starting at 2.5rms; the corresponding beam size is 24\prin \por\,17\prin. Boxes are 1\por1 arcmin$^2$ and the yellow arrow points to the cluster center. The galaxies shown are: top-left [SDG99]-SRC\,1547 (stage 1); top-right [DFL99]\,198 (stage 2); bottom-left [SDG99]-SRC\,1054 (stage 3); bottom-right [SDG99]-SRC\,1386 (stage 4). The only object in stage 5, [DFL99]\,225, presents too low FUV and \hi\ emissions to be displayed (see Figure\,\ref{Fig_FUV_HI_images}).}
   \label{Fig_FUV_HI_contours}
\end{figure*}

Considering the FUV and \hi\ properties of the studied sample we propose an evolutionary pathway consisting of five steps. 
\\ (1) {\it Pre-triggering:} galaxies display normal \hi\ and their SFR remains close to the main sequence. [SDG99]-SRC\,1461, 1547 (IDs 13, 18) are good examples. \\(2) {\it Initial SF-triggering:}  the FUV luminosity rises significantly and the corresponding morphology becomes peculiar. 
The SFR moves up from the MS but not far beyond. The \hi, though, shows only minor perturbations, if any at all; [SDG99]-SRC\,1169, [DFL99]\,198, [SDG99]-SRC\,1549 (IDs 2, 8, 19) would be in this stage. \\
(3) {\it Peak of star-formation:}  Objects in this phase reach the maximum of SF activity; they show remarkable features in FUV, like asymmetric arms and bright knots of star formation, similar to those reported by authors like \cite{Bigiel+10, Das+21, Lee+17}.  Strong disruptions in \hi\ appear in this phase. Three objects in our sample would be in this stage, [SDG99]-SRC\,1054, 1565, 1656 (IDs 1, 20, 22). \\
(4) {\it SF-fading:} The SFR fades but the galaxy still remains bright in FUV while the last generations of blue stars are still shining. The \hi\ appears disrupted and the galaxy might reach the gas deficiency threshold. Galaxies get significantly redder than in previous stages, like [SDG99]-SRC\,1386, 1584 (IDs 10, 21).\\ (5) {\it SF-quenching:} At this point the SFR fades dramatically and galaxies properly start to quench. Low FUV emission, red colors and high \hi-deficiency are the signs of this stage. It is worth noting that the full quenching process might last longer than the time needed to reach stage 5 \citep[e.g.][]{Vulcani+18, George+23}.  Figure\,\ref{Fig_FUV_HI_contours} shows examples of galaxies in the evolutionary stages 1 to 4. [DFL99]\,225 (ID 16) is the only case in our sample having reached stage 5; it is too faint in UV and \hi\ so it is not displayed.

\subsection{Studying the role of tidal interactions}
\label{subsecc:HI_kinematic}

\citetalias{Lopez-Gutierrez+22} sought for substructures that account for the blue, \hi-rich galaxies projected near the cluster core of A496. Those authors applied two independent methods searching for substructures \citep{Dressler&Shectman+88, Serna&Gerbal+96}, finding none within 1\,$R_\mathrm{200}$. Considering that present methods unveiling substructures in clusters are less effective when a group moves along the LOS, we suspect this is the case occurring in A496. With the aim of disentangling the origin of the blue, \hi-rich galaxies projected at a short cluster-centric radius (Figure\,\ref{Fig_PPS_perturbance}), we use the 2-D information provided by the \hi\ (Table\,\ref{tab_A496_galaxies_FUV_fields} and Figure\,\ref{Fig_FUV_HI_images}).  Mapping different gas phases like \hi, CO, and H$\alpha$, might provide valuable information on the plane-of-sky velocity of cluster galaxies \citep{Bravo-Alfaro+00, Vollmer+09, Yagi+10, Yagi+17, Poggianti+19, Chen+20}.
In this work, we use elements such as gas trails, one-sided compressions, and offsets (position and velocity, between optical and \hi) to infer the dominant vectors of movement.  

\begin{figure}
     \centering
\includegraphics[width=1\columnwidth]{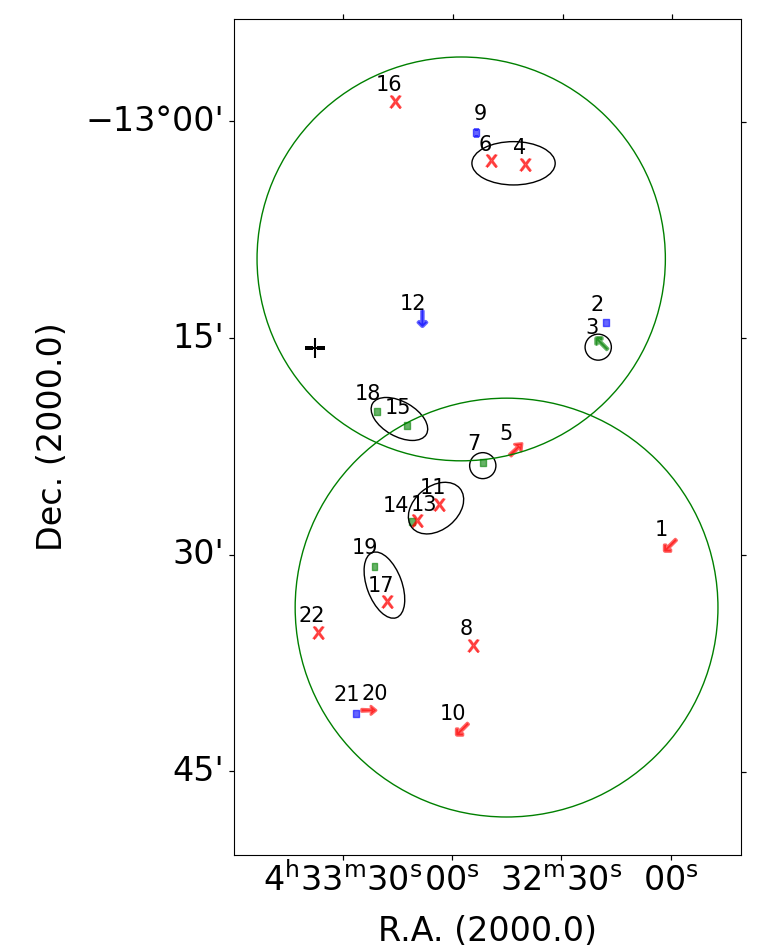}
    
    \caption{Plane of sky motions (arrows) of FUV galaxies within the UVIT fields. Crosses (points) indicate motion along the LOS, away from (towards) the observer.  The color code separates line-of-sight velocities in three slots (see text): low (blue), medium (green), and high (red). Galaxies with close companions are highlighted with black ellipses. The cD position is indicated with a black cross.}
\label{Fig_A496_distribution_kinematic}
\end{figure}

Figure \ref{Fig_A496_distribution_kinematic} resumes our results, where arrows show the movement component on the plane of the sky; crosses/points indicate motion along the LOS, away/towards the observer.  The color code separates line-of-sight velocities into three slots, relative to A496 systemic value (9\,892 \kms): \\
--blue corresponds to the low velocity range, \\
$v$ < 9548 \kms\
($\Delta v_\mathrm{rad} < -0.5\sigma_{\mathrm{cl}}$); \\ 
--green for 9548 \kms < $v$ < 10\,236 \kms \\
($-0.5\sigma_{\mathrm{cl}} < \Delta v_\mathrm{rad} < 0.5\sigma_{\mathrm{cl}}$); \\
--red for high velocity range, 10\,236 \kms\ < $v
$\\ ($0.5\sigma_{\mathrm{cl}} <\Delta v_\mathrm{rad}$).

As mentioned before, the sample of FUV galaxies in A496 is dominated by objects with radial velocities larger than the systemic value of A496 (see Figure\,\ref{Fig_Histograma_A496_Mass_and_Velocity} and Figure\,\ref{Fig_PPS_perturbance}).
In order to provide statistical support to this result we applied the random {\it shuffling} test, which is a robust statistical method used to determine if an observed pattern in a dataset has a significant effect or a chance result.  The two samples to be compared are the line-of-sight velocities of the FUV galaxies and the corresponding values for the cluster members.  For congruence, we selected only those members lying within the two UVIT fields, avoiding any positional bias on the velocity distribution. We defined the null hypothesis, $H_o$: {\it the sample of FUV galaxies do not have larger velocity than the cluster population}. If we confirmed this statement, the FUV objects would have the same distribution of radial velocities as all the cluster members. We calculated the test by comparing the median values, helping to reduce the effect of outliers. We randomly {\it shuffle} the data and repeat the process through 10\,000 iterations. As a result, the p-value is systematically below 0.05 and, in fact, very close to 0.01. We therefore reject the null hypothesis. Similar results are obtained by comparing the mean of the same samples. We conclude, with a very high likelihood, that line-of-sight velocities of FUV galaxies are, for real, systematically higher than the bulk of cluster members. To accomplish the statistical test we used a public Python routine developed by one of the authors (GAM, 
{\it CompareByShuffles}\footnote{https://gitlab.com/gmamon/python-codes}). \\

\noindent 
$\bullet${\bf Correlations with high radial velocities}\\

We analyzed the joint FUV and \hi\ properties of the FUV galaxies as a function of their velocities relative to the cluster. In principle, objects with high infalling velocities could show major disruptions due, for instance, to RPS effects. We seek any possible correlation between FUV/\hi\ disruptions vs. relative velocity but only a slight trend appears, with more massive galaxies having larger velocities relative to the cluster compared with low-mass counterparts (see Fig.\,\ref{Fig_Histograma_A496_Mass_and_Velocity}). Most of these {\it massive} objects present high sSFR and/or FUV disruptions, being classified in evolutionary stages 2 or beyond. These galaxies with stellar mass above 10$^9$\,\msolar\ and $\Delta v \geq 1\sigma$ are identified with IDs 2, 8, 12, 16, 20, 22.  Interestingly, Figure\,\ref{Fig_A496_distribution_kinematic} shows that these objects have no companions, some of them projected at large cluster-centric distances.  \\

\noindent 
$\bullet$ {\bf The presence of galaxy companions}\\

It is known that tidal interactions play a key role in galaxy evolution, triggering starburst episodes under specific conditions. With the aim to explore the role played by gravitational mechanisms in the evolutionary sequence described above, we applied a two steps strategy. First, we seek for possible companions whose interaction could explain -at least in part- the observed SF activity.  Second, we inspected for possible disturbances in the stellar body of the low-mass galaxies. With this aim,
we carried out a preliminary asymmetry analysis based on $r$ and $z$-band images, looking for obvious asymmetries on the old stellar component. These are known to be produced by tidal interactions rather than by RPS \citep{Boselli&Gavazzi+06}. We produced an axial symmetric model and inspected the residual after subtraction from the original image, following \cite{Venkatapathy+17}.

We search for possible neighbors around all our FUV objects by applying quite conservative criteria: a projected separation $\leq$\,100\,kpc\ and a |$\Delta v$|\,$\leq$\,500\,\kms; the latter is the typical range of velocity dispersion observed in different types of groups. Under such assumptions, eleven of the 22 FUV galaxies have (at least) one companion, typically another FUV galaxy. Figure\,\ref{Fig_A496_distribution_kinematic} indicates (with black ellipses) those objects having a close neighbor. We observe a very interesting pattern: galaxies with masses above 10$^9$\msolar, showing high SFRs and \hi\ disruptions, do {\it not} have close neighbors. These are the same galaxies with high relative velocity listed in the previous paragraph, with objects ID 2 and 12 joining this sub-sample. Not having close companions, all these objects would be candidates to be under strong RPS regime. The only exception is the galaxy [SDG99]-SRC\,1547, with a stellar mass of $\sim$ 3 \por10$^9$\msolar, having  [SDG99]-SRC\,1482 as a close neighbor (ID 18, 15 respectively); both are in evolutionary stage 2.

The sample of FUV objects with close neighbors is dominated by galaxies with stellar masses below 10$^9$\msolar; these are identified with ID 4, 6, 11, 13, 14, 15, 17.  In these cases, tidal interactions must play an important role raising the SF-activity and producing gas disruptions. We show two examples of galaxies with clear asymmetries seen in {\it r}-band (Figure\,\ref{Fig_asymmetries}); $z$-band frames being of reduced quality compared with $r$, are not shown. One additional fact should be considered when evaluating tidal interactions. The criteria we used to find possible {\it interactors} are quite conservative, as the velocity range does not include flyby-interactions. These might occur between objects with relative velocities well beyond the applied criteria of 500\,\kms. This might concern a low-mass galaxy like [SDG99]-SRC\,1366 (ID 9), one of the brightest in FUV and a very striking object in this sample, having a possible neighbor with relative velocity of 700\,\kms.

\begin{figure}
     \begin{subfigure}[tbp]{0.23\textwidth}
 \includegraphics[width=\columnwidth]{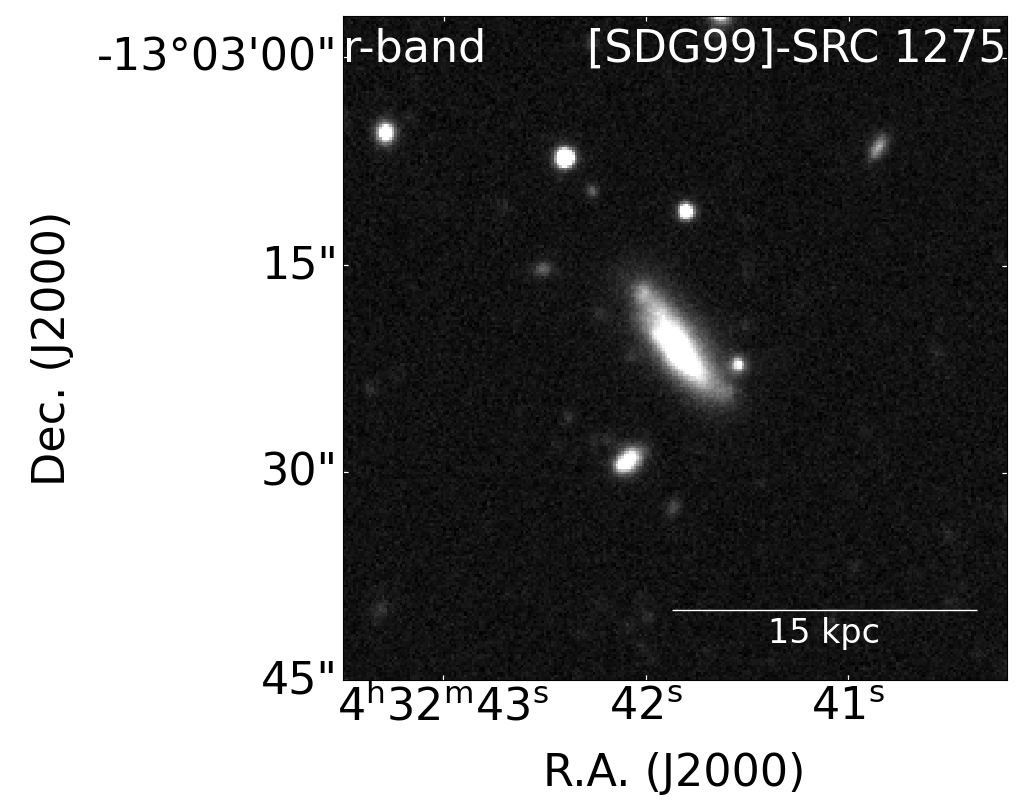}
  \end{subfigure}
  \begin{subfigure}[tbp]{0.23\textwidth}
 \includegraphics[width=\columnwidth]{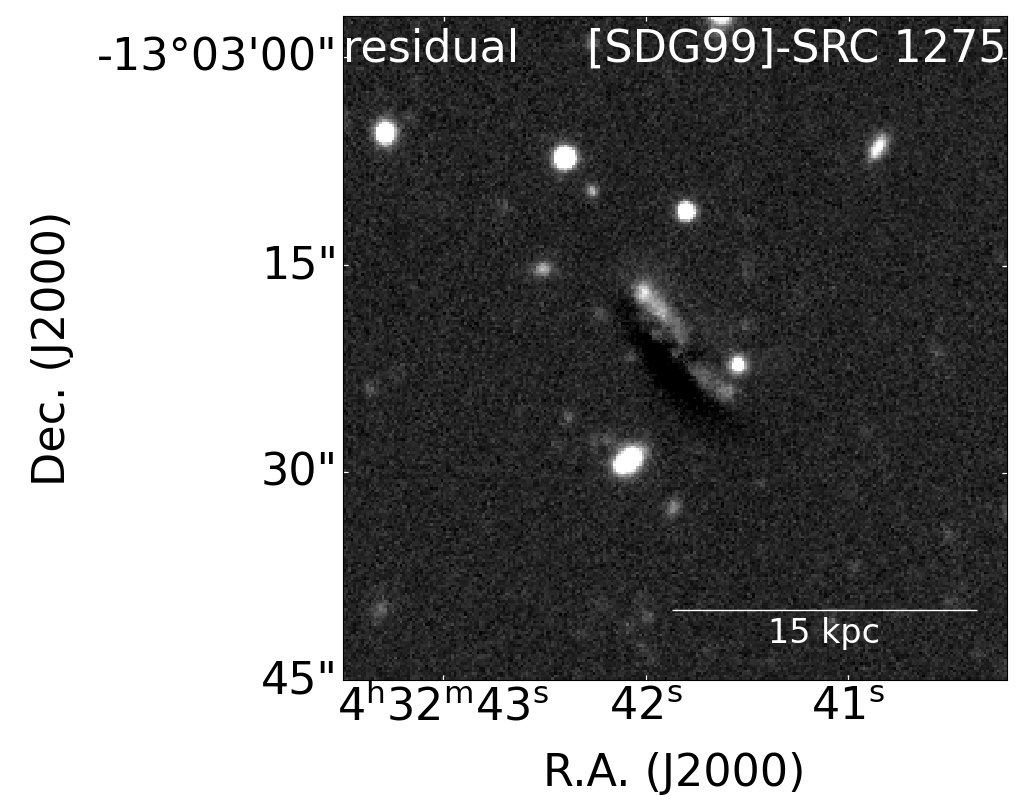}
  \end{subfigure}
  \\
   \begin{subfigure}[tbp]{0.23\textwidth}
 \includegraphics[width=\columnwidth]{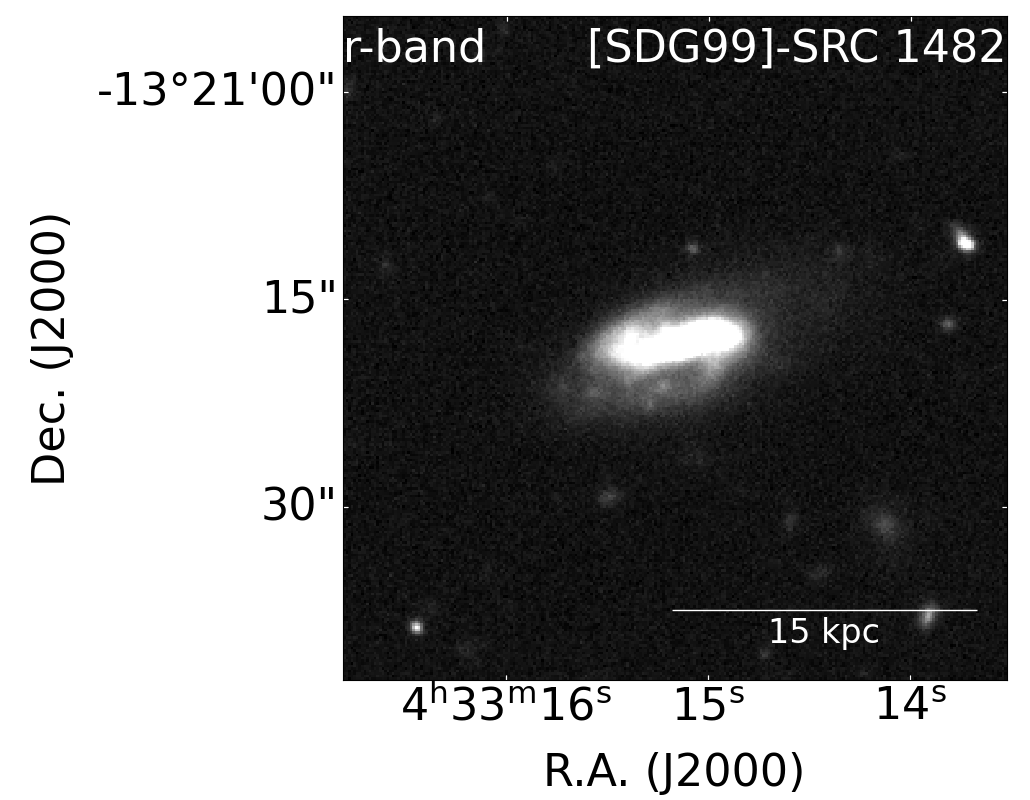}
  \end{subfigure}
  \begin{subfigure}[tbp]{0.23\textwidth}
 \includegraphics[width=\columnwidth]{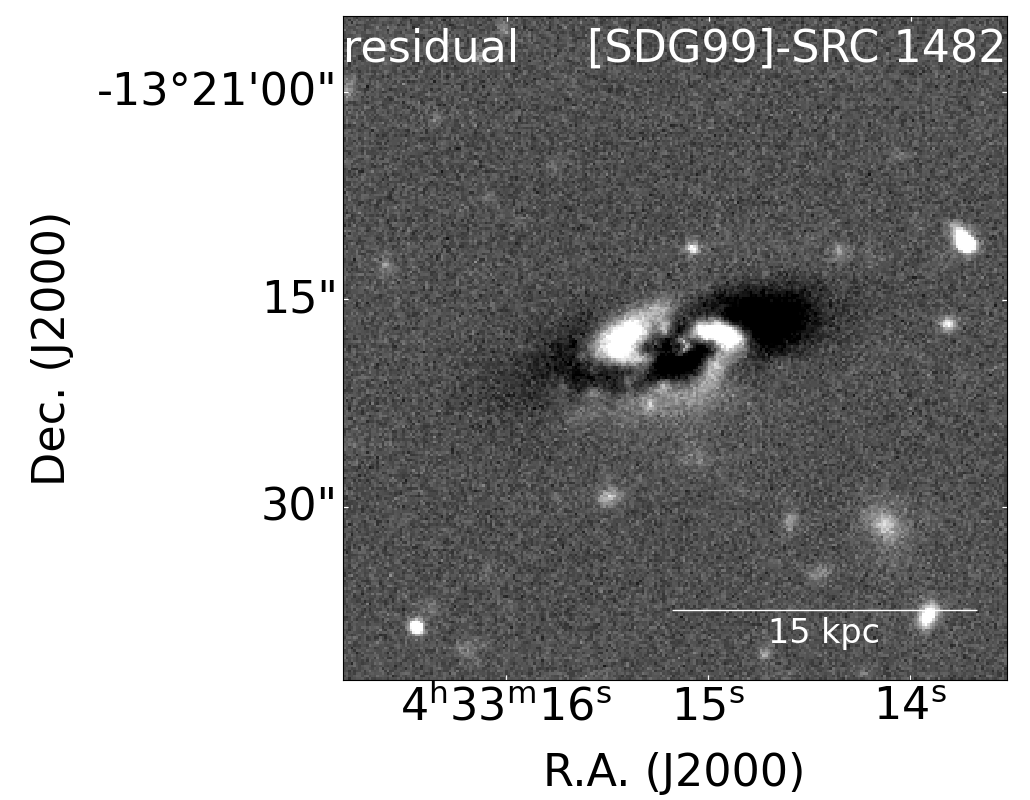}
  \end{subfigure}
   \caption{Two examples of low-mass FUV galaxies showing clear asymmetries in the $r$-band (left panels). Residuals after subtracting an axial symmetric BMODEL are shown (right panels). [SDG99]-SRC\,1275 (ID 4) is on top, and [SDG99]-SRC\,1482 (ID 15) on the bottom.}
   \label{Fig_asymmetries}
\end{figure}

This trend observed in A496, with more massive FUV galaxies being isolated, and low-mass counterparts having close companions, must be further studied on the basis of larger samples of galaxies in clusters with different environment conditions. Finally, the two galaxies with the lowest stellar-mass in our sample (04.56-13.38 and [SDG99]-SRC\,1584; ID 5, 21) do not follow this trend. They have no close companions and present the highest sSFR among all the FUV galaxies in this study. They must be under the effects of the cluster environment and constitute candidates to be {\it very young galaxies}, deserving a deeper study on their own.

\section{Summary and Conclusions}
\label{secc:conclusions}

We report the detection of 22 FUV galaxies within two 28$’$ fields observed with UVIT-AstroSat in the relaxed cluster A496. All these FUV galaxies have been previously detected in \hi, with one exception, and many of them are intriguingly projected onto (or close to) the cluster virialized zone. We combined the study of SF-activity with the \hi\ as a tracer for environmental effects. Our principal conclusions are summarized as follows:\\

\noindent
$\bullet$ From the 22 FUV galaxies, ten are below the stellar mass threshold of 10$^9$\msolar. We estimated the SFR normalized by stellar mass (sSFR$_\mathrm{FUV}$) finding that half of the sample shows SF-activity significantly above the main sequence. We do not find any positive correlation between SF-activity and stellar mass. In fact, we have found that the highest sSFR values correspond to galaxies with stellar masses below 10$^9$\msolar.\\

\noindent
$\bullet$ We confirm a moderate correlation between \hi-content and $L_\mathrm{FUV}$. Cold gas is the primary fuel to build new stars and constitutes a good tracer for SF-activity in the absence of more direct indicators, like molecular and ionized gas. \\

\noindent
$\bullet$We traced PPS diagrams to reconstruct the infalling history of the FUV galaxies. Globally speaking, they are distributed across the PPS and do not cluster like the bulk of cluster members, which is consistent with an infalling population. However, more than half of the FUV sample is projected within the virialized zone, most of them with rather normal \hi\ content. We conclude that Abell 496 could be feeding with fresh, gas-rich objects from the foreground. However, a larger sample of galaxies and an homogeneous volume coverage across the cluster, is needed to settle this question. \\

\noindent
$\bullet$ All FUV galaxies in this study retain significant amounts of \hi\ gas. We find that less than half of the studied sample shows strong disruptions in \hi, while almost all are classified as FUV-peculiar. We conclude that the gas depletion time-scale, i.e. a few times 10$^8$\,yr, constitutes an upper limit to the age of the observed disruption/infalling process.  This supports that most galaxies in the studied sample have not reached their first pericenter passage, which needs more than 1\,Gyr. \\

\noindent
$\bullet$ We sought possible companions of the FUV galaxies, within a radius of 100\,kpc, and $\Delta v \lesssim 500$\,\kms. Interestingly, all galaxies with stellar masses above 10$^9$\msolar\ and being in advanced infalling stages (2 or beyond, see below) have no close neighbors. This supports the idea that RPS dominates this subsample. The opposite occurs with the low-mass objects, where most of the FUV dwarfs have at least one companion. Therefore, the triggering of SF activity for low-mass galaxies could be related to tidal interactions. A preliminary asymmetry analysis of the stellar component, carried out on $r, z$-bands, supports this hypothesis.\\

\noindent
$\bullet$ After combining the FUV and \hi\ data in A496, we propose an evolutionary sequence consisting of five stages occurring since the beginning of their first infall, and lasting a few times 10$^{8}$\,yr: (1) {\it Pre-triggering:} galaxies are undisturbed in \hi, with normal SFR along the MS. (2) {\it Initial SF-triggering:} galaxies show minor \hi\ disruptions but SFR values move significantly upwards from the MS. (3) {\it Peak of star-formation:} the \hi\ becomes more perturbed and SFR reaches its peak; FUV emission might show strong asymmetries. (4) {\it Star-formation fading:} the \hi\ can be very disrupted; FUV emission fades significantly. (5) {\it SF-quenching:} SFR drops below the main sequence and objects become red passive, and strongly \hi-deficient. \\

Combining multi-wavelength data, like the present FUV and \hi\ study, shows to be a powerful tool for studying galaxy evolution in nearby clusters.  Our results emphasize the importance of investigating a reliable star-formation indicator plus the cold gas to infer the orbital histories and the cycle of gas. In A496 we observe a dual evolutionary path, where the SF-triggering and gas removal of more massive FUV galaxies (>\,10$^9$\msolar), are dominated by the global cluster environment, most likely through ram pressure stripping. For these objects, pre-processing does not appear to play a major role. On the contrary, we report that most of the low-mass FUV objects (<\,10$^9$\msolar), have close companions. This strongly suggests that tidal interactions should not be underestimated as an important player driving the transformation of infalling galaxies.\\

We emphasize that the UVIT fields analyzed in this study were selected ad-hoc, containing a large number of blue galaxies. Therefore, the studied sample of FUV galaxies is not necessarily representative of the cluster. A larger FUV coverage is strongly recommended to carry out a statistical analysis. It is equally important to extend this study to other clusters with different physical (environment) properties.

\section*{Acknowledgements}
The authors thank an anonymous referee for her/his very helpful comments, which helped to improve this paper. M.M.L.G. and H.B.A. acknowledge diverse funding from CONACyT and from DAIP-UG, in Mexico. We thank the anonymous referee for valuable comments that significantly improved the paper. M.M.L.G. is supported by KIAS Individual Grant (PG101201) at Korea Institute for Advanced Study (KIAS), in Republic of Korea. H.B.A. acknowledges the  {\it CNRS/Sorbonne-Univ., Institut d'Astrophysique de Paris,} for the hospitality during a working visit. We acknowledge J. Fritz, B. Cervantes, C.A. Caretta, and R. Coziol for useful comments improving this work. YLJ acknowledges support from the Agencia Nacional de Investigaci\'on y Desarrollo (ANID) through Basal project FB210003, FONDECYT Regular projects 1241426 and 123044, and  Millennium 
Science Initiative Program NCN2024\_112. \\

%%%%%%%%%%%%%%%%%%%%%%%%%%%%%%%%%%%%%%%%%%%%%%%%%%
\section*{Data Availability}
Data available on request.



%\bibliographystyle{mnras}

\bibliographystyle{mnras}
\bibliography{referencias}

%\begin{thebibliography}{99}
%\end{thebibliography}

%%%%%%%%%%%%%%%%%%%%%%%%%%%%%%%%%%%%%%%%%%%%%%%%%%

%%%%%%%%%%%%%%%%% APPENDICES %%%%%%%%%%%%%%%%%%%%%

\appendix 
\label{secc:apendix}

\section{The FUV images and \hi\ maps of FUV galaxies}
%%%%%%%%%%%%%%%%%%%%%%%%%%%%%%%%%%%%%%%%%%%%%%%%%%

The atlas of 22 FUV galaxies is reported in the present work.  The FUV images are shown in the left panels and \hi\ maps, overlaid on optical CFHT-$ugr$ images, in the right panels.

%%%%%%%%%%%%%%% FUV and HI images %%%%%%%%%%%%%%%%%%%%

\begin{figure*}
\centering
\begin{subfigure}[tbp]{0.23\textwidth}
 \includegraphics[width=\columnwidth]{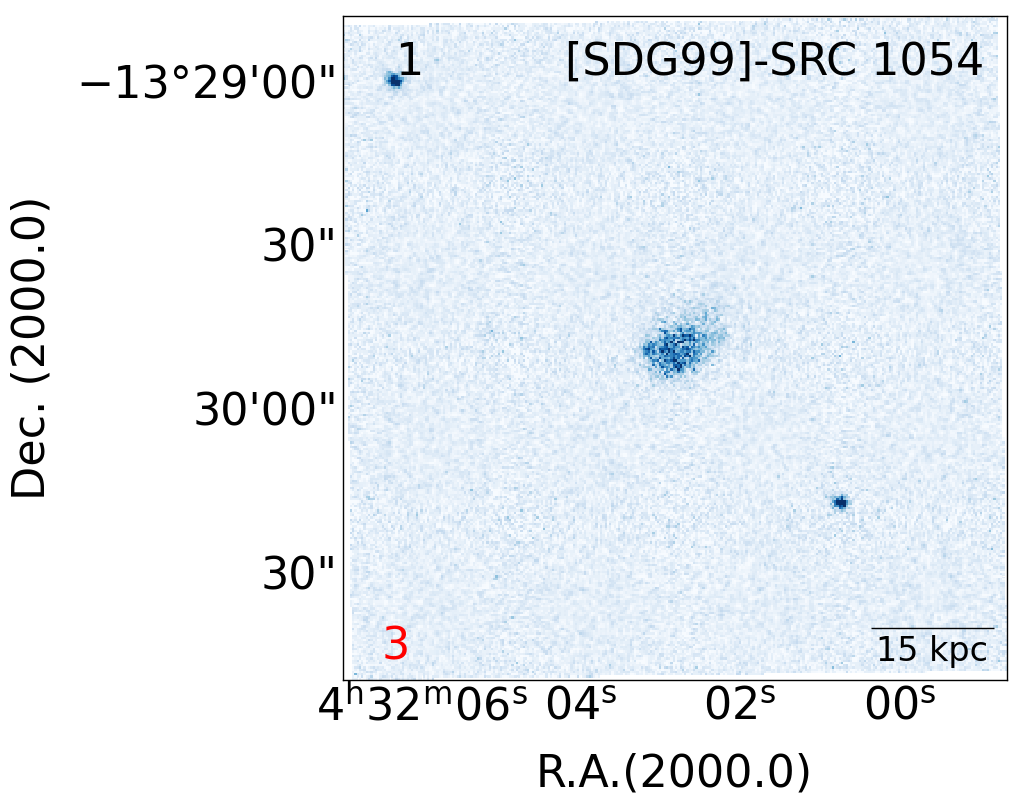}
  \end{subfigure}
  \begin{subfigure}[tbp]{0.23\textwidth}
 \includegraphics[width=\columnwidth]{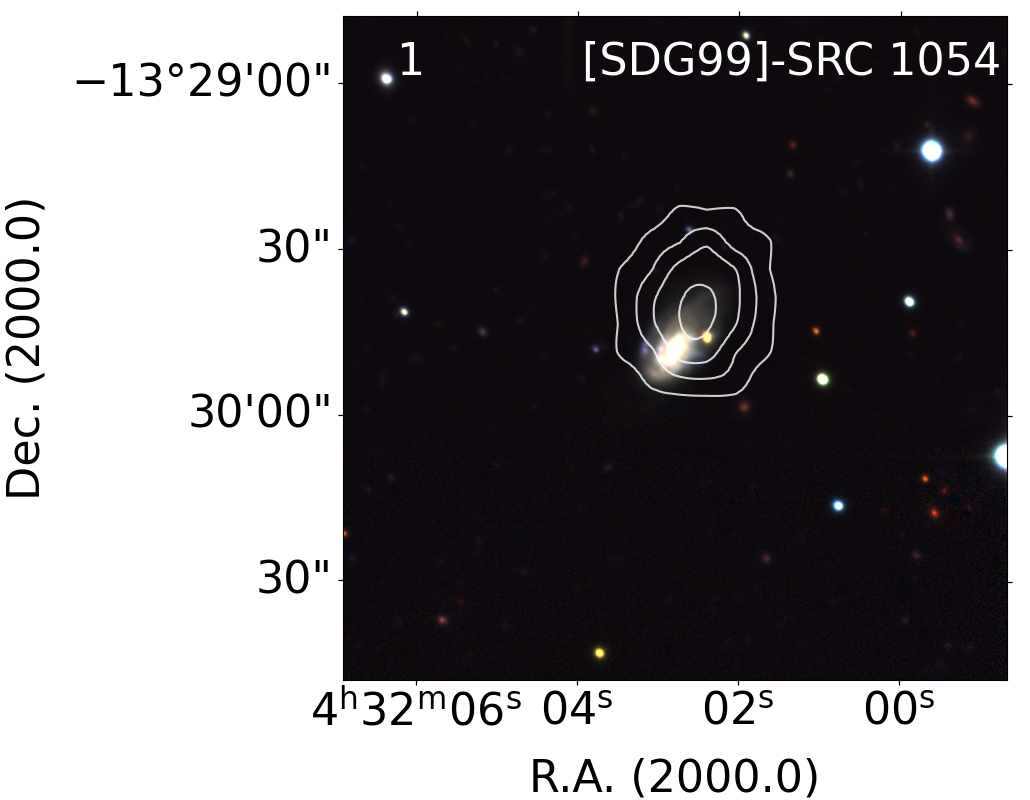}
  \end{subfigure}
 \begin{subfigure}[tbp]{0.23\textwidth}
 \includegraphics[width=\columnwidth]{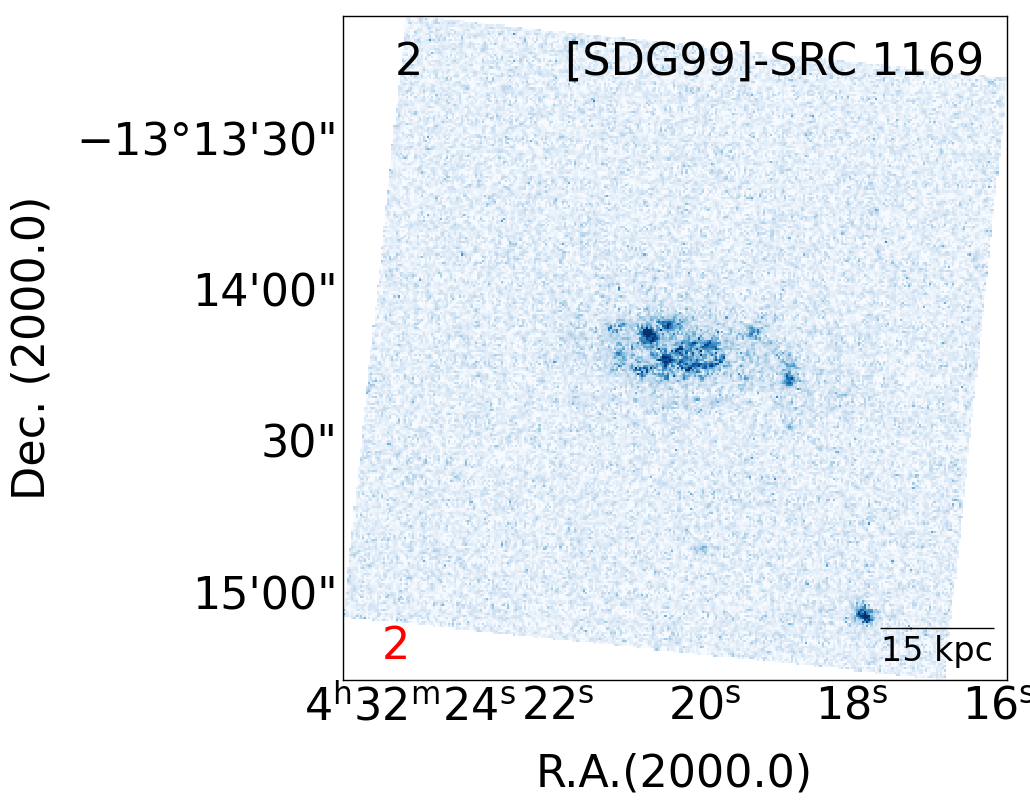}
  \end{subfigure}
  \begin{subfigure}[tbp]{0.23\textwidth}
 \includegraphics[width=\columnwidth]{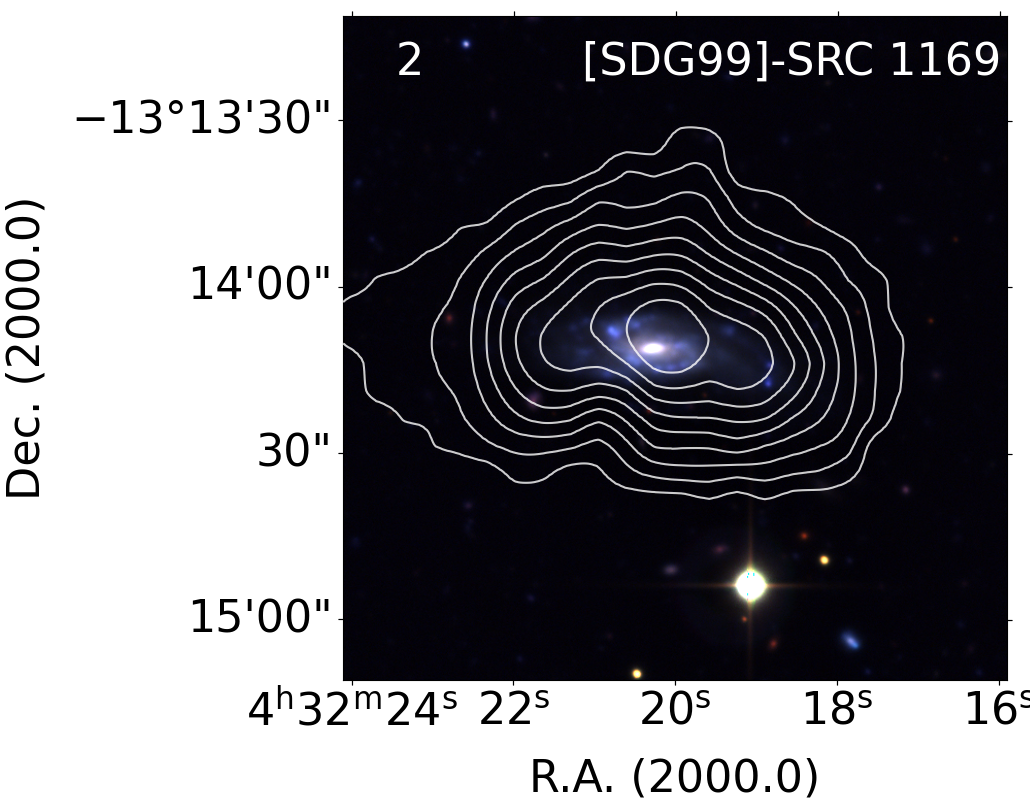}
  \end{subfigure}
  \\
  \begin{subfigure}[tbp]{0.23\textwidth}
 \includegraphics[width=\columnwidth]{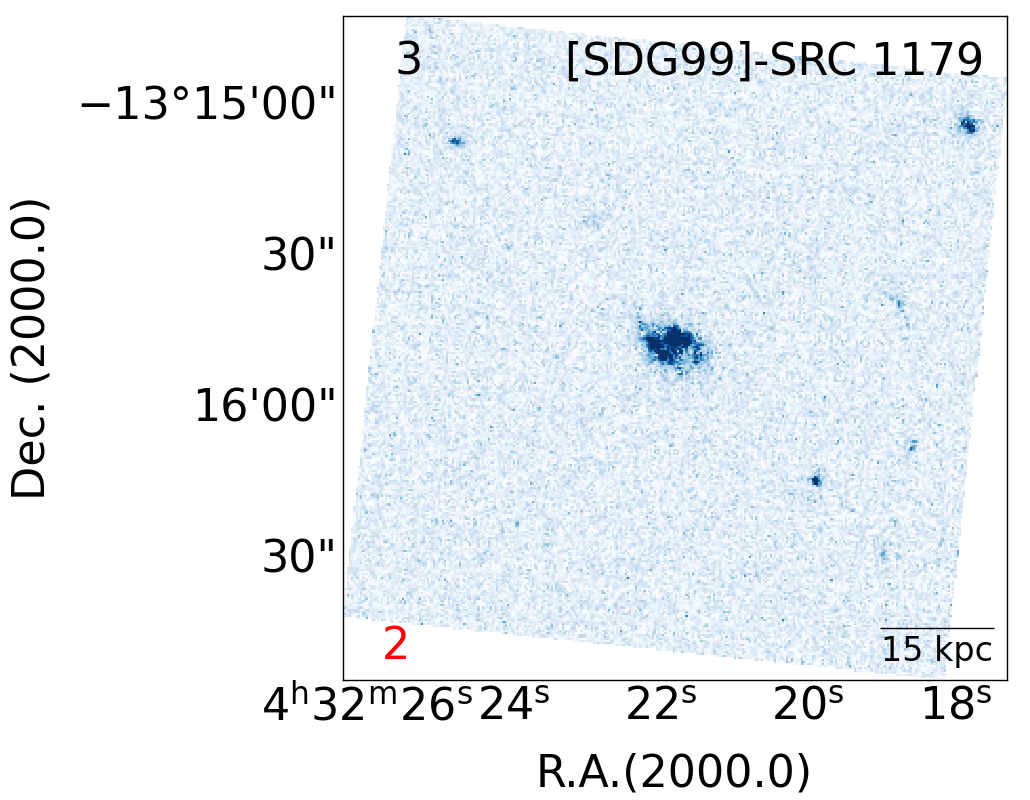}
  \end{subfigure}
  \begin{subfigure}[tbp]{0.23\textwidth}
 \includegraphics[width=\columnwidth]{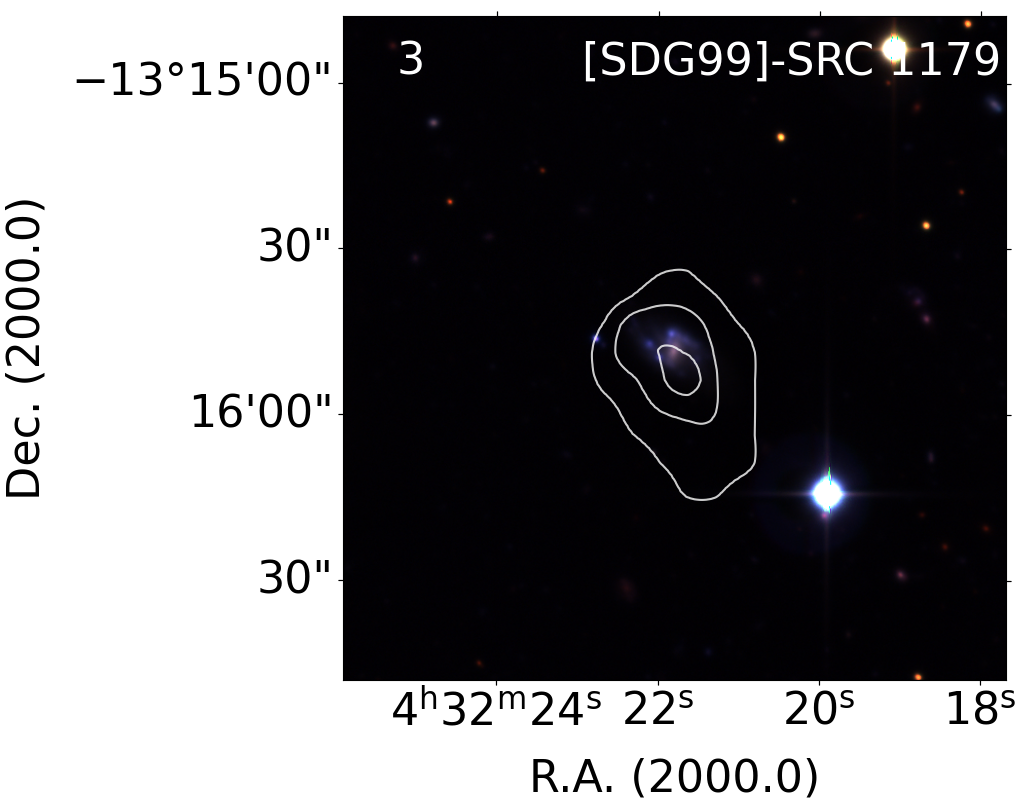}
  \end{subfigure}
   \begin{subfigure}[tbp]{0.23\textwidth}
 \includegraphics[width=\columnwidth]{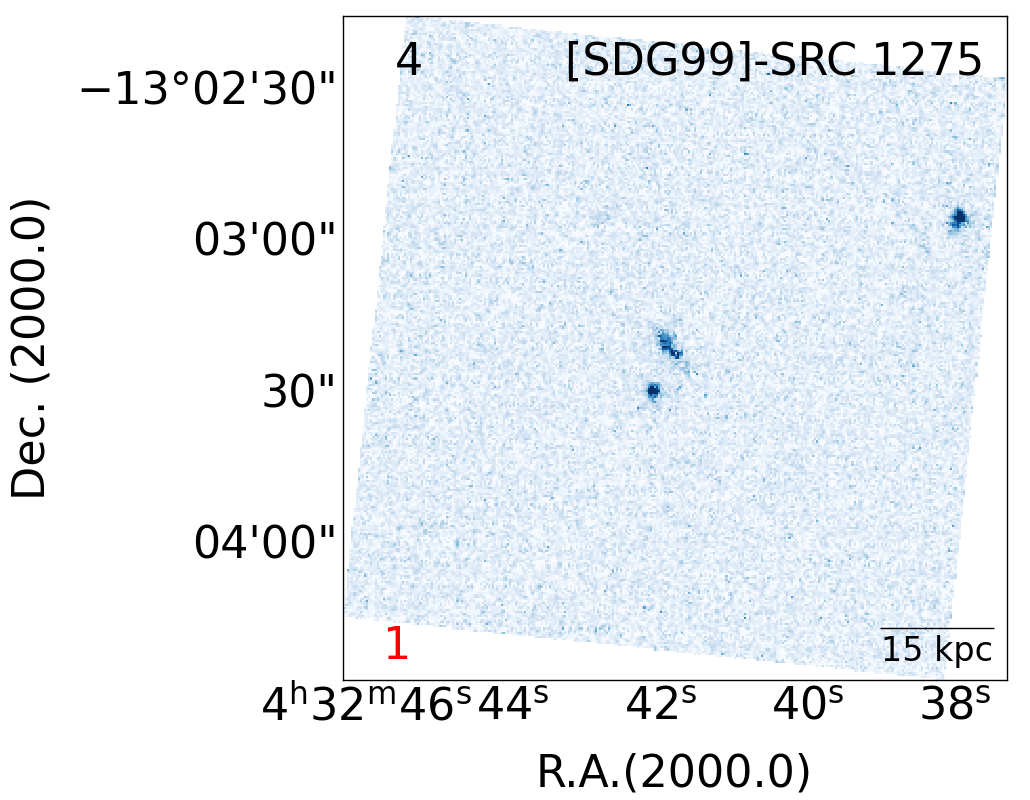}
  \end{subfigure}
   \begin{subfigure}[tbp]{0.23\textwidth}
 \includegraphics[width=\columnwidth]{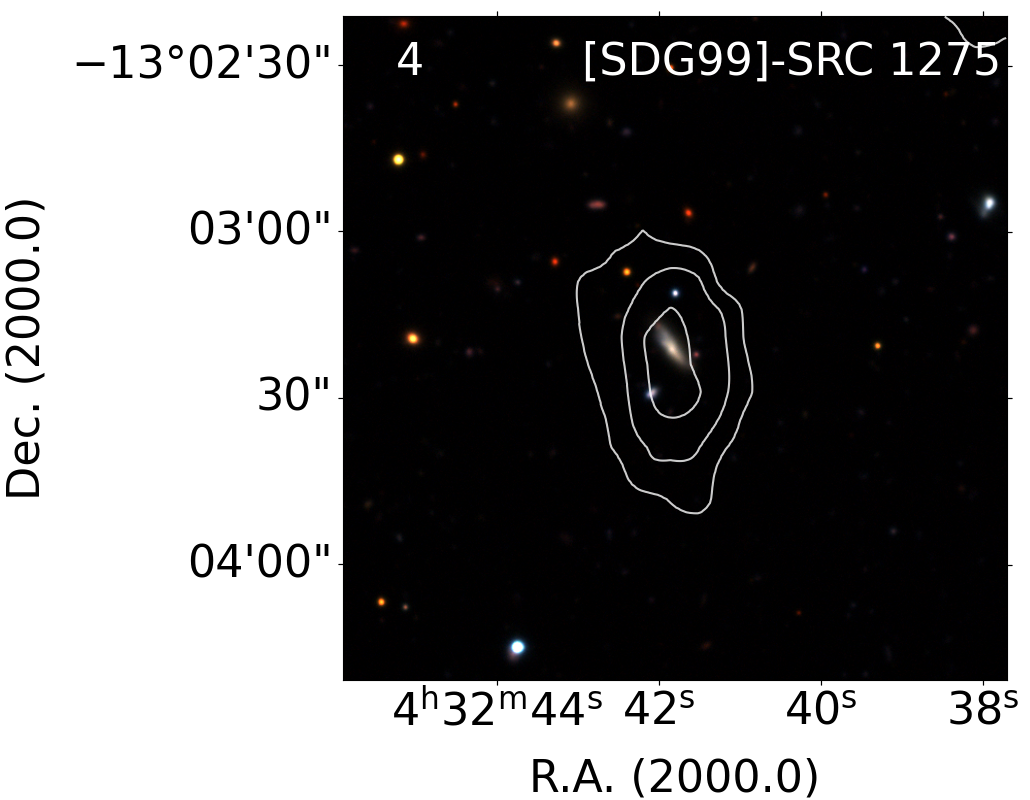}
  \end{subfigure}
  \\
  \begin{subfigure}[tbp]{0.23\textwidth}
 \includegraphics[width=\columnwidth]{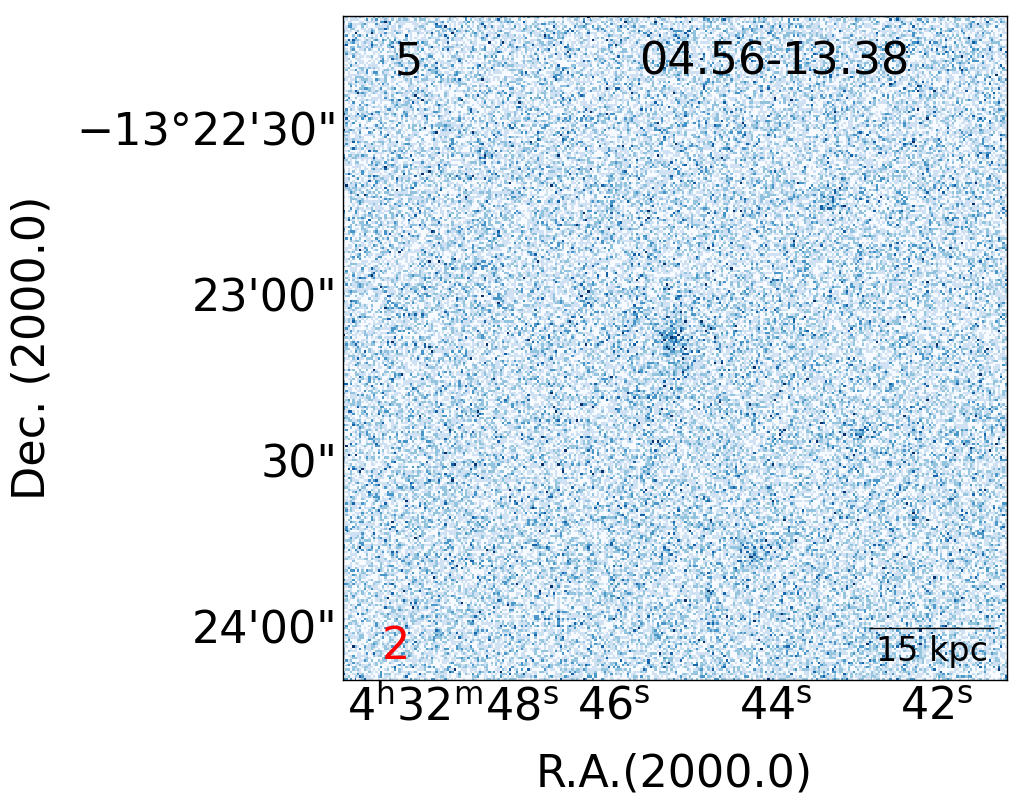}
 \end{subfigure}
 \begin{subfigure}[tbp]{0.23\textwidth}
 \includegraphics[width=\columnwidth]{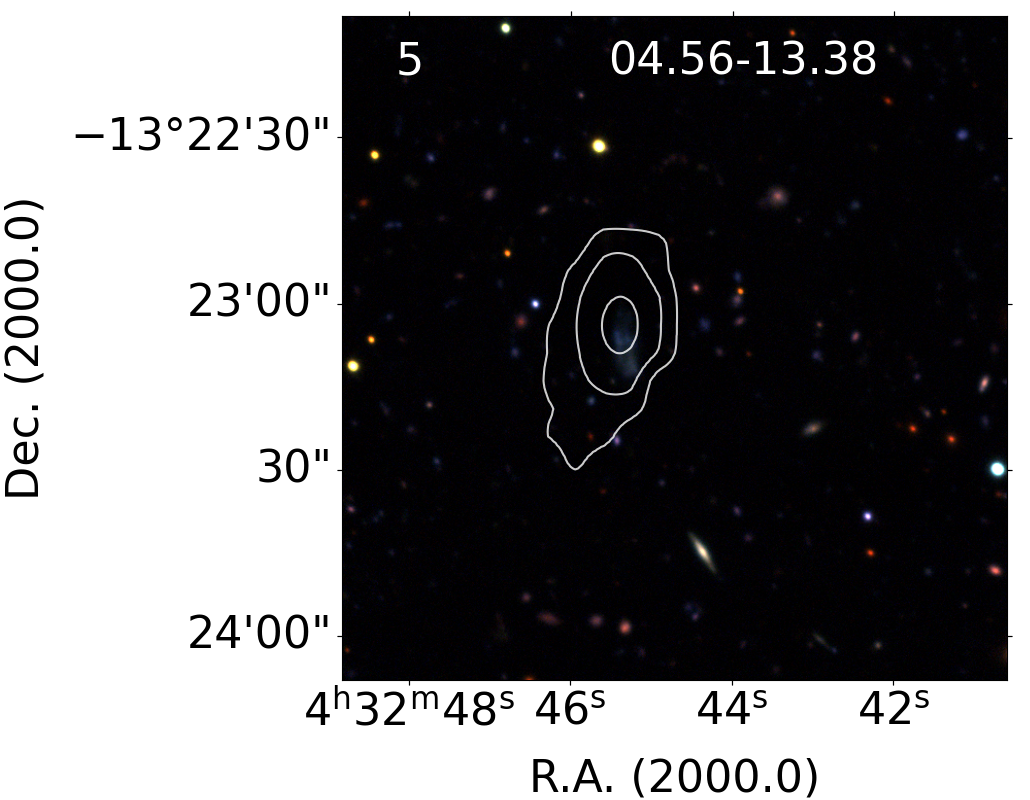}
  \end{subfigure}
   \begin{subfigure}[tbp]{0.23\textwidth}
 \includegraphics[width=\columnwidth]{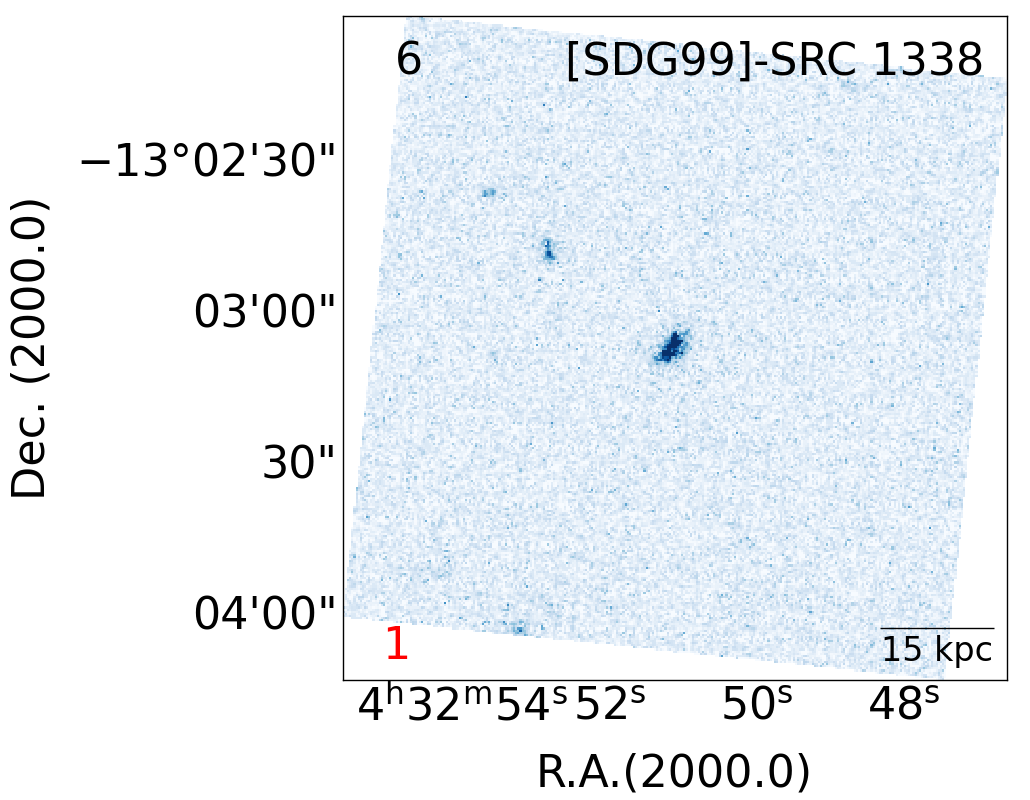}
  \end{subfigure}
     \begin{subfigure}[tbp]{0.23\textwidth}
 \includegraphics[width=\columnwidth]{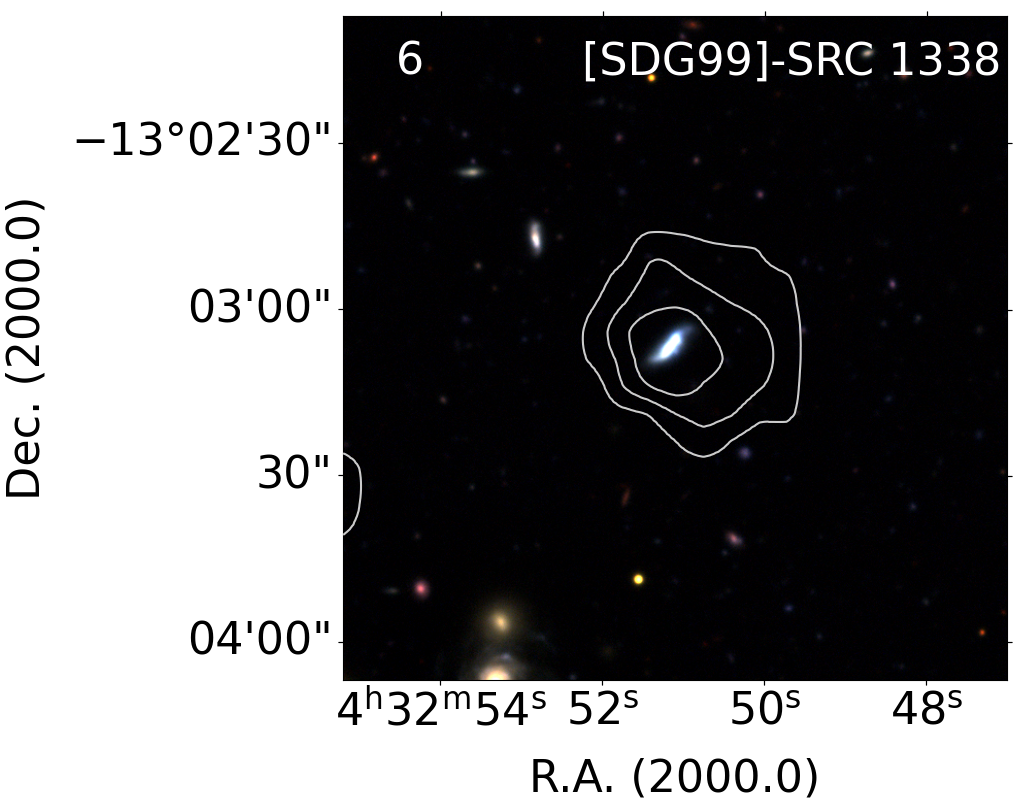}
  \end{subfigure}
  \\
   \begin{subfigure}[tbp]{0.23\textwidth}
 \includegraphics[width=\columnwidth]{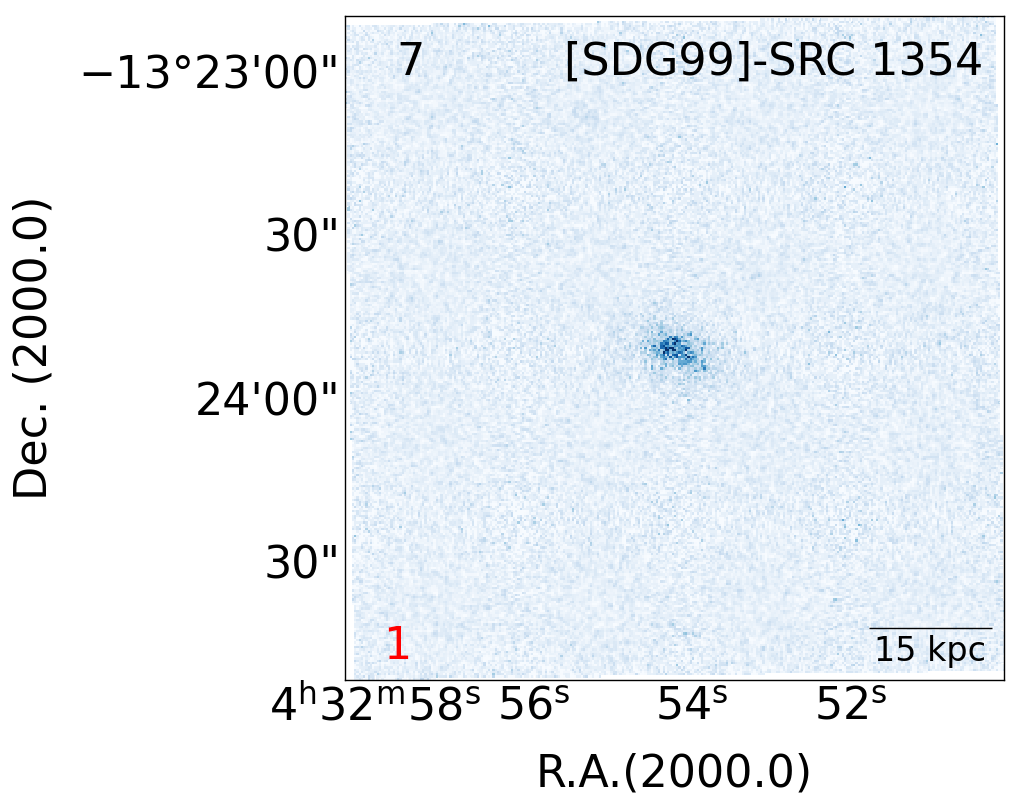}
  \end{subfigure}
  \begin{subfigure}[tbp]{0.23\textwidth}
 \includegraphics[width=\columnwidth]{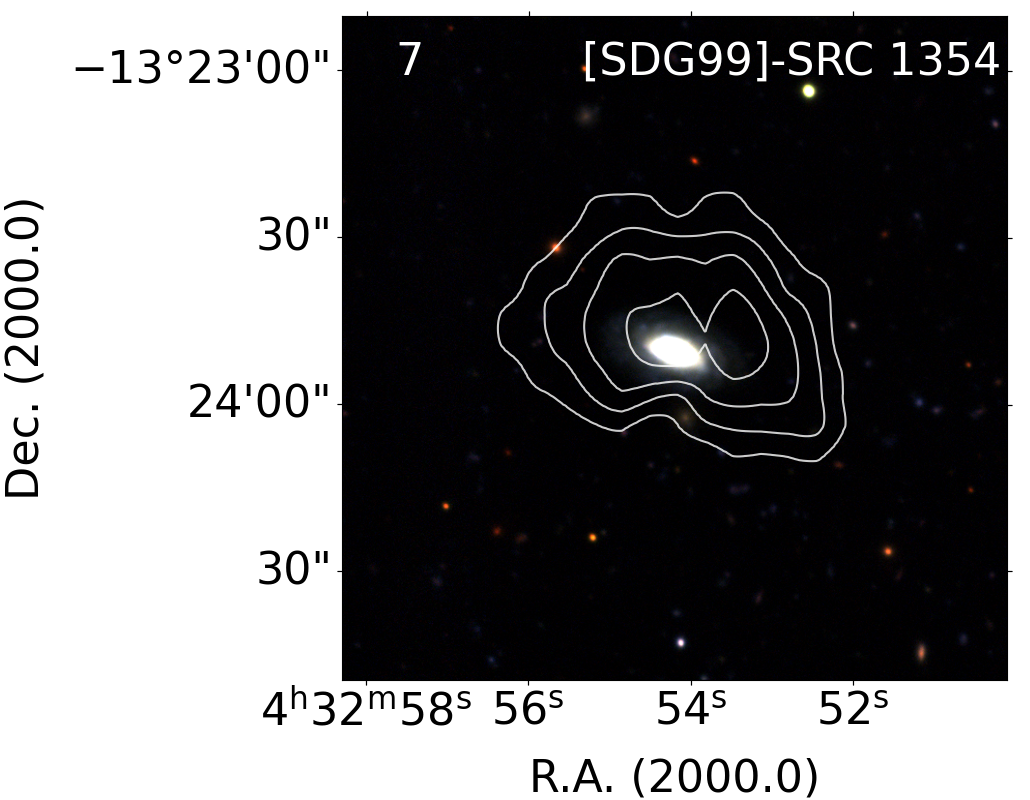}
  \end{subfigure}
   \begin{subfigure}[tbp]{0.23\textwidth}
 \includegraphics[width=\columnwidth]{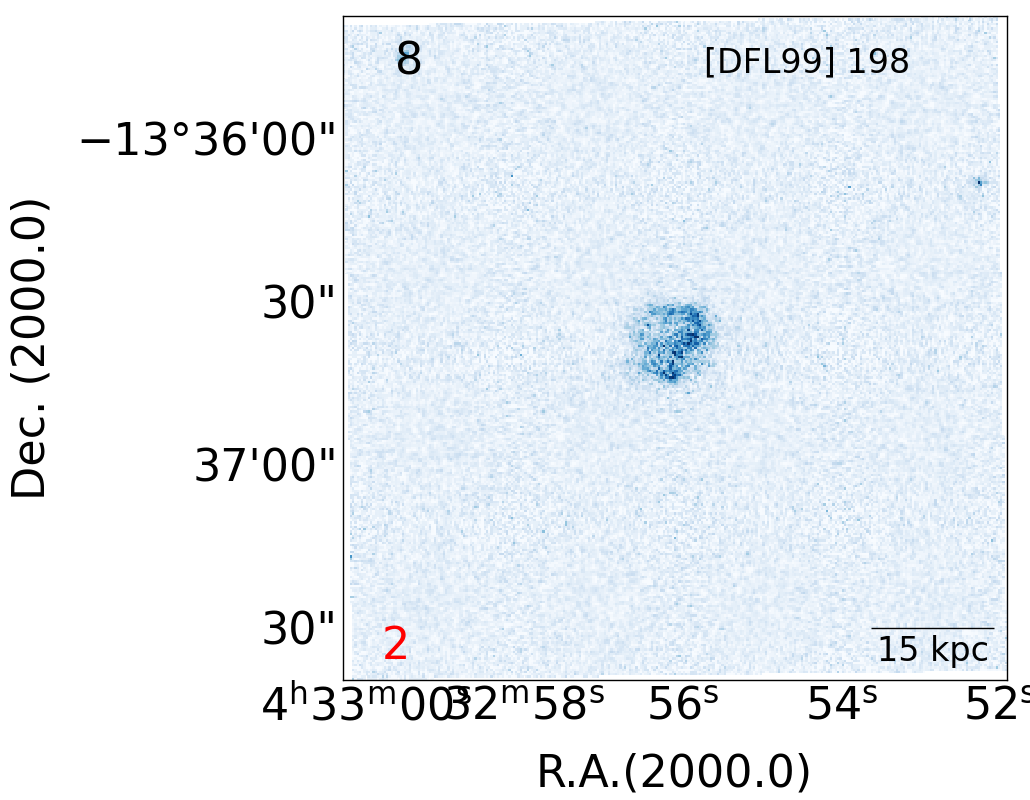}
  \end{subfigure}
   \begin{subfigure}[tbp]{0.23\textwidth}
 \includegraphics[width=\columnwidth]{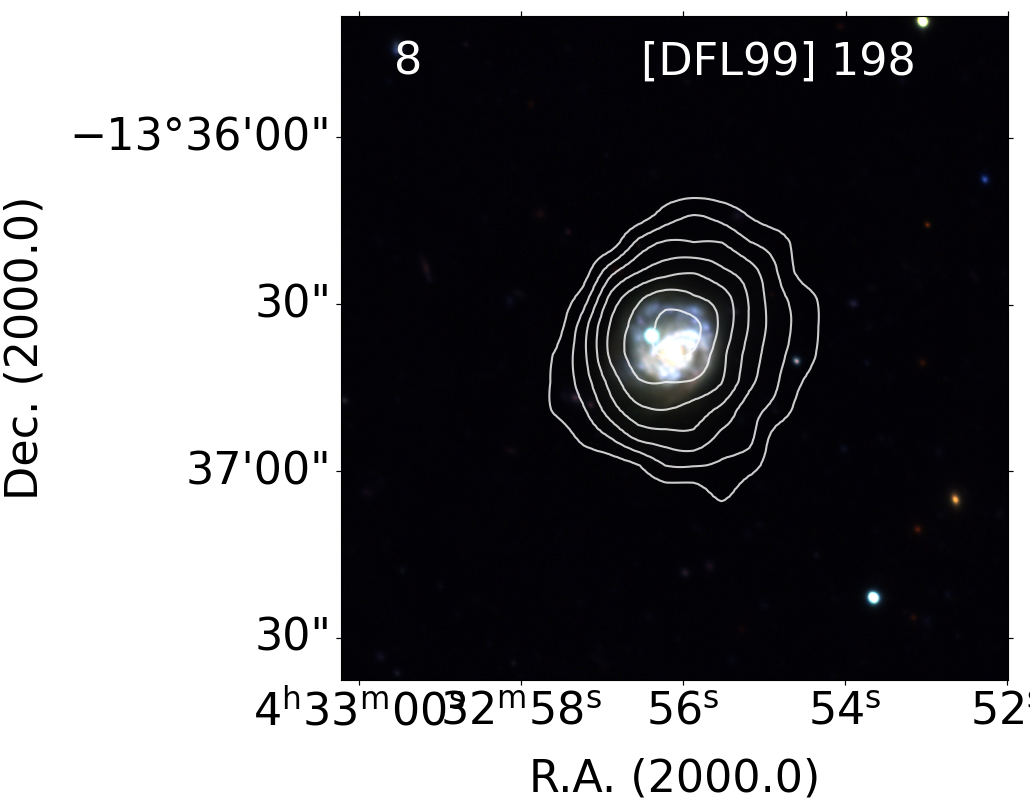}
  \end{subfigure}
  \\
  \begin{subfigure}[tbp]{0.23\textwidth}
 \includegraphics[width=\columnwidth]{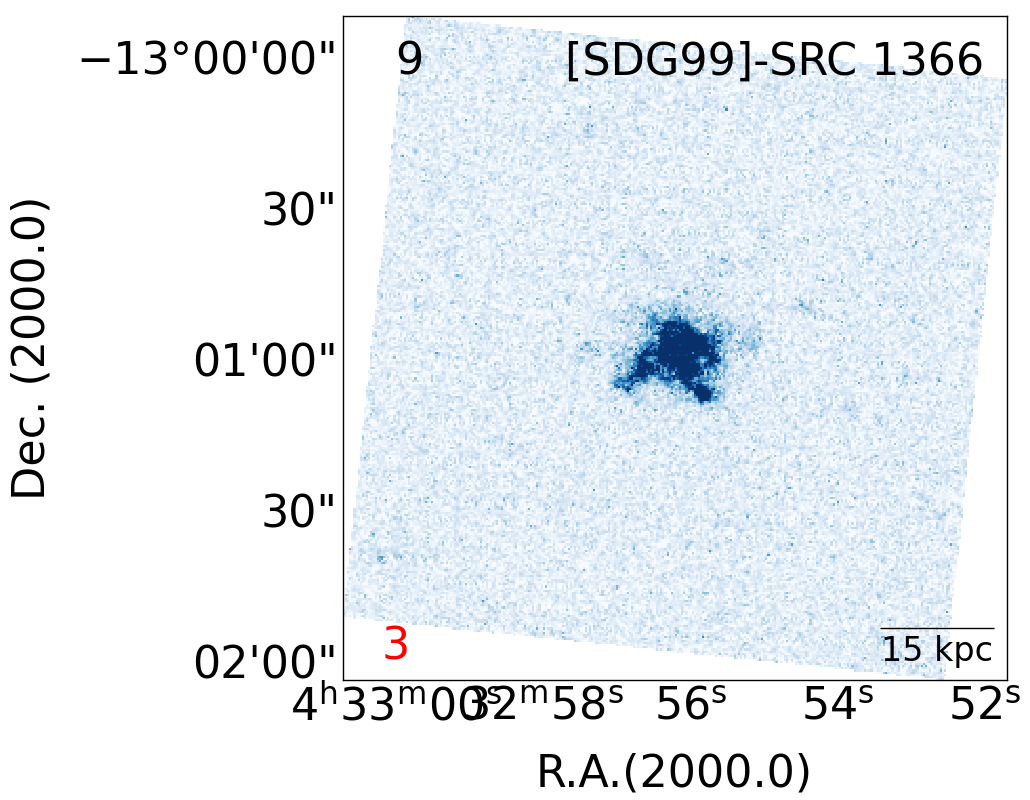}
  \end{subfigure}
   \begin{subfigure}[tbp]{0.23\textwidth}
 \includegraphics[width=\columnwidth]{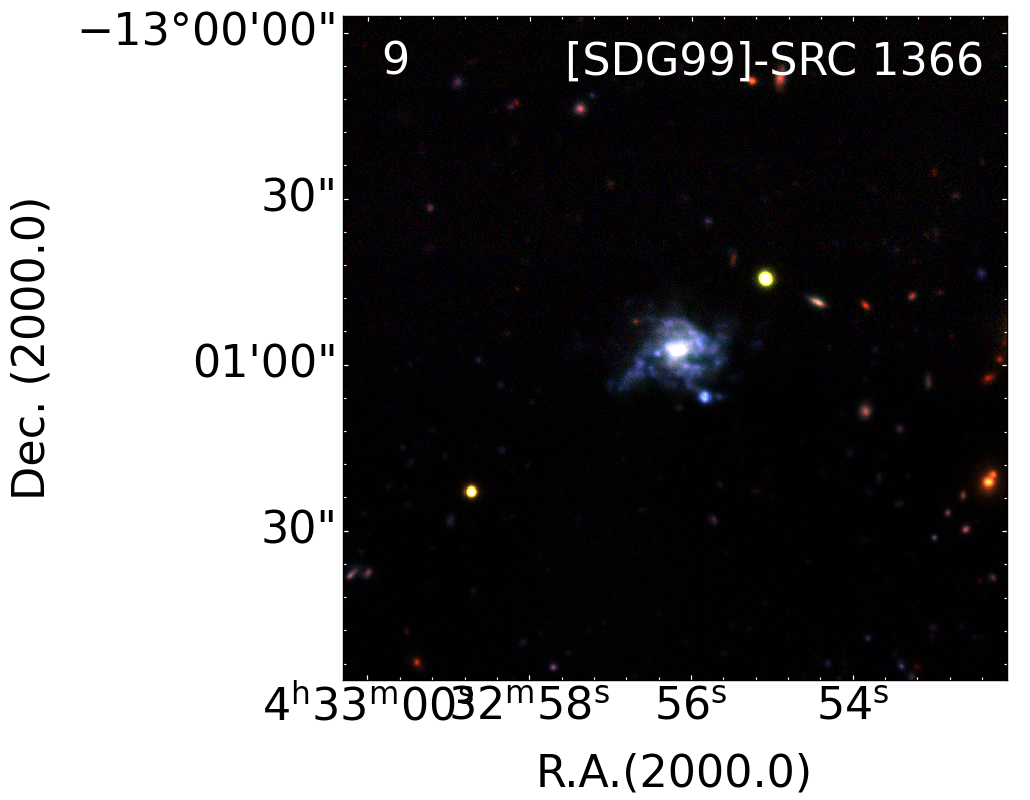}
  \end{subfigure}
  \begin{subfigure}[tbp]{0.23\textwidth}
 \includegraphics[width=\columnwidth]{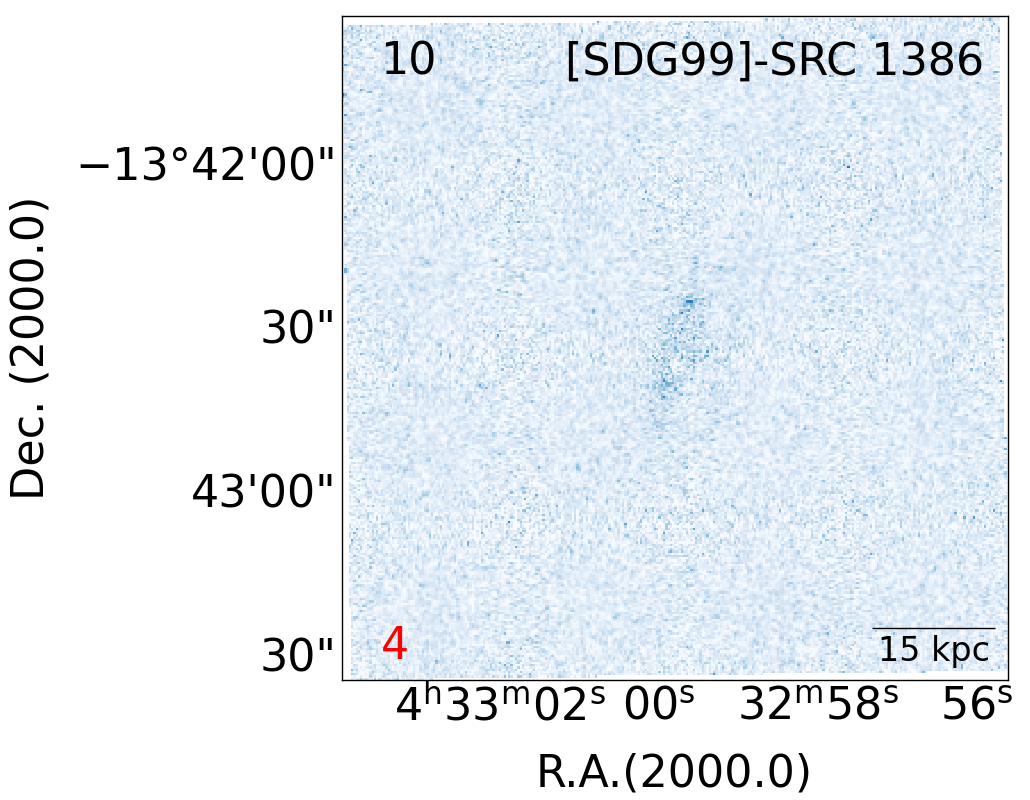}
  \end{subfigure}
   \begin{subfigure}[tbp]{0.23\textwidth}
 \includegraphics[width=\columnwidth]{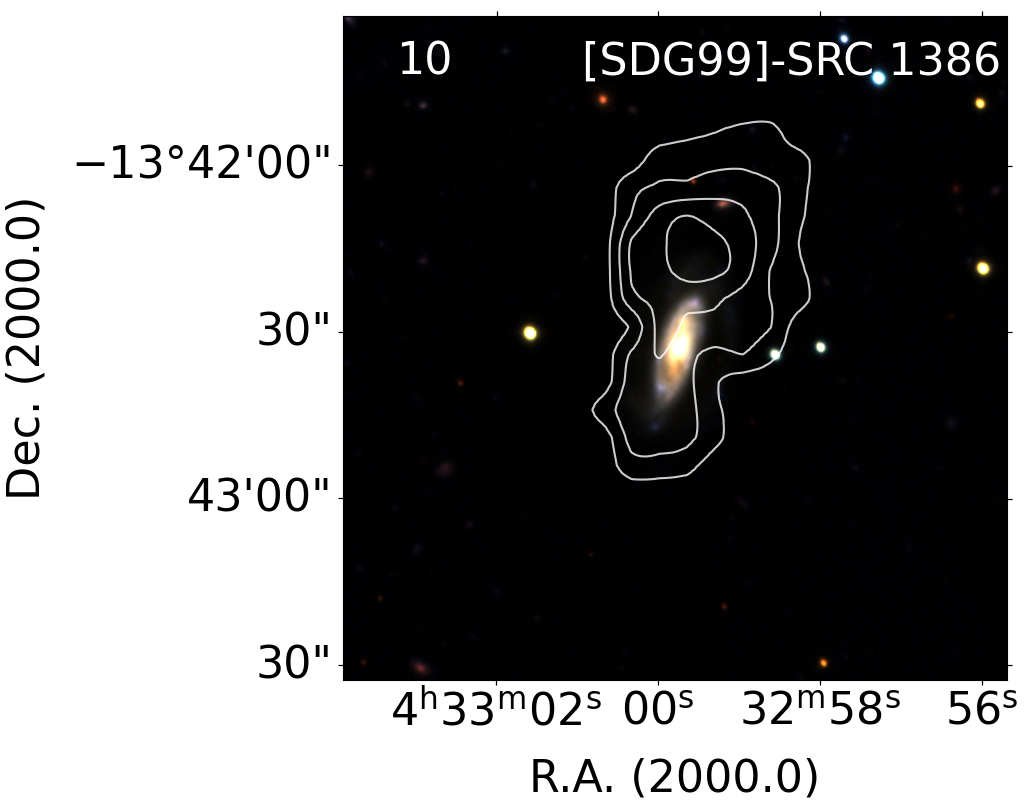}
  \end{subfigure}
  \\
    \begin{subfigure}[tbp]{0.23\textwidth}
 \includegraphics[width=\columnwidth]{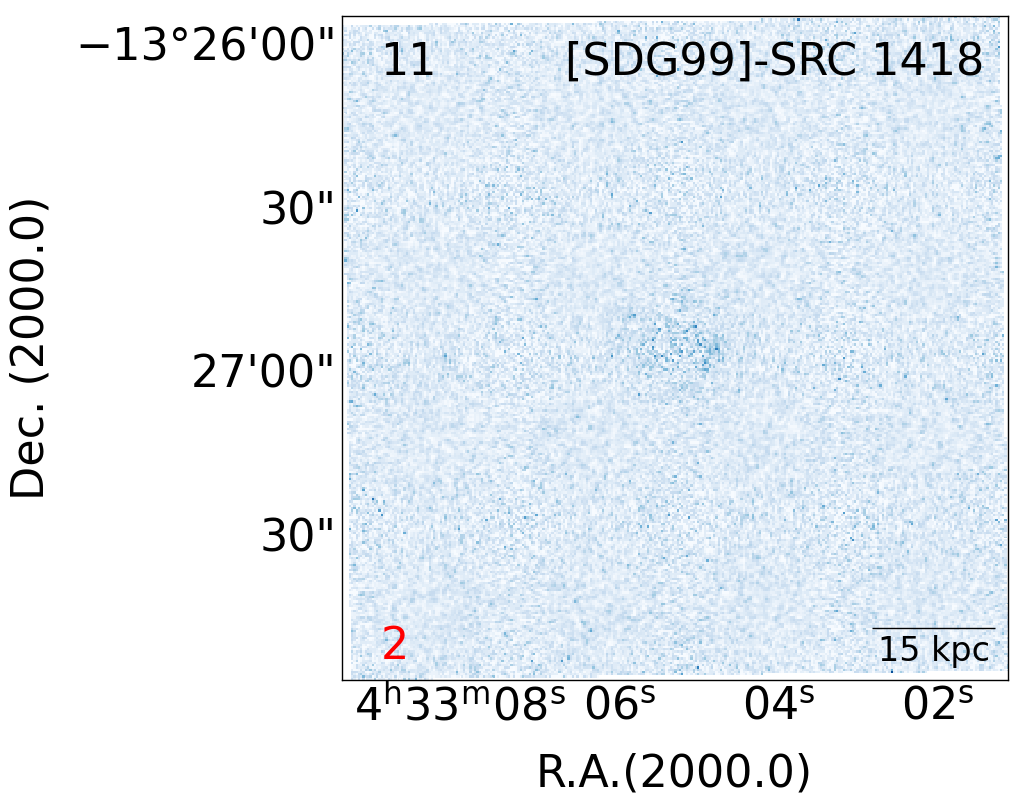}
  \end{subfigure}
   \begin{subfigure}[tbp]{0.23\textwidth}
 \includegraphics[width=\columnwidth]{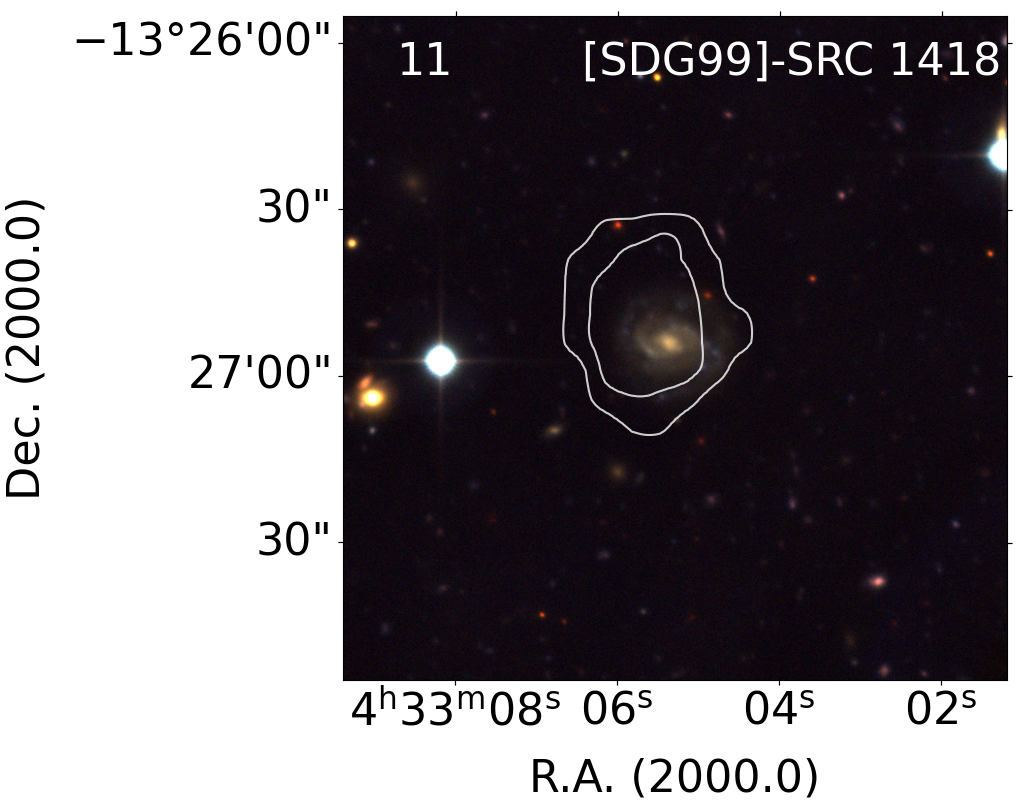}
  \end{subfigure}   
   \begin{subfigure}[tbp]{0.23\textwidth}
 \includegraphics[width=\columnwidth]{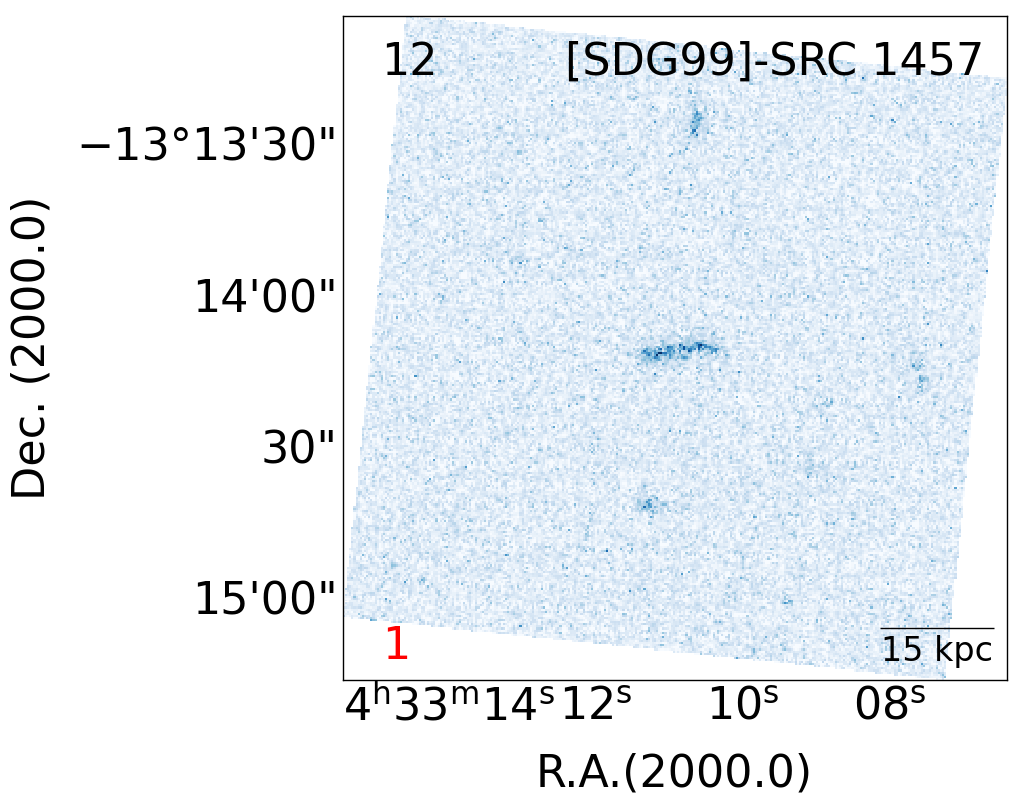}
  \end{subfigure}
  \begin{subfigure}[tbp]{0.23\textwidth}
 \includegraphics[width=\columnwidth]{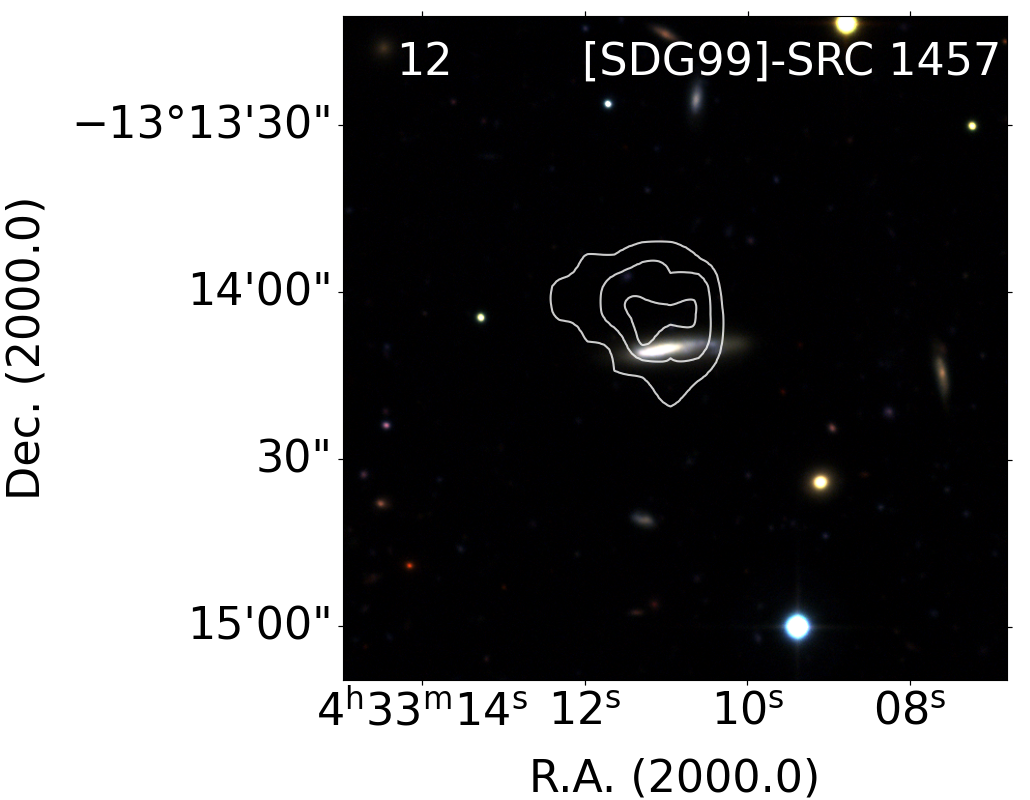}
  \end{subfigure}
  \\
  \hspace{0.5cm}
  \raggedright
   \begin{subfigure}[tbp]{0.23\textwidth}
 \includegraphics[width=\columnwidth]{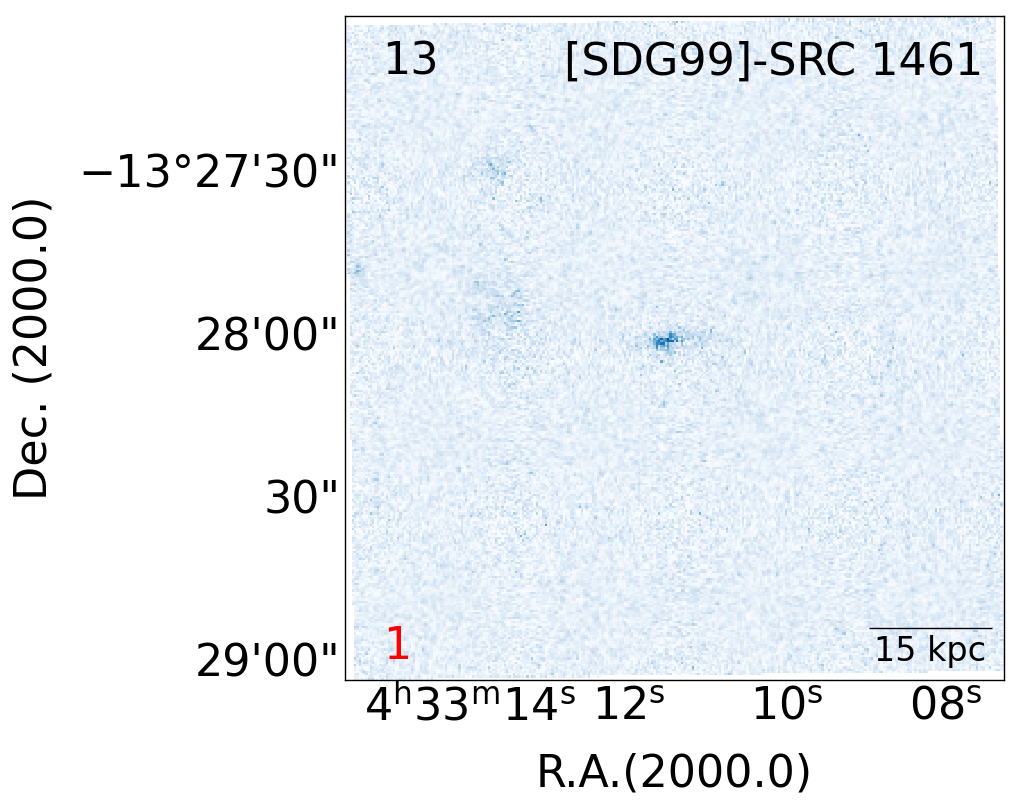}
 \end{subfigure}
  \begin{subfigure}[tbp]{0.23\textwidth}
 \includegraphics[width=\columnwidth]{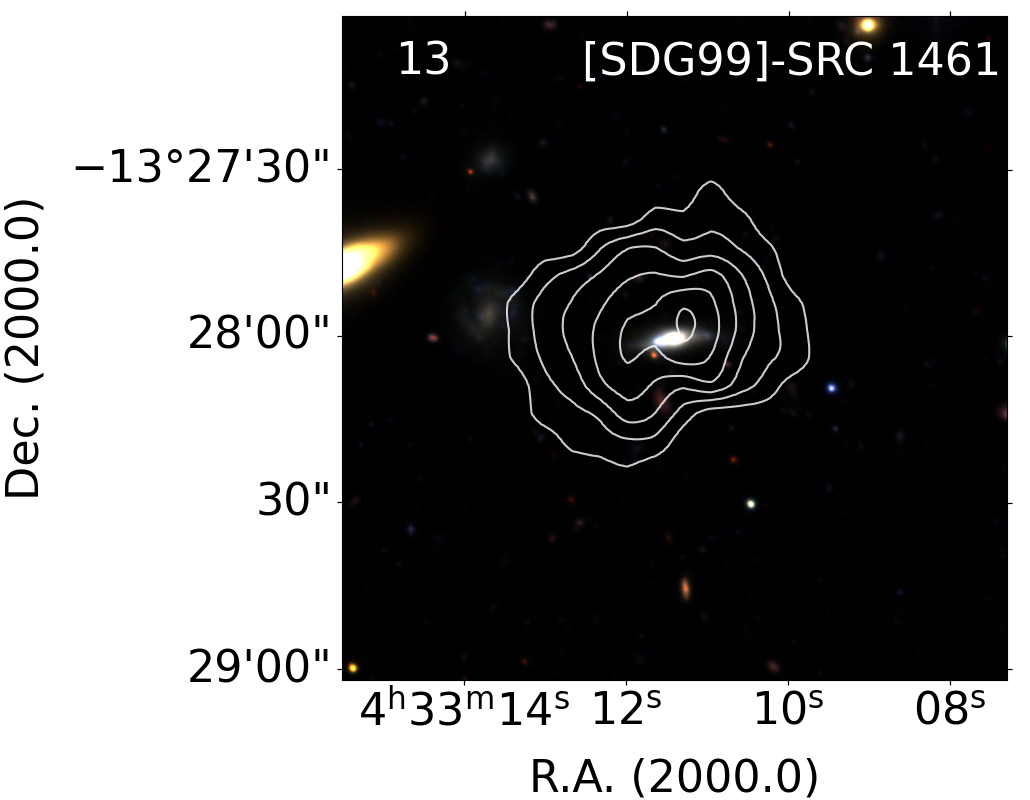}
 \end{subfigure}
 \begin{subfigure}[tbp]{0.23\textwidth}
 \includegraphics[width=\columnwidth]{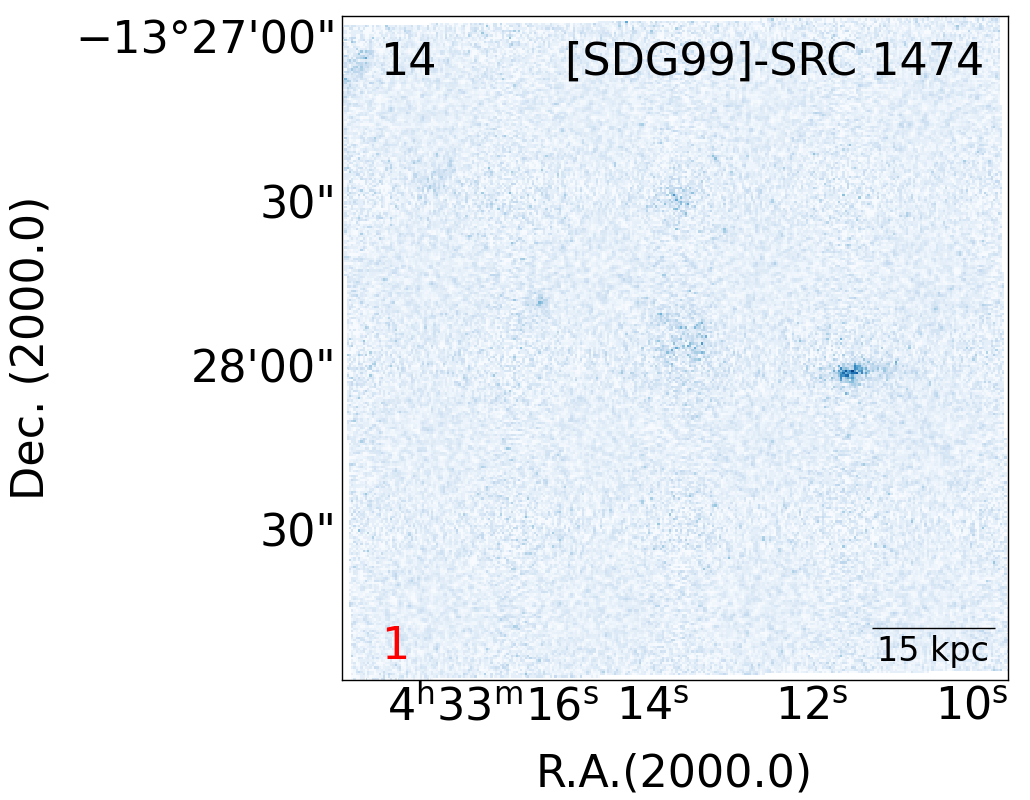}
  \end{subfigure}
  \begin{subfigure}[tbp]{0.23\textwidth}
 \includegraphics[width=\columnwidth]{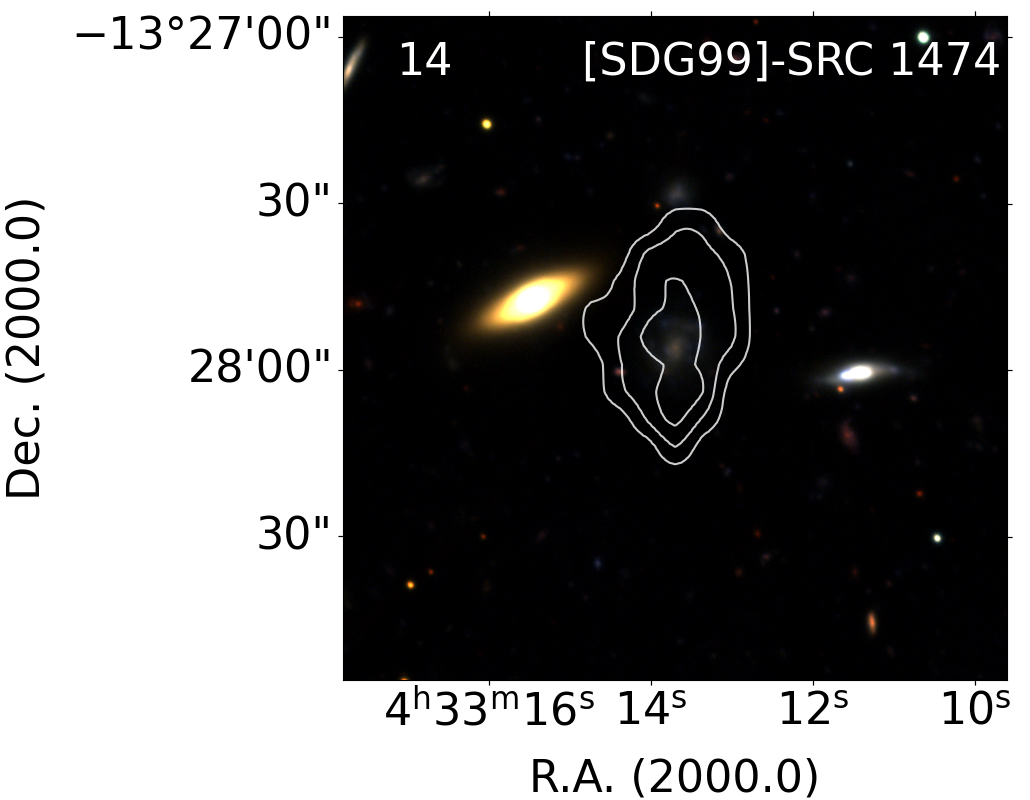}
  \end{subfigure}
\contcaption{}
%\label{Fig_FUV_HI_images}
\end{figure*}

 \begin{figure*}\ContinuedFloat
 \centering
   \begin{subfigure}[tbp]{0.23\textwidth}
 \includegraphics[width=\columnwidth]{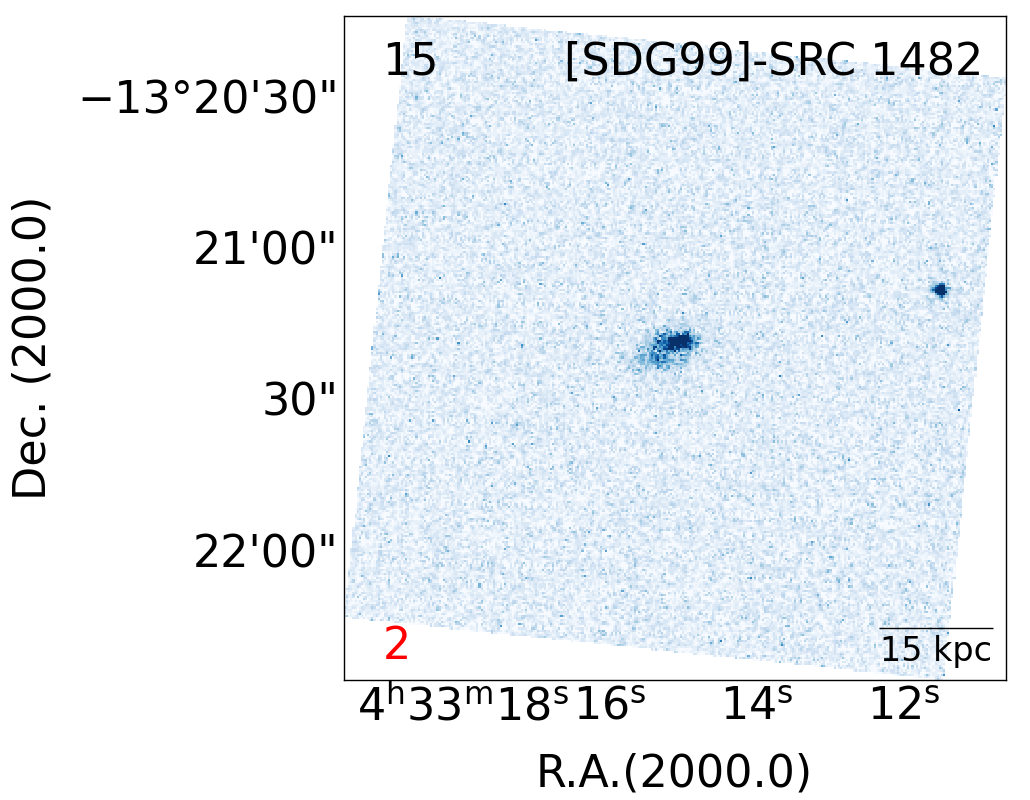}
  \end{subfigure}
   \begin{subfigure}[tbp]{0.23\textwidth}
 \includegraphics[width=\columnwidth]{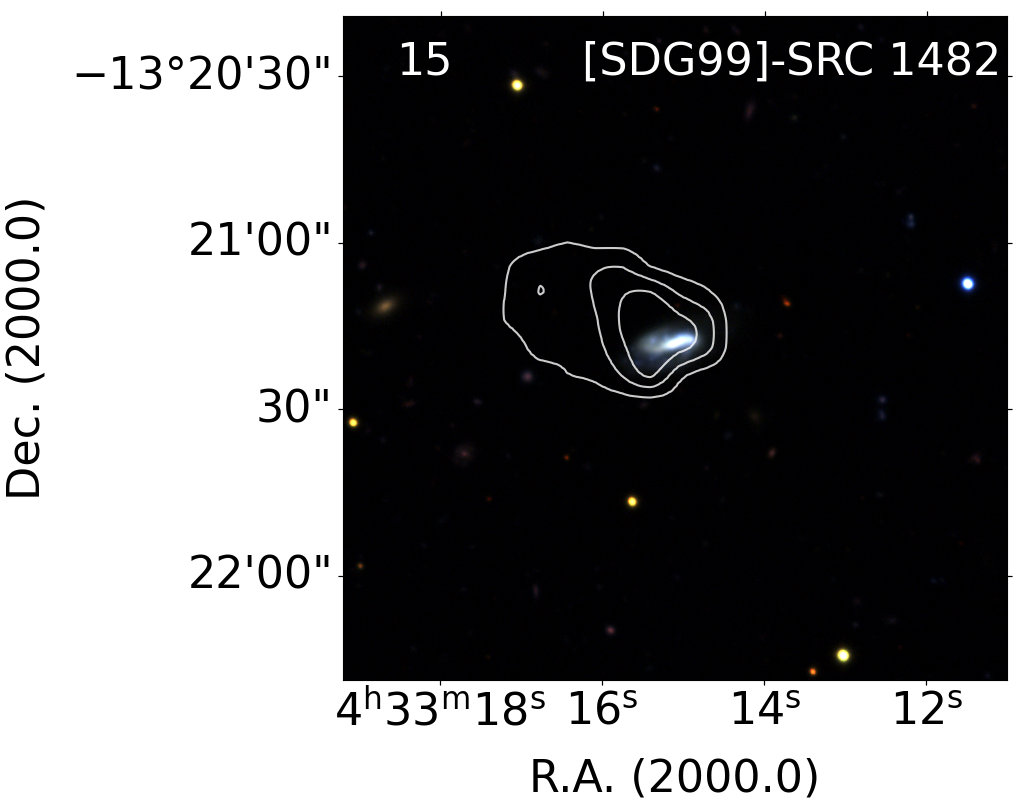}
  \end{subfigure}
  \begin{subfigure}[tbp]{0.23\textwidth}
 \includegraphics[width=\columnwidth]{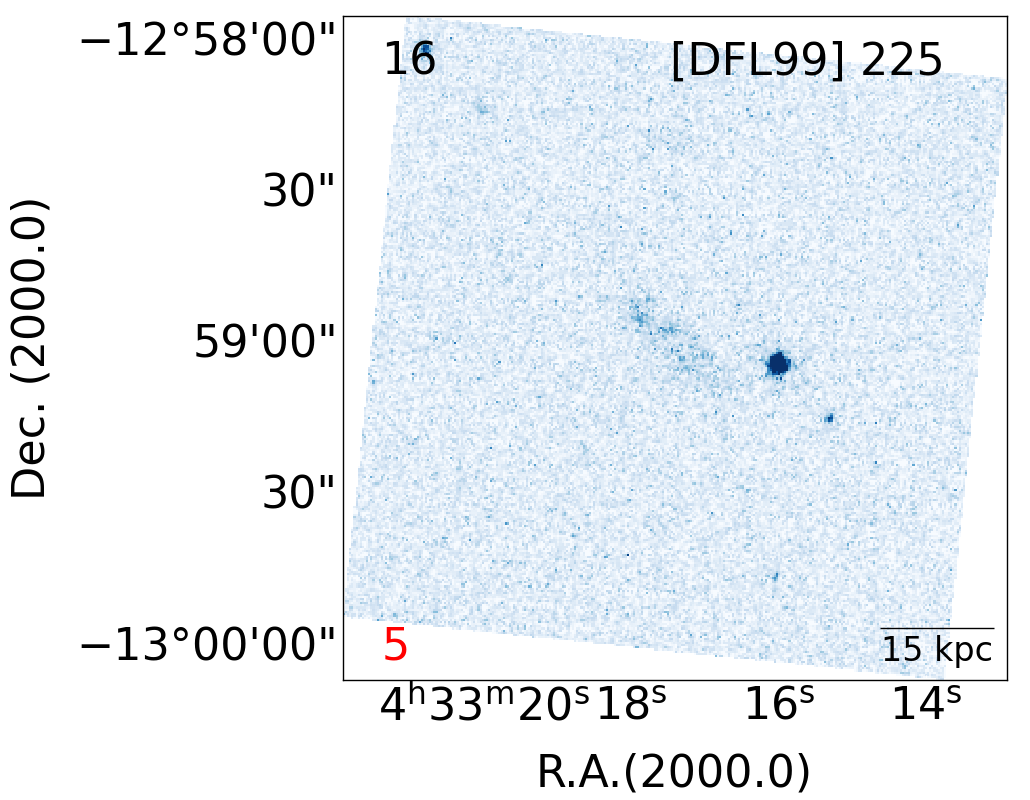}
  \end{subfigure}
  \begin{subfigure}[tbp]{0.23\textwidth}
 \includegraphics[width=\columnwidth]{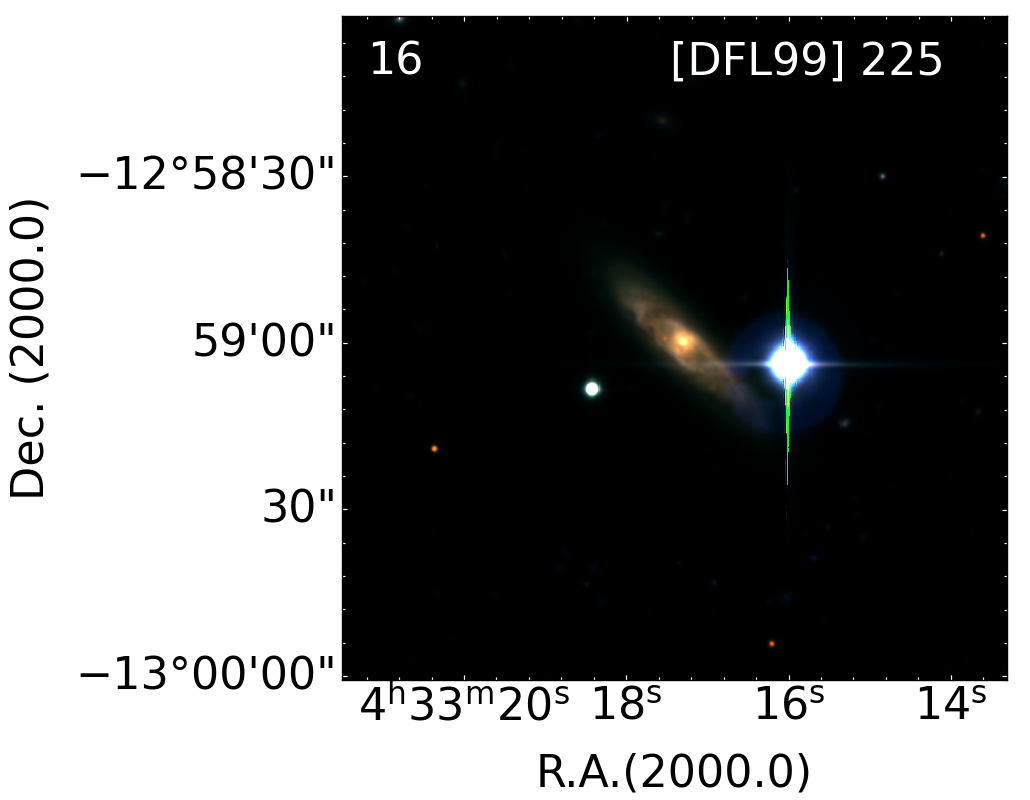}
 \end{subfigure}
 ¸\\
   \begin{subfigure}[tbp]{0.23\textwidth}
 \includegraphics[width=\columnwidth]{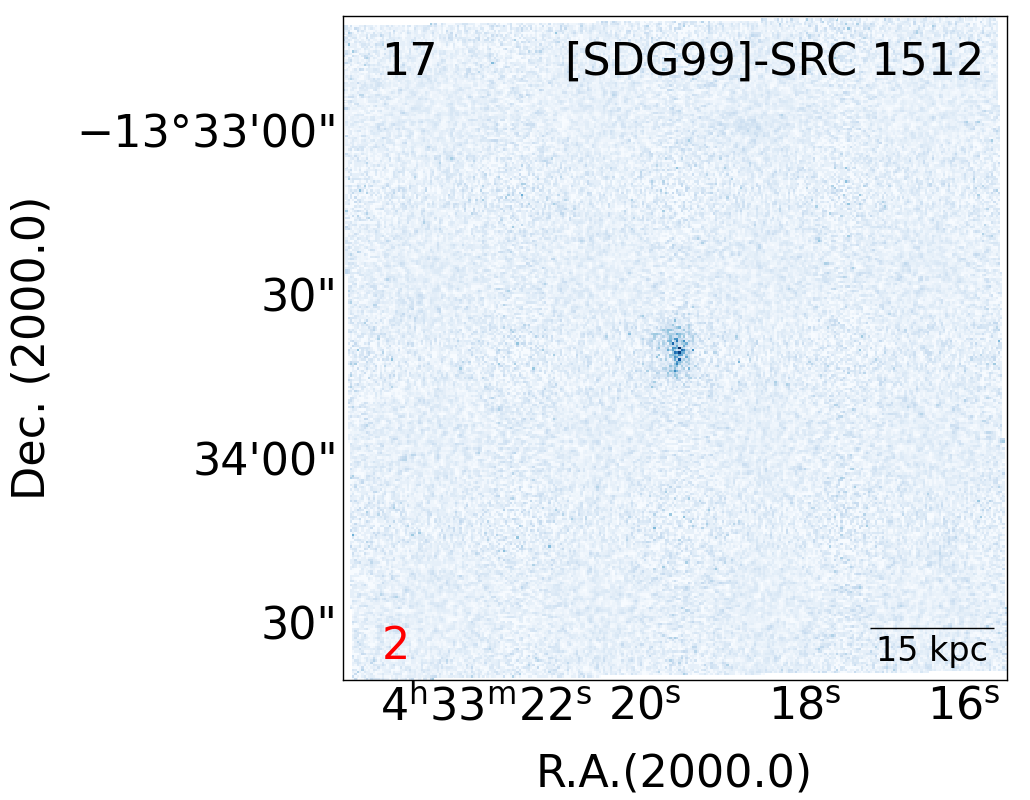}
  \end{subfigure}
  \begin{subfigure}[tbp]{0.23\textwidth}
 \includegraphics[width=\columnwidth]{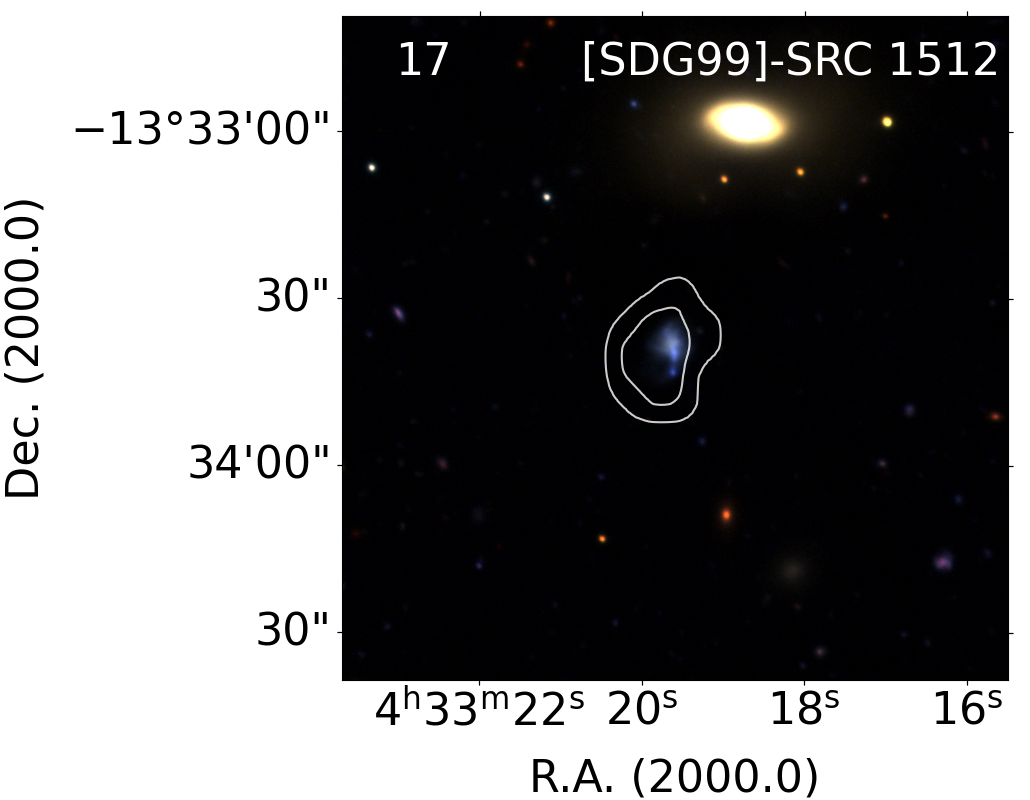}
  \end{subfigure}
   \begin{subfigure}[tbp]{0.23\textwidth}
 \includegraphics[width=\columnwidth]{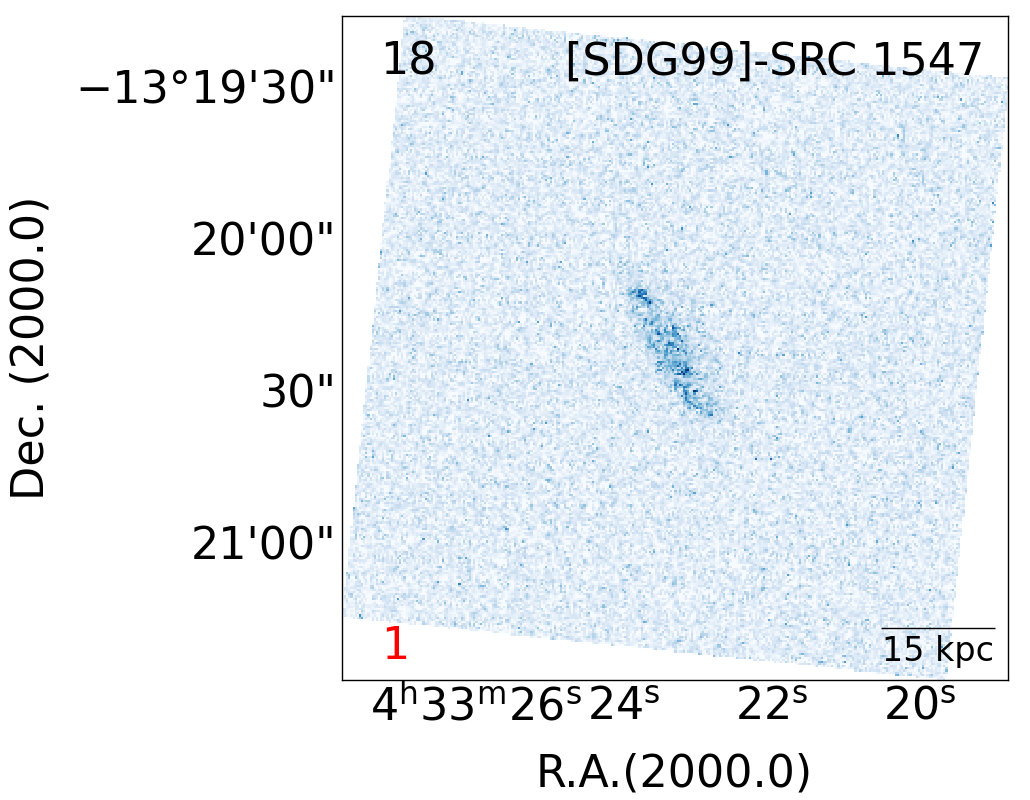}
  \end{subfigure}
   \begin{subfigure}[tbp]{0.23\textwidth}
 \includegraphics[width=\columnwidth]{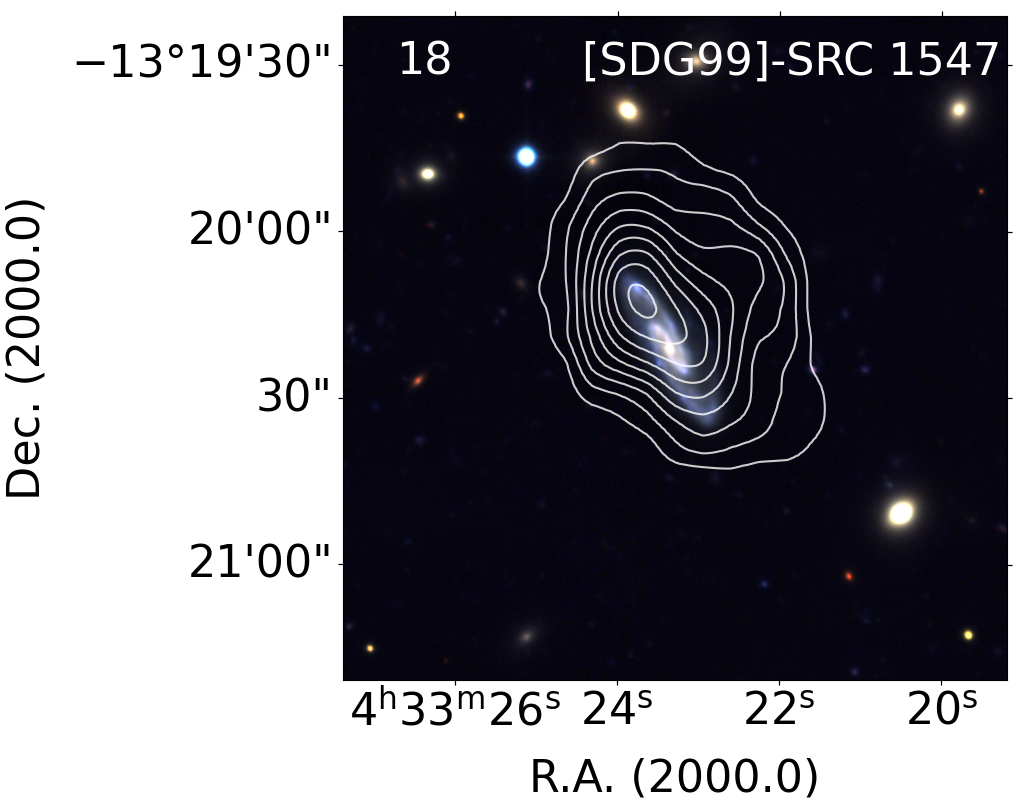}
  \end{subfigure}
  \\
     \begin{subfigure}[tbp]{0.23\textwidth}
 \includegraphics[width=\columnwidth]{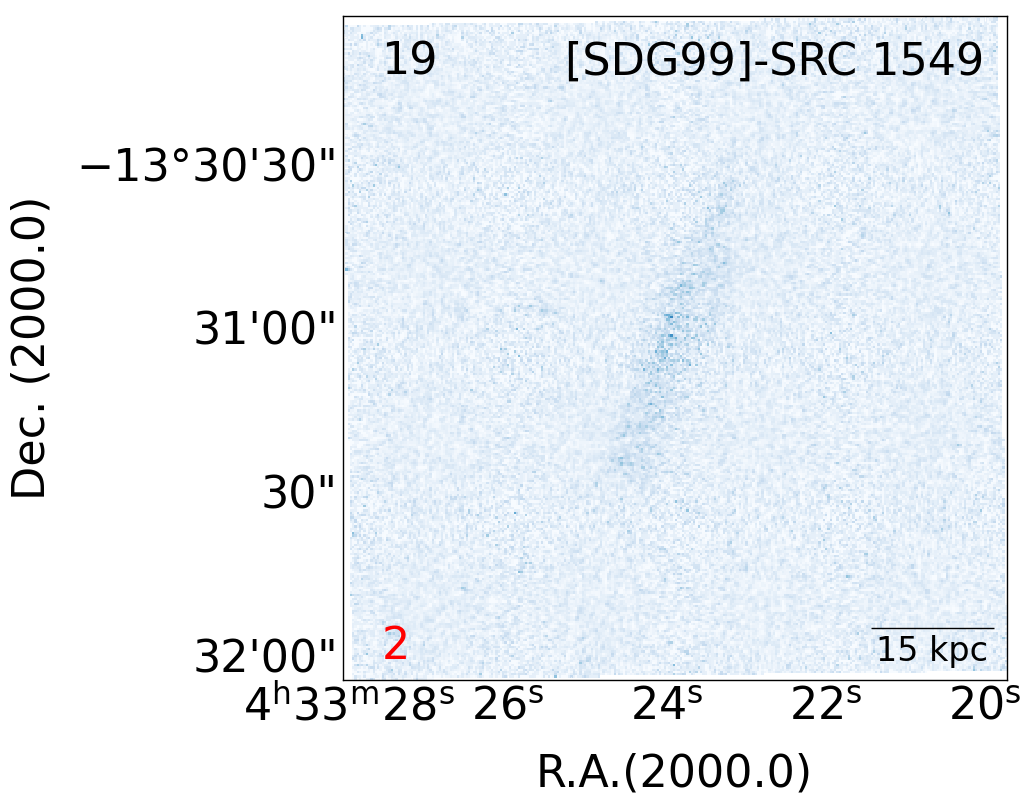}
  \end{subfigure}
   \begin{subfigure}[tbp]{0.23\textwidth}
 \includegraphics[width=\columnwidth]{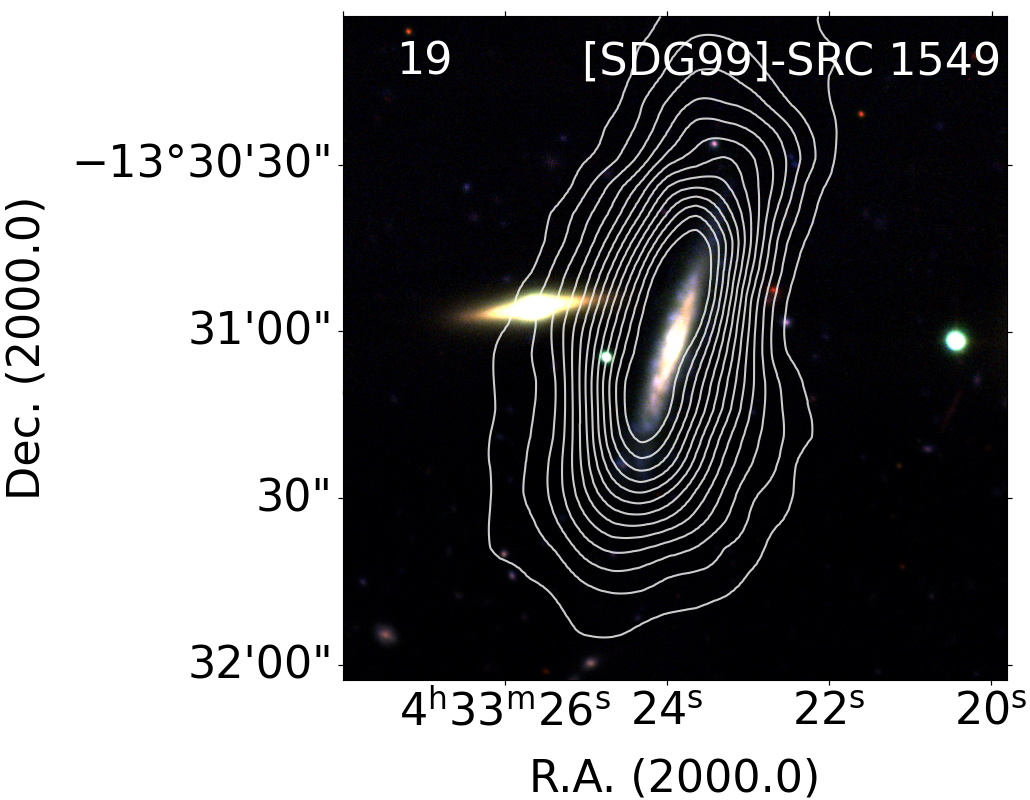}
  \end{subfigure}
   \begin{subfigure}[tbp]{0.23\textwidth}
 \includegraphics[width=\columnwidth]{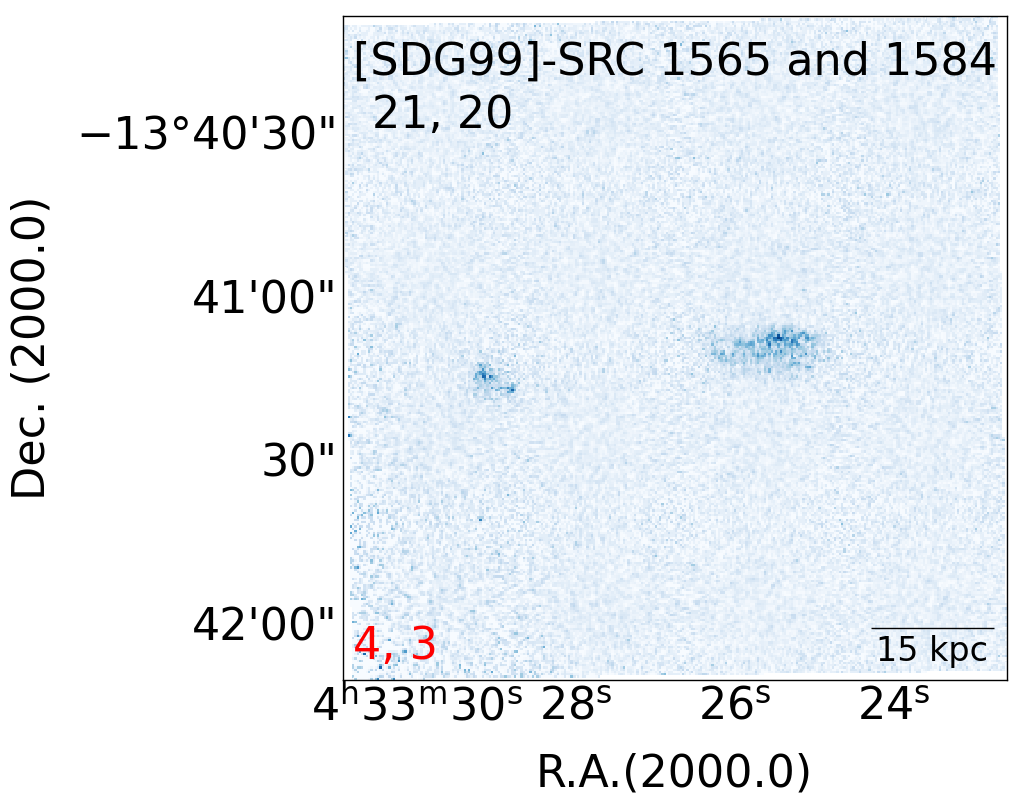}
  \end{subfigure}
  \begin{subfigure}[tbp]{0.23\textwidth}
 \includegraphics[width=\columnwidth]{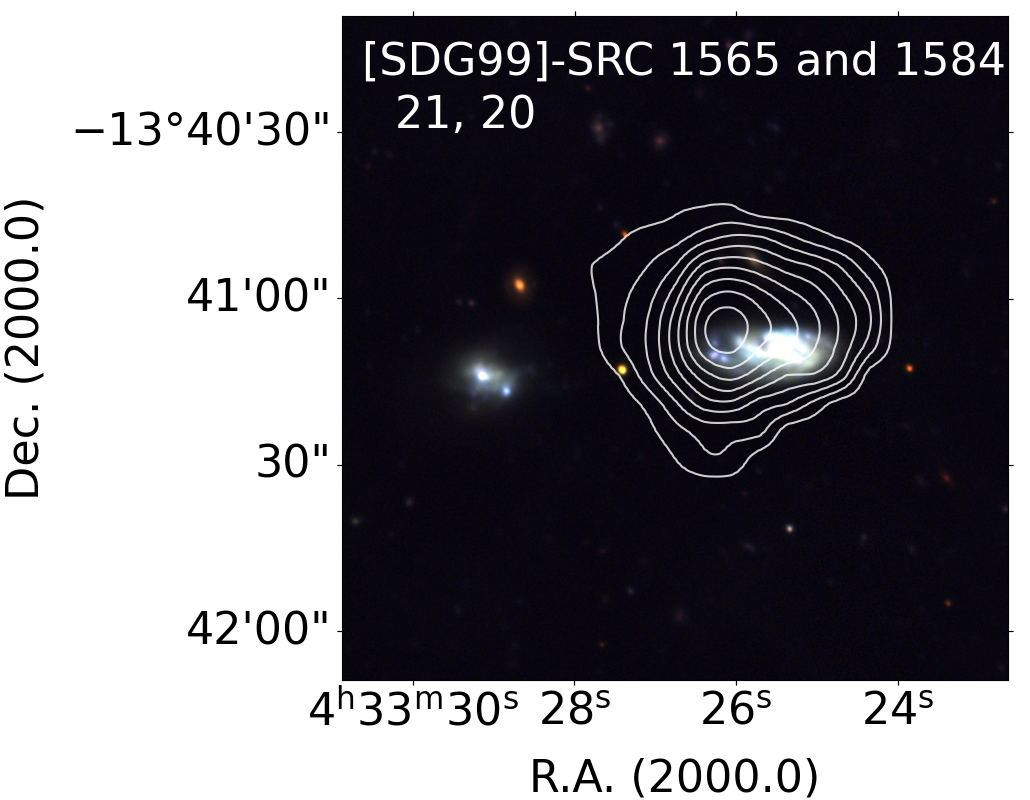}
  \end{subfigure}
  \\
  \hspace{0.5cm}
  \raggedright
   \begin{subfigure}[tbp]{0.23\textwidth}
 \includegraphics[width=\columnwidth]{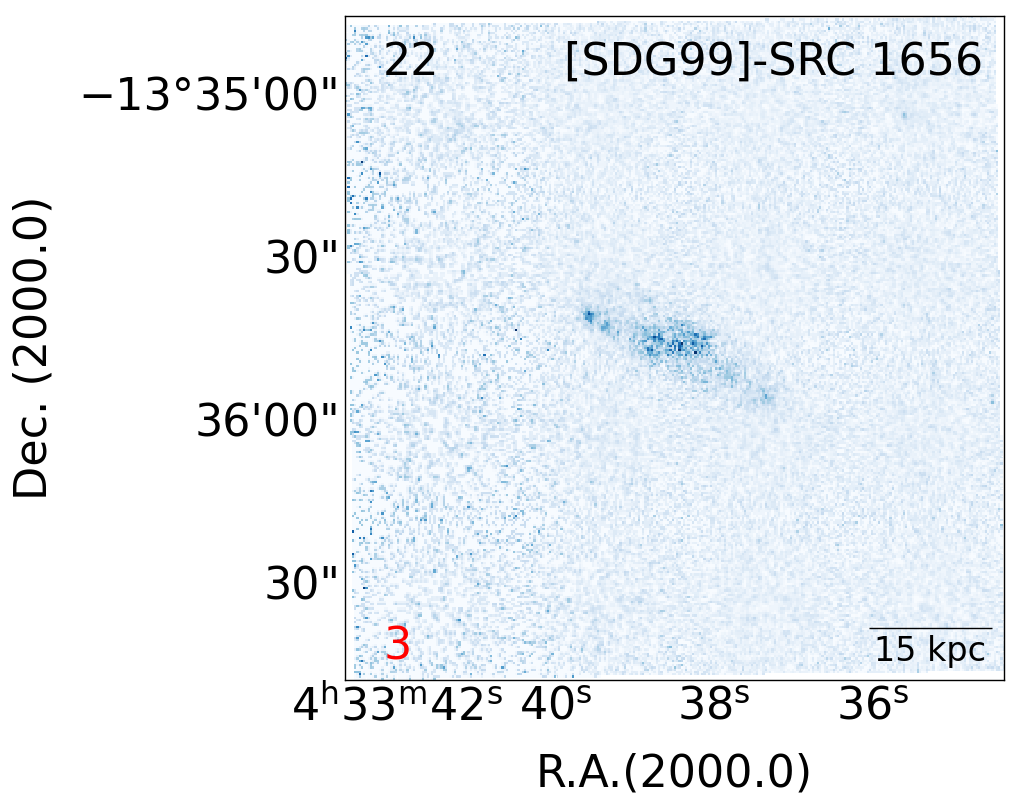}
  \end{subfigure}
  \begin{subfigure}[tbp]{0.23\textwidth}
 \includegraphics[width=\columnwidth]{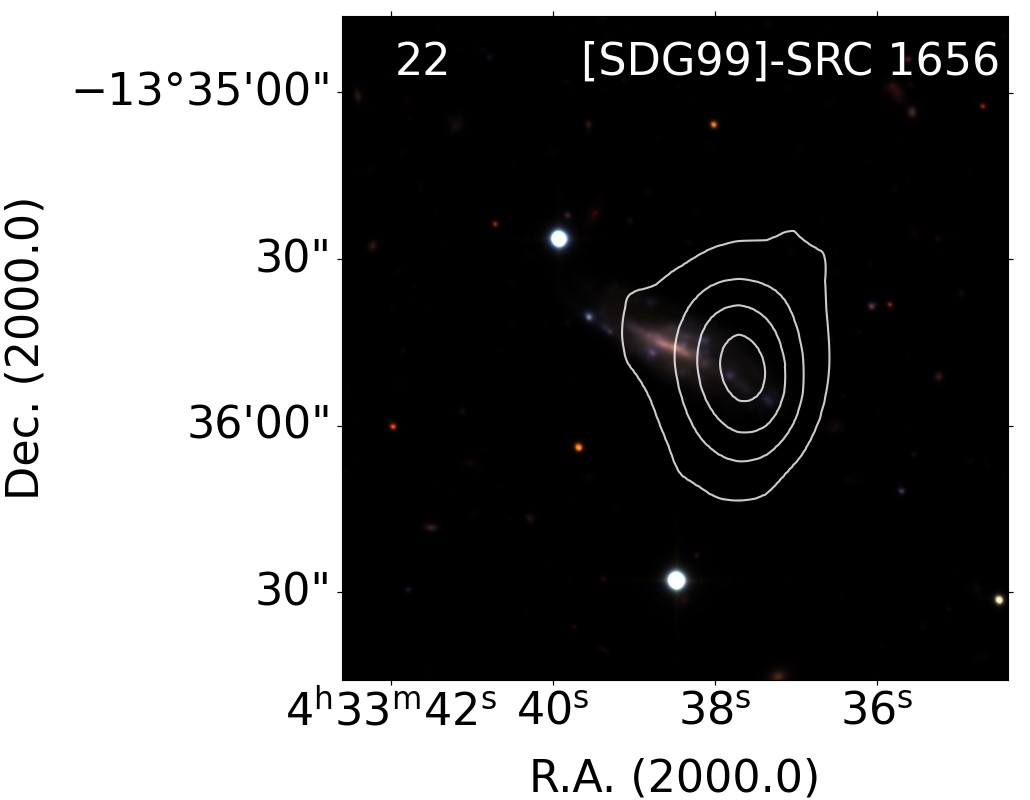}
  \end{subfigure}
  \caption{The UVIT-FUV images (left panels) and the \hi-maps (right panel) overlaid on optical frames for the 22 detected galaxies in A496. The galaxy names are displayed in the upper-right corner, the galaxy IDs appear in the upper-left corner, and the evolutionary stage number is indicated in red in the bottom-left corner as shown in Tables\,\ref{tab_A496_FUV_galaxies} and \ref{tab_A496_galaxies_FUV_fields}. The white contours show the \hi\ distribution, spaced by 2.5$\,\cdot \,rms$, having a spatial resolution of 24\prin \por\,17\prin. Fields are 2\por2 arcmin$^2$. The \hi-maps for [SDG99]-SRC\,1366, [DFL99]\,225 and [SDG99]-SRC\,1584 could not be traced due to their very faint \hi\ emission.}
 \label{Fig_FUV_HI_images}
\end{figure*}

\bsp	% typesetting comment
\label{lastpage}
\end{document}